\documentclass{article}

\usepackage{amssymb}

\usepackage[authoryear]{natbib}
\usepackage{xurl}
\usepackage{graphicx}
\usepackage[export]{adjustbox}
\usepackage{authblk}
\usepackage{booktabs}
\usepackage{dcolumn}
\usepackage{scalerel}
\usepackage{amsmath}
\usepackage{amsthm}
\usepackage{multicol}
\usepackage{algorithm}
\usepackage{algorithmic}
\usepackage{caption}
\usepackage{subcaption}
\DeclareCaptionLabelFormat{andtable}{#1~#2  \&  \tablename~\thetable}

\usepackage{float}
\restylefloat{figure}

\newtheorem{definition}{Definition}[section]
\newtheorem{proposition}{}[section]

\begin{document}

\title{Ridge regression with adaptive additive rectangles and other piecewise functional templates}

\author{Edoardo Belli \\
        edoardo.belli@polimi.it \and Simone Vantini \\ 
                                     simone.vantini@polimi.it}

\affil{MOX - Modeling and Scientific Computing, Department of Mathematics, Politecnico di Milano, Italy}

\date{}

\maketitle

\begin{abstract}
We propose an $L_{2}$-based penalization algorithm for functional linear regression models, where the coefficient function is shrunk towards a data-driven shape template $\gamma$, which is constrained to belong to a class of piecewise functions by restricting its basis expansion. In particular, we focus on the case where $\gamma$ can be expressed as a sum of $q$ rectangles that are adaptively positioned with respect to the regression error. As the problem of finding the optimal knot placement of a piecewise function is nonconvex, the proposed parametrization allows to reduce the number of variables in the global optimization scheme, resulting in a fitting algorithm that alternates between approximating a suitable template and solving a convex ridge-like problem. The predictive power and interpretability of our method is shown on multiple simulations and two real world case studies.
\end{abstract}

\section{Introduction}  
As the the dimensionality of data sets kept increasing in the last decades, it is no surprise that variable selection in linear models is still a major topic of discussion between statistical and machine learning researchers. The main objective of these methods is to reduce prediction error by decreasing the variance of the predictions, at the expense of an increase in bias, while at the same time allowing for an interpretable model by discarding and/or shrinking the least relevant variables. Arguably the most intuitive approach is best subset selection, where between the $p$ variables, we retain only the subset of size $k<p$ that minimizes the error. Despite that, the combinatorial nature of these techniques is prone to higher variability and computational cost if compared to the shrinkage approach \citep{book_esl}, which is usually based on a penalized least squares formulation. The fact that ridge regression \citep{ridge} uniformly shrinks all the regression coefficients often leads to nonsparse solutions that do not have a clear interpretation, even when the prediction error is low. By penalizing the $\ell_1$ norm of the coefficients' vector, the lasso \citep{lasso} can be seen as a convex relaxation of best subset selection, and is able to produce sparse models often without sacrificing the predictive performance. There is a vast literature that encompasses multiple aspects of these types of methods, including nonconvex penalties like the SCAD \citep{scad} and MCP \citep{mcp}, refer to \cite{book_sls} for an extensive view on sparsity in multivariate linear models. 

In the context of functional data analysis (FDA) \citep{book_fda,book_nonparafda}, the concept of sparsity has multiple connotations \citep{chapter_fdasparsity} and here we will refer to the sparsity of the regression function $\beta$, which selects the subsets of the domain where the predictors have no effect on the response. From the theoretical point of view, the scalar on function linear model is ill-posed and there is the need to introduce some form of regularization as an identifiability constraint. The the two main approaches involve either restricting the number of basis functions for the expansion of $\beta$, like in the case of designer bases and functional principal components \citep{fda_pred}, or fixing a rich enough basis while applying a suitable penalization on $\beta$, usually an $L_2$ penalty on its $m$-th derivative to impose smoothness \citep{fda_splineflm,fda_splineerrors,fda_smoothsplines,fda_rkhs} as well as sparsity inducing ones like the group SCAD \citep{fda_L1reg}. Hybrid approaches have also been proposed, where the number of basis functions is adaptively chosen while an additional penalty is included \citep{psplinesgenreg,fda_sparseest}, or sparsity is imposed on the $m$-th derivatives of $\beta$ \citep{fda_flirti}. In this work we will adopt the penalization approach and in particular we express $\beta$ as a finite expansion on a simple grid basis with $p$ dense and equispaced knots on the input data:

\begin{equation*}
  \beta(t) = \sum_{j=1}^{p} \beta_{j} b_{j}(t)  \hspace{1.5cm}
  b_{j}(t) = 
  \begin{cases}
  1 & \hspace{0.1cm} \text{if} \hspace{0.2cm} \frac{j-1}{p} < t \leq \frac{j}{p} \\
  0 & \hspace{0.1cm} \text{otherwise} 
  \end{cases}
\end{equation*}

which is a common choice that allows to use any multivariate method to compute the coefficients $\beta_{j}$ of the expansion and consequently recover $\beta$. While at glance this may seem a very constrained setting, as the functional covariates could have been sampled on different grids with noise and missing values, a standard practice is to estimate the true functions individually by means of interpolation or smoothing, which are then evaluated on the same dense and equispaced grid to obtain the actual discrete data set. This estimation process is adequate when the raw data has been collected with high resolution techniques, as in chemometrics, environmental or electrical engineering applications. If however the raw data contains few and exceptionally sparse observations over $t$, as in the case of longitudinal studies, the previous estimation approach fails as there is not enough information to recover the single function individually. A solution to this problem is to use functional principal components with mixed effects \citep{fda_smoothsplnested,fda_pcasparse} or local smoothing \citep{fda_sparselongdata} in order to jointly leverage the information of the full raw data set for the estimation of each functional covariate. 

In this work, we are interested in those penalization approaches that beyond pure sparsity, allow to impose some degree of structure to the regression coefficients/function, like joint sparsity and smoothness as in the case of the elastic net \citep{elasticnet} and other $\ell_1 / \ell_2$ combined approaches \citep{smoothlasso}, imposing an order between the coefficients \citep{orderedlasso} and especially the fused lasso \citep{fusedlasso}, which allows to recover a piecewise constant regression function. Although not in the scope of our investigation, starting from the group lasso \citep{grouplasso}, multiple works have dealt with the assumption of a known grouping structure between the covariates, including sparsity and competing variables within the groups \citep{sparsegrouplasso,groupexpolasso,multi_exclusivelasso,reg_exclusivelasso}, overlapping groups \citep{cap_penalty,grouplassooverlap} and supervised clustering-based groups \citep{clusteredlasso}. In principle, our method could be used as a preprocessing step to estimate such grouping structure. It is also worth to note that other non penalty-based approaches force $\beta$ to be convex by selecting the maximum between $K$ hyperplanes, where each hyperplane is obtained by fitting a linear model using only a subset of the observations \citep{convex_pwlin,cap_partit}.
 
From a practical standpoint, the type of penalty can be seen as an hyperparameter that has to be selected, either by cross-validation or by domain knowledge, and the availability of different options allows researchers and engineers to tackle a multitude of $p>>N$ scenarios. However, current methods are not flawless, as in the case of the lasso where the number of non-zero coefficients that can be recovered is at most $N$, which can be a limiting factor and may exclude important variables, especially in the presence of highly collinear functional data, where the irrepresentable condition is often violated  and selection consistency is not guaranteed \citep{varsellassograph,lassoirrepr}. Therefore, the design of new penalties is still an open problem, and our proposal aims to add further modeling flexibility to the user, by allowing a parsimonious and yet adaptive formulation of the shape $\gamma$ that we want to impose on $\beta$, by expressing $\gamma$ as a finite sum of a few user-defined piecewise basis functions like a rectangular pulse. In particular, our method is based on the nonzero centered $L_2$ penalty \citep{ridgeprior,ridgefusion,genridgeinvcov,targetedridge} and uses a global optimization scheme in order to adaptively find a suitable and possibly sparse shape template that will be the target of the shrinkage, with the objective to reduce bias. 

The article is organized as follows: in Section 2 we explain our method together with the optimization details, Section 3 contains multiple simulation studies and two real world applications, with concluding remarks in Section 4.

\section{Adaptive additive piecewise functional templates}
Let $\mathcal{D}=\lbrace (x_i,y_i) \rbrace_{i=1}^{N}$ be the training set with random i.i.d. functions $x_i \in L^{2}(I)$ and responses $y_i \in \mathbb{R}$, we study the scalar on function regression model: 

\begin{equation}
\label{eq:linearmodel}
y_{i} = \beta_{0} + \int_{I} x_i(t)\beta(t)dt + \epsilon_{i}
\end{equation}

\noindent where $\beta_{0} \in \mathbb{R}$ is the intercept and $\epsilon_{i} \sim \mathcal{N}(0,\sigma^{2})$ random i.i.d. errors. Without loss of generality, we will assume that the regressors have been standardized and that $I=[-1,1]$. For a given template function $\gamma:[-1,1] \rightarrow \mathbb{R}$, we estimate the linear model by solving the following problem: 

\begin{equation}
\label{eq:L2}
\underset{\scaleto{\beta_{0},\beta}{6pt}}{\scaleto{min}{7pt}} \:\:\: \sum_{i=1}^{N} \left[ y_{i}  -\beta_{0} - \int_{I}x_{i}(t)\beta(t)dt \right]^{2} + \lambda \int_{I} \big[\beta(t) -\gamma(t)\big]^{2}dt
\end{equation}

\noindent which is an $L_2$-penalized least squares with $\beta$ shrunk towards $\gamma$ instead of zero, and $\lambda>0$ the hyperparameter that controls the amount of shrinkage. While the template function $\gamma$ can be directly provided by the user, in this work we propose an adaptive global optimization approach for the choice of $\gamma$, which we restrict to belong to the following class of functions $\Gamma$:

\begin{multicols}{2}	
  \begin{equation*}
  \Gamma = \left\{ \gamma \! :[-1,1] \rightarrow \mathbb{R} \; \Bigg| \; \gamma(t)=\sum_{j=1}^{q} A_{j}g_{j}\left(t,t_{0j},T_{j}\right) \right\}
  \end{equation*} 
  \\
  \begin{align*}
   & q \in \mathbb{N}^{+} \\
   & A_{j} \in \mathbb{R} \\
   & t_{0j} \in [-1,1] \\
   & T_{j} \in (0,2] \\
   \hspace{3cm}
  \end{align*}
\end{multicols}

\noindent where each of the $g_{j}\left(t_{0j},T_{j}\right):[-1,1] \rightarrow \mathbb{R}$ can be an arbitrary but known piecewise function such that the knots that define the intervals depend on $t_{0j}$ and $T_{j}$. This means that $\gamma$ can be expressed as a finite expansion with piecewise basis functions that are chosen adaptively. In fact, $g_{j}$ could be any user-defined function parametrized by $t_{0j}$ and $T_{j}$, but we will focus on the case where $g_{j}=g_{i}$ for $j\neq i$, dropping the subscript for ease of notation, with $g$ being a rectangular function:

\begin{equation*}
g\left(t,t_{0},T\right) = 
\begin{cases}
1 & \: \text{if} \: \left|\frac{t-t_{0}}{T}\right| \leq \frac{1}{2} \\
0 & \: \text{otherwise} 
\end{cases}
\hspace{1cm}
\includegraphics[valign=c,width=0.45\textwidth,trim={0.7cm 0cm 1.58cm 1.43cm},clip]{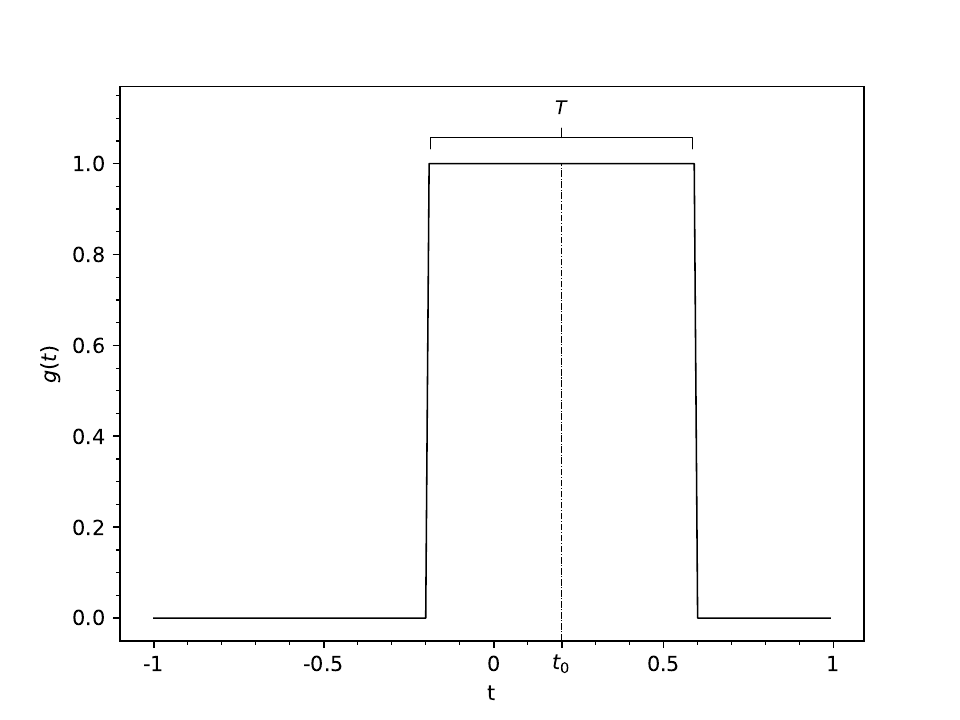}
\end{equation*}

\noindent The choice of the rectangular shape allows correlated parts of the domain to be equally included in the model and has a straightforward interpretation. While this template based approach could appear to be too rigid, expressing $\gamma$ as a sum of $q$ adaptively positioned rectangles seems to provide the flexibility needed to avoid sacrificing prediction error, maintaining at the same time a sound regularization effect. Moreover, once the expansion of $\gamma$ is optimized, the impact on $\beta$ is further controlled by selecting $\lambda$ by cross-validation, as the regression function is not expanded with respect to the same restricted basis, but instead uses the rich enough grid basis. This is an important aspect of our method, as optimizing $\gamma$ is equivalent to the problem of selecting the optimal number and position of the knots in free-knot splines, which is known to be nonconvex \citep{goodapproxfreeknot} even for univariate functions. Multiple heuristics have been proposed to tackle this problem, including nonlinear optimization \citep{freeknots,penfreeknot}, using the lasso to determine the knots \citep{knotselectlasso} or fixing a large number of equidistant knots while penalizing the squared finite differences of the coefficients of adjacent B-splines \citep{psplines}. Many successful approaches are built on algorithmic stepwise addition/removal of knots \citep{flexsmoothadd,pmethod,mars}, with locally adaptive splines \citep{lars} that are based on total variation penalization on derivatives and enjoy optimal rates of convergence in bounded variation function classes, also for linear splines that saturate (are constant) outside the data range \citep{satursplines}. The stepwise addition approach has also been used in FDA for the estimation of the mean and the covariance function of a set of curves \citep{fda_freeknotsplines}. In the case of trend filtering \citep{trendfilter} the positions of the knots are adaptively chosen by specifying a suitable discrete difference operator of some order and solving a generalized lasso problem. In fact, for first order differences/constant trend, this reduces to the fused lasso with univariate inputs and pure fusion penalty. Moreover, the trend filtering estimate in the constant or piecewise-linear case is shown to be equal to the adaptive spline estimate \citep{polyestimtrendfilter}. Our approach is instead based on derivative-free optimization, which has already been used to select the knot vector for B-splines \citep{adasplgenetic}, for density-based approximation of univariate piecewise linear functions \citep{pwlga} and in supervised dimensionality reduction for classification problems with a functional regressor \citep{fda_interpdimred}, where the basis for the expansion of the input functions is chosen between a set of simple shapes, through a stochastic search algorithmic procedure that exploits the class membership. Altough in a different context, this last work is similar in spirit to ours, as it leverages both interpretable shapes and derivative-free optimization to select them. In particular, both approaches avoid working at the full knot level and allow to define any number of black-box basis shapes, but we provide a different parametrization for the shapes and a different algorithm for the optimization. As previously mentioned, while the types of basis $g_{j}$ could be considered as hyperparameters, we will focus on rectangular shapes only. The number of basis functions $q$ is fixed before the optimization but is instead selected by cross-validation, usually between $Q=\{ 1,2,3 \}$ to aid interpretability, as higher values of $q$ would lead to a less restricted $\gamma$, while at the same time increasing the computational cost. This results in a total of $3q$ variables in the global optimization problem, which are $A_{j}$, $t_{0j}$ and $T_{j}$ for $j$=$1,\dots,q$. In fact, we will show that the variables $A_{j}$ that control the heights of the rectangles can be optimized in closed form, further reducing the number of variables to $2q$, where for the $j$-th rectangle, $t_{0j}$ controls the horizontal shift with respect to the origin and $T_{j}$ determines the support.

\subsection{Fitting algorithm}   
In order to fit the model, we propose a two step iterative algorithm that is based both on global optimization and $L_2$-penalized least squares. The iterative part of the algorithm aims to mitigate the fact that the heuristic steps could provide inadequate values for $\gamma$, allowing for some degree of correction. 

Given the number of shapes $q$, which is fixed a priori and chosen with grid search, let $A$=$\left(A_{1},\dots,A_{q}\right)^{\top}$, $t_{0}$=$\left(t_{01},\dots,t_{0q}\right)^{\top}$ and $T$=$\left(T_{1},\dots,T_{q}\right)^{\top}$, the first step consists in solving the following problem:  

\begin{equation}
\label{eq:hardinit}
\underset{\scaleto{A,t_{0},T}{6pt}}{\scaleto{min}{7pt}} \:\:\: \sum_{i=1}^{N} \left[ y_{i}  -\bar{y} - \int_{I}x_{i}(t)\sum_{j=1}^{q} A_{j}g\left(t,t_{0j},T_{j}\right)dt \right]^{2} 
\end{equation}

\noindent where $\bar{y}=N^{-1}\sum_{i=1}^{N}y_{i}$ and $N$ is the size of the training set. This is a least squares formulation in which the regressor function is hard-constrained to belong to $\Gamma$. The objective function of Problem \ref{eq:hardinit} is nonconvex but marginally convex in $A$, therefore we approximate the solution with a heuristic approach that will be described in the next subsection. The resulting shape template $\gamma^{init}$ is now recovered from $\left(A^{*},t_{0}^{*},T^{*}\right)$ and Problem \ref{eq:L2} is solved with $\gamma=\gamma^{init}$ and a given $\lambda$, obtaining $\tilde{\beta}$. With the intention of improving on the current solution $\tilde{\beta}$, which depends on the initial approximation of $\gamma$, we can iterate the process by reshaping $\gamma$ towards $\tilde{\beta}$, through the following problem: 

\begin{equation}
\label{eq:reshape}
\underset{\scaleto{A,t_{0},T}{6pt}}{\scaleto{min}{7pt}} \sum_{i=1}^{N} \left[ y_{i}  -\bar{y} - \int_{I}x_{i}(t)\sum_{j=1}^{q} A_{j}g\left(t,t_{0j},T_{j}\right)dt \right]^{2} + \lambda \int_{I} \Big[ \tilde{\beta}(t) - \sum_{j=1}^{q} A_{j}g\left(t,t_{0j},T_{j}\right) \Big]^{2}dt 
\end{equation}

\noindent where $\tilde{\beta}$ is fixed from the previous step. As in the case of Problem \ref{eq:hardinit}, the objective function of Problem \ref{eq:reshape} is also nonconvex but marginally convex in $A$, and the optimization details will also be discussed in the next subsection. Once the new shape template $\gamma$ is recovered, we can solve again Problem \ref{eq:L2} (with the same $\lambda$) to obtain a refined $\tilde{\beta}$. These last two steps can be repeated for any number of iterations, with a small reduction of the training loss as the termination condition, and/or a maximum number of iterations. While it is clear that there is no guarantee to find a good shape template, if $\gamma$ is not suitable for the regression task at hand, the chosen $\lambda$ will be small so as to avoid shrinking $\beta$ towards a shape that would result in a poor fit, as $\lambda$ is selected with grid search after the initialization step, regardless of the number of iterations that we run. Beyond visual inspection of $\tilde{\beta}$, this allows to use the resulting value of $\lambda$ as a further indicator of the reasonableness of the shape $\gamma$ for the current problem. To summarize, let $Q$ be the maximum number of shapes from which $q$ is to be selected and let $M$ be the length of the grid of values from which $\lambda$ is selected, we fit the linear regression model by running Algorithm \ref{alg:algo} on the data set $\mathcal{D}$, obtaining the fitted values $\tilde{\beta}^{*}$ and $\tilde{\beta_{0}}^{*}$.

\begin{algorithm}
\caption{Ridge Regression with Adaptive Additive Functional Templates}
\label{alg:algo}
\begin{algorithmic}[1]
\FOR{$q=1,\dots,Q$}
\STATE Solve Problem \ref{eq:hardinit} with $x_{i} \in \mathcal{D}$ 
\ENDFOR
\STATE $\rightarrow$ $\gamma^{init}_{1}, \dots ,\gamma^{init}_{Q}$
\STATE K-fold split $\mathcal{D}$  in  $\mathcal{D}_{k-train}$ and $\mathcal{D}_{k-val}$:
\begin{ALC@g}
\FOR{$q=1,\dots,Q$}
\FOR{$m=1,\dots,M$}
\STATE Solve Problem \ref{eq:L2} with $\gamma$=$\gamma^{init}_{q}$, $\lambda$=$\lambda_{m}$, $x_{i} \in \mathcal{D}_{k-train}$
\STATE $\rightarrow$ $\tilde{\beta}_{qm}$, $\tilde{\beta_{0}}_{qm}$
\WHILE{Error on $\mathcal{D}_{k-train}$ decreases}
\STATE Solve Problem \ref{eq:reshape} with $\tilde{\beta}$=$\tilde{\beta}_{qm}$  
\STATE $\rightarrow$ $\gamma_{qm}$
\STATE Solve Problem \ref{eq:L2} with with $\gamma$=$\gamma_{qm}$, $\lambda$=$\lambda_{m}$, $x_{i} \in \mathcal{D}_{k-train}$
\STATE $\rightarrow$ $\tilde{\beta}_{qm}$, $\tilde{\beta_{0}}_{qm}$
\ENDWHILE
\STATE Test  $\tilde{\beta}_{qm}$, $\tilde{\beta_{0}}_{qm}$  on  $\mathcal{D}_{k-val}$
\ENDFOR 
\ENDFOR
\end{ALC@g}
\STATE $\rightarrow$ $q^{*}$, $m^{*}$
\STATE Refit on the full training set $\mathcal{D}$:
\STATE Solve Problem \ref{eq:L2} with $\gamma$=$\gamma^{init}_{q^{*}}$, $\lambda$=$\lambda_{m^{*}}$, $x_{i} \in \mathcal{D}$
\STATE $\rightarrow$ $\tilde{\beta}^{*}$, $\tilde{\beta_{0}}^{*}$
\WHILE{Error on $\mathcal{D}$ decreases}
\STATE Solve Problem \ref{eq:reshape} with $\tilde{\beta}$=$\tilde{\beta}^{*}$  
\STATE $\rightarrow$ $\gamma^{*}$
\STATE Solve Problem \ref{eq:L2} with with $\gamma$=$\gamma^{*}$, $\lambda$=$\lambda_{m^{*}}$, $x_{i} \in \mathcal{D}$
\STATE $\rightarrow$ $\tilde{\beta}^{*}$, $\tilde{\beta_{0}}^{*}$
\ENDWHILE
\RETURN $\tilde{\beta}^{*}$, $\tilde{\beta_{0}}^{*}$
\end{algorithmic}
\end{algorithm}

\subsection{Optimization and computational details}
Finding the global optimum of a nonconvex function is in general NP-hard and multiple approaches have been proposed to compute approximate solutions in different contexts. When the objective function is of real variables, smooth, and cheap to evaluate, derivative-based methods can leverage gradients to find local minima. A more general approach, which is also suitable for combinatorial problems, is the derivative-free one, where black-box evaluations of the objective function are used in conjunction with different stochastic selection and/or sampling schemes. In practice, the two approaches are often implemented jointly, where derivative-free techniques are used for an initial search of the solution space and identification of prospect points, followed by local gradient-based optimization \citep{globolocal_survey}. Another distinction that is often made is the one between single point methods, where a single solution is refined during the iterations, and population-based methods, where the selection schemes rely on a pool of candidate solutions. In this work we propose a hybrid approach based on differential evolution (DE) \citep{de} with a closed form solution for a subset of the variables. DE is a derivative-free population-based algorithm that generates new points by perturbation of existing solutions, which is suitable for problems with real variables. In particular, at each iteration three distinct points with indexes $i,j,k$ are randomly selected from the population, and a new candidate is computed by adding to point $k$ the scaled difference between points $i$ and $j$. There are multiple variants of this base scheme, and for an in depth analysis refer to \cite{book_de}. The choice of DE was motivated by its success in multiple areas, including clustering \citep{declust} and global plus local search \citep{delocal}.

Regarding our method, we already mentioned that the objectives in Problem \ref{eq:hardinit} and \ref{eq:reshape} are nonconvex, as they try to optimize the knot placement of a piecewise function. Restricting $\gamma$ in $\Gamma$ reduces the number of variables to $A_{j}$, $t_{0j}$ and $T_{j}$ for $j$=$1,\dots,q$, instead of directly optimizing the full knot vector, which is often high dimensional. Recall the following simplified definitions of marginal convexity and marginally optimum coordinate:

\begin{definition}[Marginal convexity]
\label{marginalconvexity}
A function of two variables $J(x,y):\mathbb{R}^{p}\times\mathbb{R}^{q} \rightarrow \mathbb{R}$ is marginally convex in $x \in \mathbb{R}^{p}$ if $\: \forall y \in \mathbb{R}^{q}$, the function $J_{y}(x):\mathbb{R}^{p} \rightarrow \mathbb{R}$ is convex.
\end{definition}

\begin{definition}[Marginally optimum coordinate]
\label{marginallyoptimumcoordinate}
We say that $\tilde{x}\in \mathbb{R}^{p}$ is marginally optimal with respect to $\tilde{y}\in \mathbb{R}^{q}$ if $\: \forall x \in \mathbb{R}^{p}$, $J(\tilde{x},\tilde{y})\leq J(x,\tilde{y})$.
\end{definition}

\noindent The property of marginal convexity lays the foundation for the alternating minimization principle, which is often used in nonconvex optimization problems where the objective function is marginally convex with respect to the single variables, but not jointly convex with respect to all variables \citep{nonconvexoptml}. Unfortunately our setting is quite different, as the objective functions of Problem \ref{eq:hardinit} and \ref{eq:reshape} are not marginally convex in $t_{0}$ and $T$ but only in $A$. In fact, the variables $A_{j}$ for $j$=$1,\dots,q$ do not control the position of the knots, but only the height of the $q$ shape templates $g_{j}$, and in particular both problems are quadratic in $A$, allowing us to optimize $T$ and $t_{0}$ with DE while recovering the marginally optimal $A$ in closed form, as shown in \ref{closedform1} and \ref{closedform2}. From the asymptotical standpoint, running Algorithm \ref{alg:algo} involves solving Problem \ref{eq:L2} and computing the marginally optimal $A$, which can both be done in $O(p^3)$. In practice however, it is arduous to provide a formal analysis, as the total cost is vastly dominated by the global optimization part, which overall depends on the total number of DE iterations. In particular, while the initialization of $\gamma$ is done only $Q$ times, Problem \ref{eq:reshape} is instead solved multiple times to select the optimal values for $\lambda$ and $q$. For this reason, we allocate the major part of the DE computational budget to the initalization of $\gamma$, as the following alternating iterations depend on a suitable initial value. Finally, it is worth to note that all the for loops in Algorithm \ref{alg:algo} can be executed in parallel, as the problems are separable with respect to $q$ and $\lambda$. As for the implementation, our global optimization scheme heavily relies on the \textit{Nevergrad} library by Facebook Research \citep{nevergrad}.

\section{Applications}
In this section we show the performance of our proposed method (AATR as \textit{adaptive additive template ridge}) in multiple simulation studies and two real world applications. We compare AATR with other functional linear models with known penalties like the lasso, fused lasso, ridge, roughness penalty, elastic net, elastic SCAD, elastic MCP, and we also include the minimum norm least squares solution (mnlstsq) as a reference. Besides AATR and the roughness penalized model, for which we provide our own code, all the other implementations are from \textit{scikit-learn} \citep{scikit} except for the fused lasso which is available in the CRAN package \textit{genlasso} \citep{package_genlasso} and the elastic SCAD and elastic MCP that are available in the CRAN package \textit{ncvreg} \citep{package_ncvreg}.

\subsection{Simulation studies}
\begin{figure}[H]
  \centering
    \begin{subfigure}[c]{0.27\textwidth}
      \includegraphics[width=\textwidth]{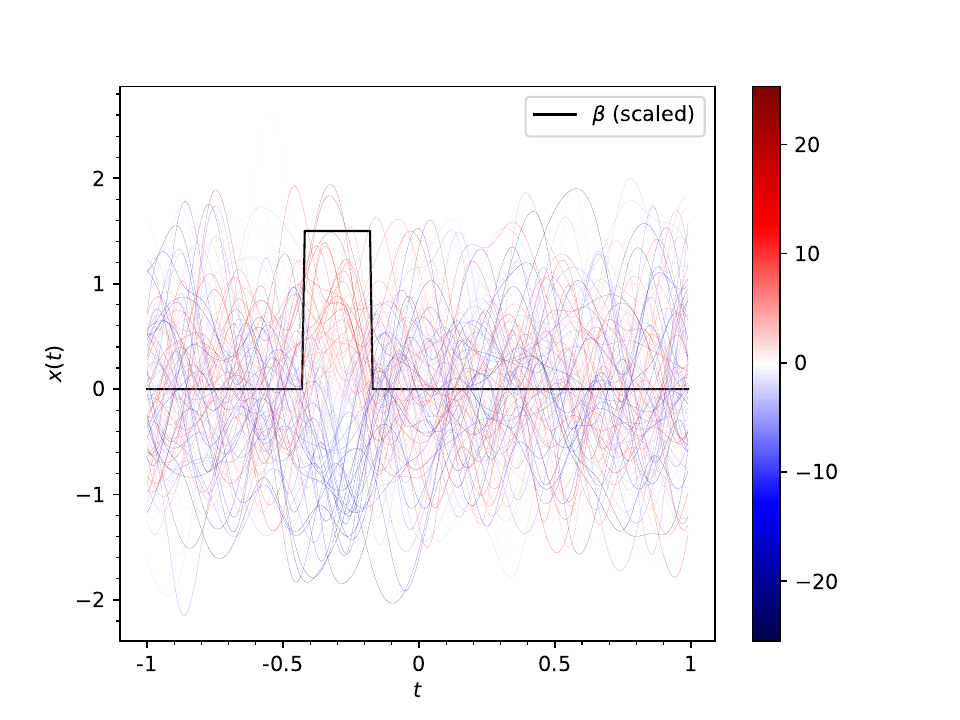}
      \caption{$x_{i}$ - independent}
    \end{subfigure}
    \begin{subfigure}[c]{0.27\textwidth}
      \includegraphics[width=\textwidth]{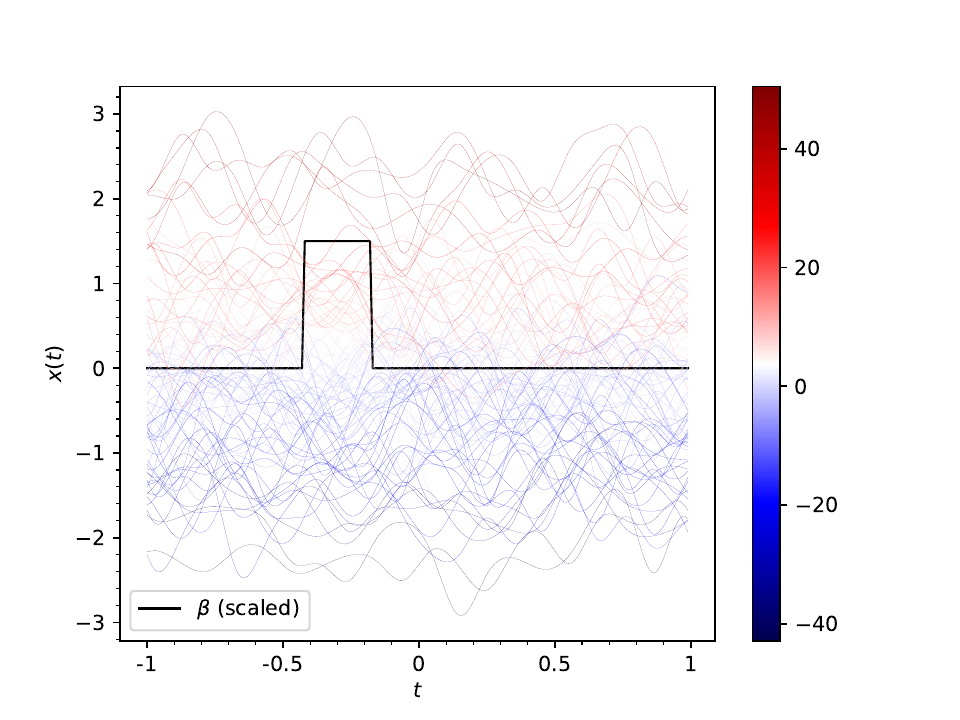}
      \caption{$x_{i}$ - dependent}
    \end{subfigure} 
    \begin{subfigure}[c]{0.27\textwidth}
      \includegraphics[width=\textwidth]{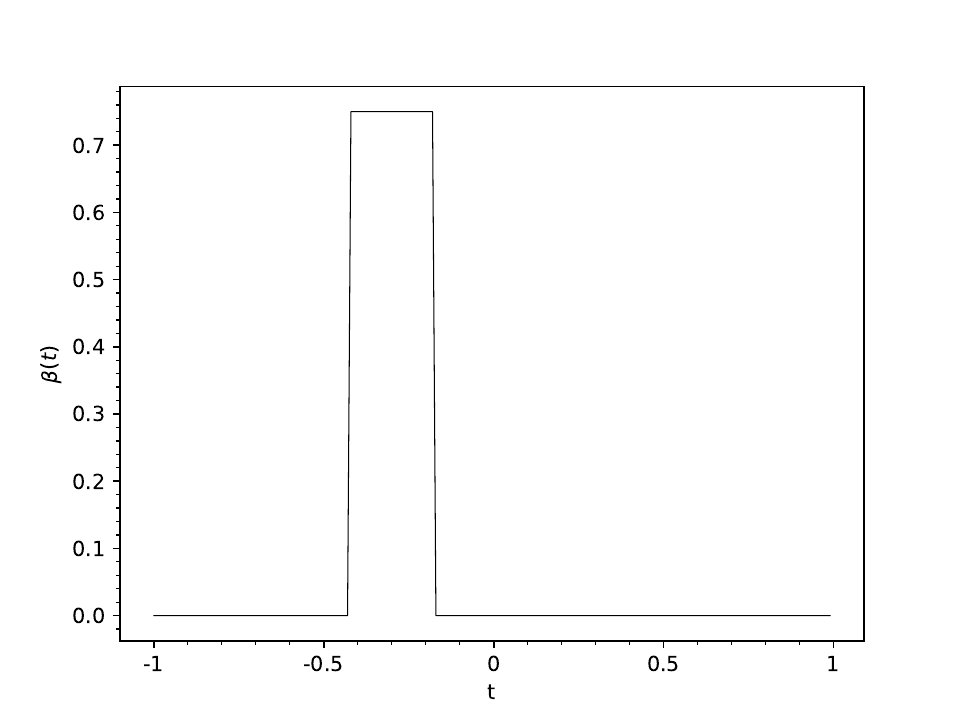}
      \caption{true $\beta$}
    \end{subfigure} 
  \caption{Simulated data with independent and highly dependent spline coefficients, true $\beta$ with $q$=1}
  \label{fig:data_1rect}
\end{figure}

\begin{figure}[H]
  \centering
  	\begin{subfigure}[c]{0.22\textwidth}
      \includegraphics[width=\textwidth]{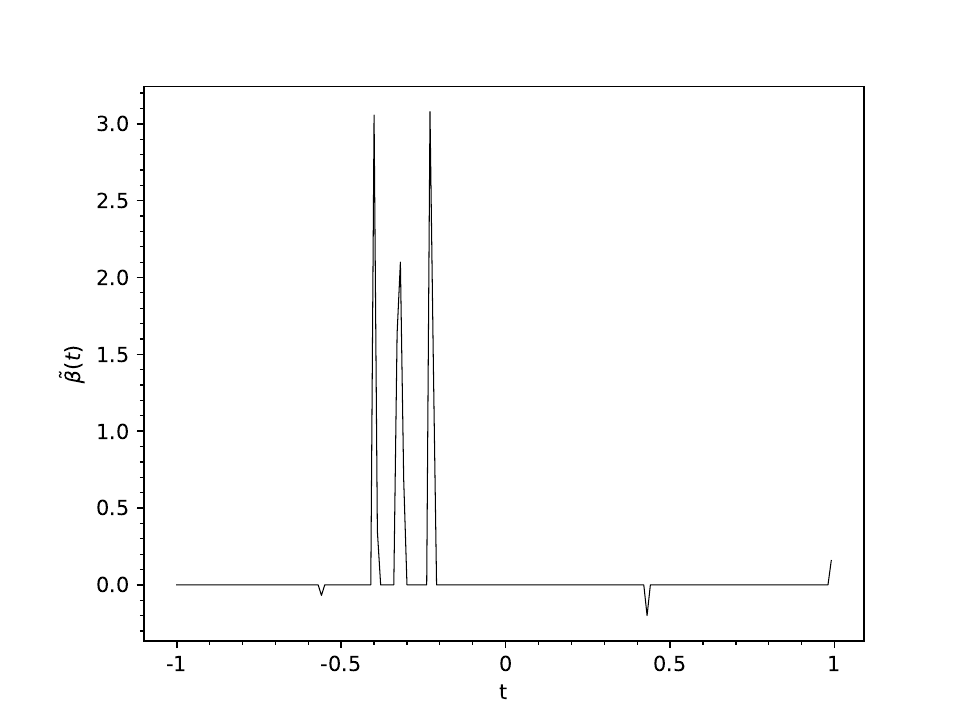}
      \caption{lasso}
    \end{subfigure}
    \begin{subfigure}[c]{0.22\textwidth}
      \includegraphics[width=\textwidth]{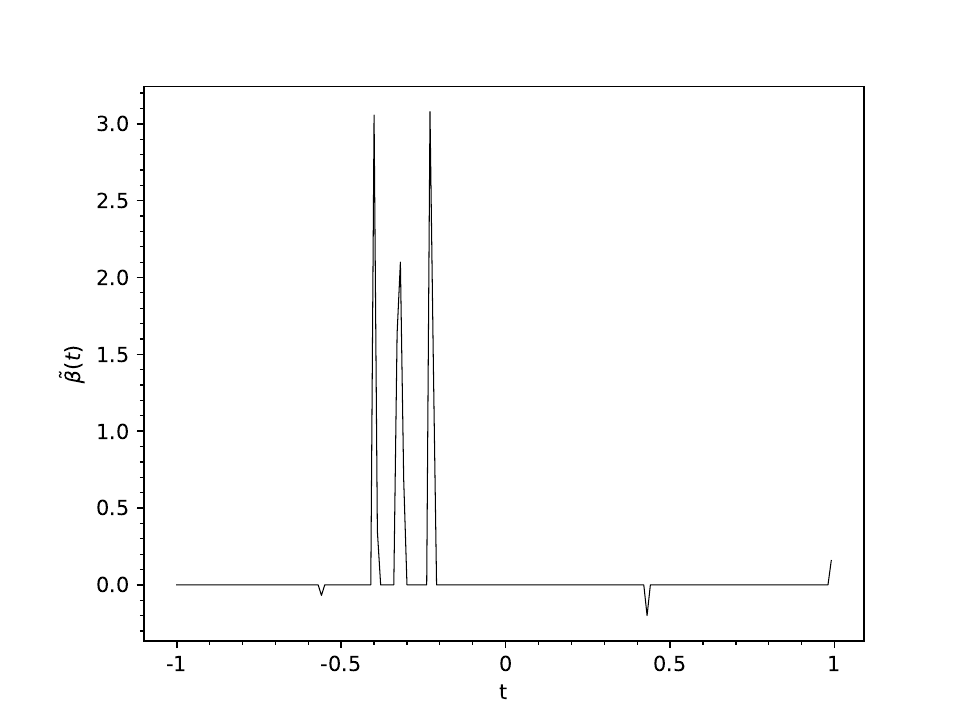}
      \caption{elastic net}
    \end{subfigure}
    \begin{subfigure}[c]{0.22\textwidth}
      \includegraphics[width=\textwidth]{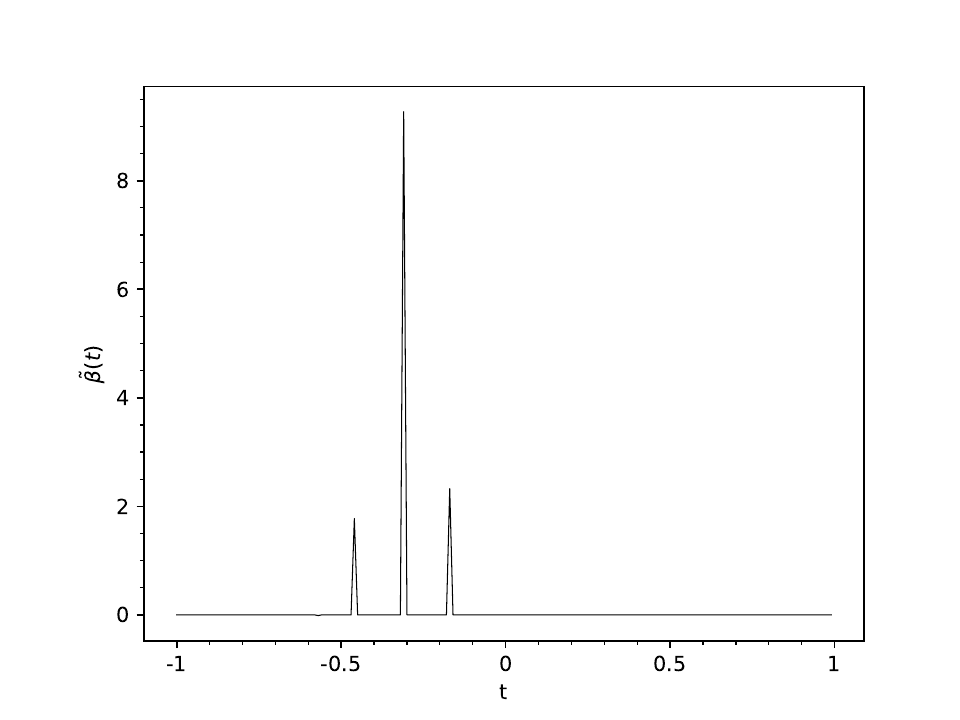}
      \caption{elastic SCAD}
    \end{subfigure}
    \begin{subfigure}[c]{0.22\textwidth}
      \includegraphics[width=\textwidth]{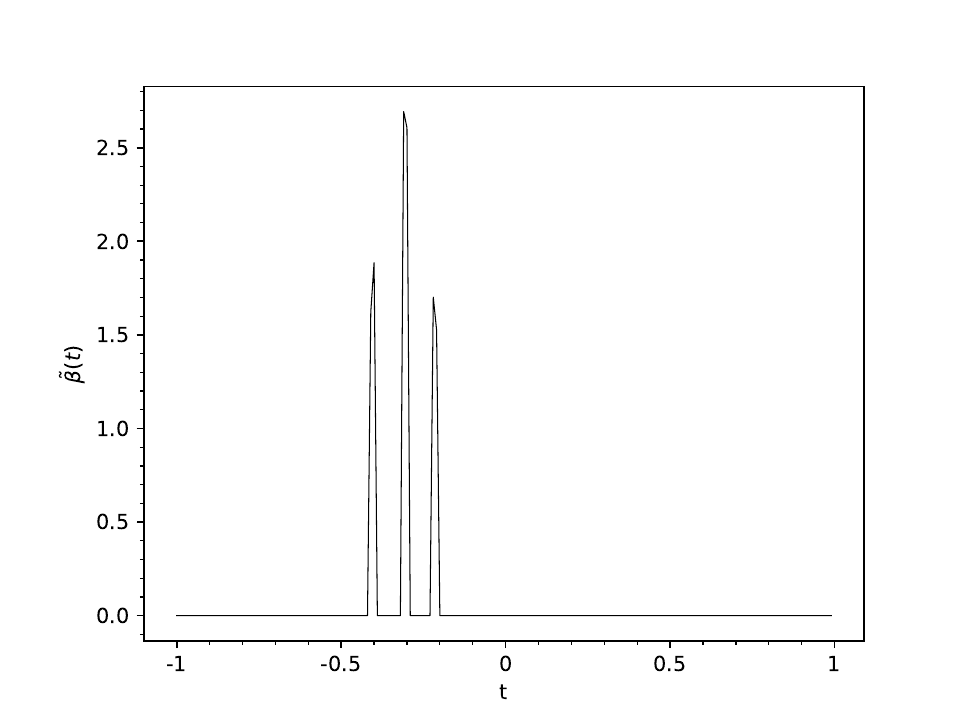}
      \caption{elastic MCP}
    \end{subfigure}

    \begin{subfigure}[c]{0.22\textwidth}
      \includegraphics[width=\textwidth]{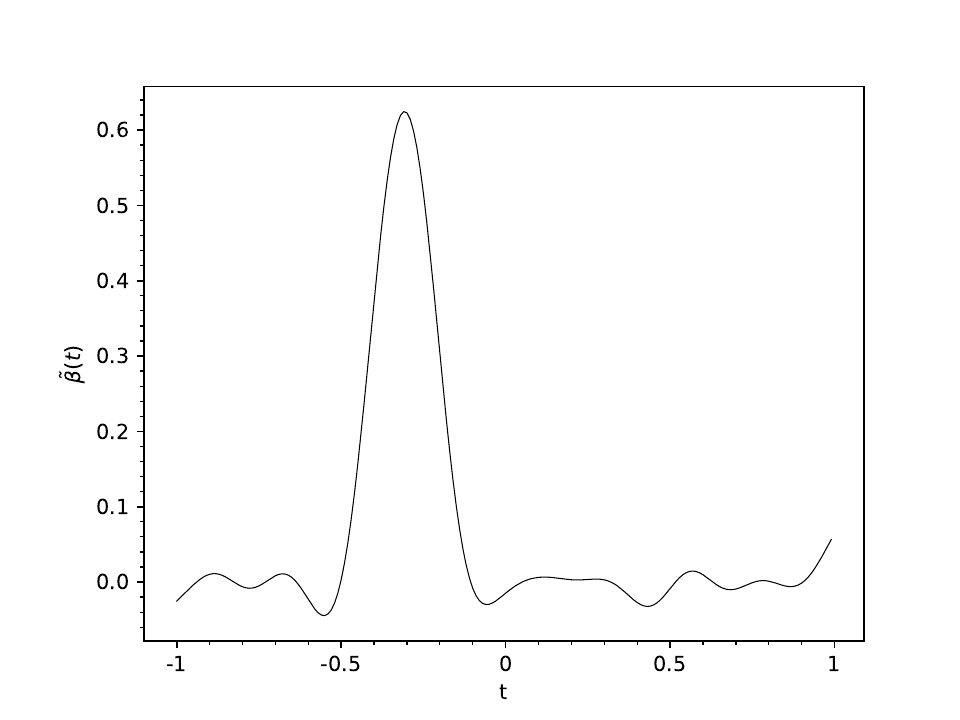}
      \caption{roughness}
    \end{subfigure}    
    \begin{subfigure}[c]{0.22\textwidth}
      \includegraphics[width=\textwidth]{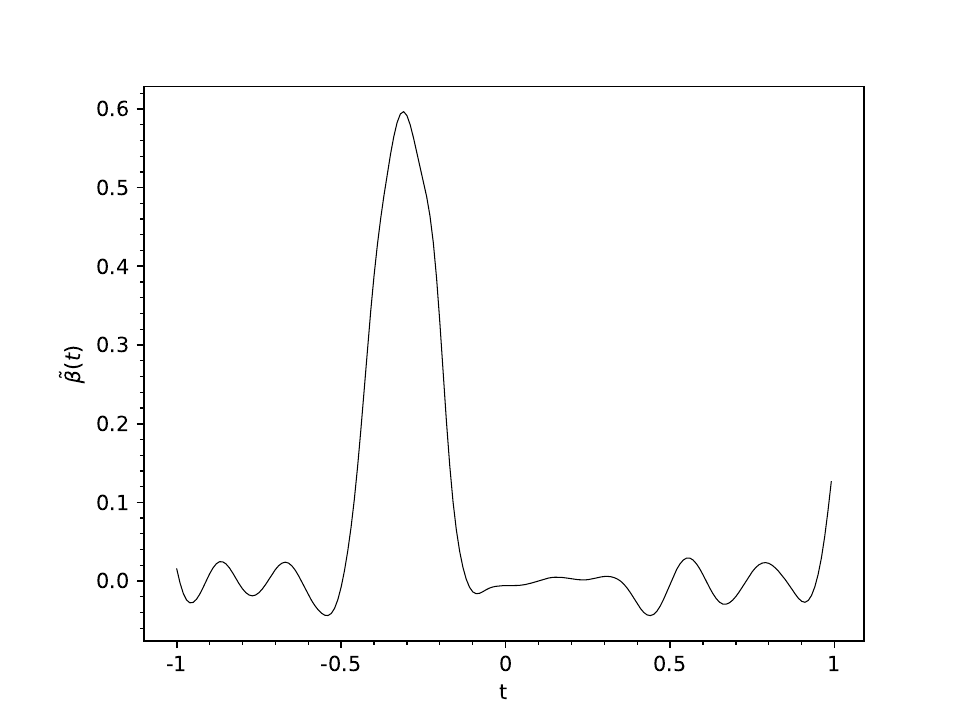}
      \caption{ridge}
    \end{subfigure}
    \begin{subfigure}[c]{0.22\textwidth}
      \includegraphics[width=\textwidth]{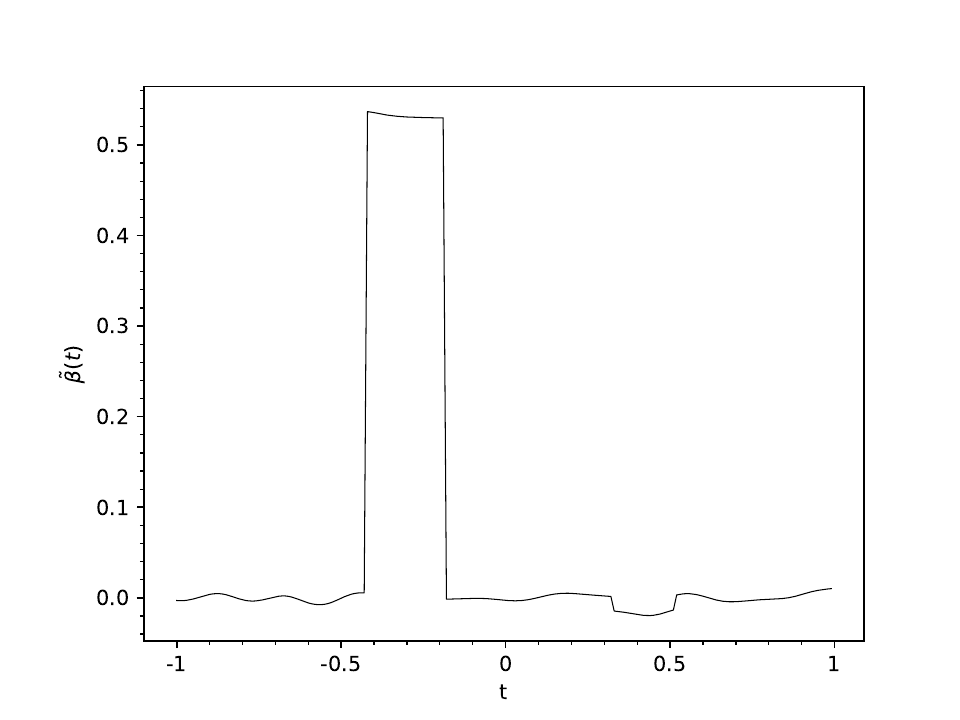}
      \caption{AATR}
    \end{subfigure}
    \begin{subfigure}[c]{0.22\textwidth}
      \includegraphics[width=\textwidth]{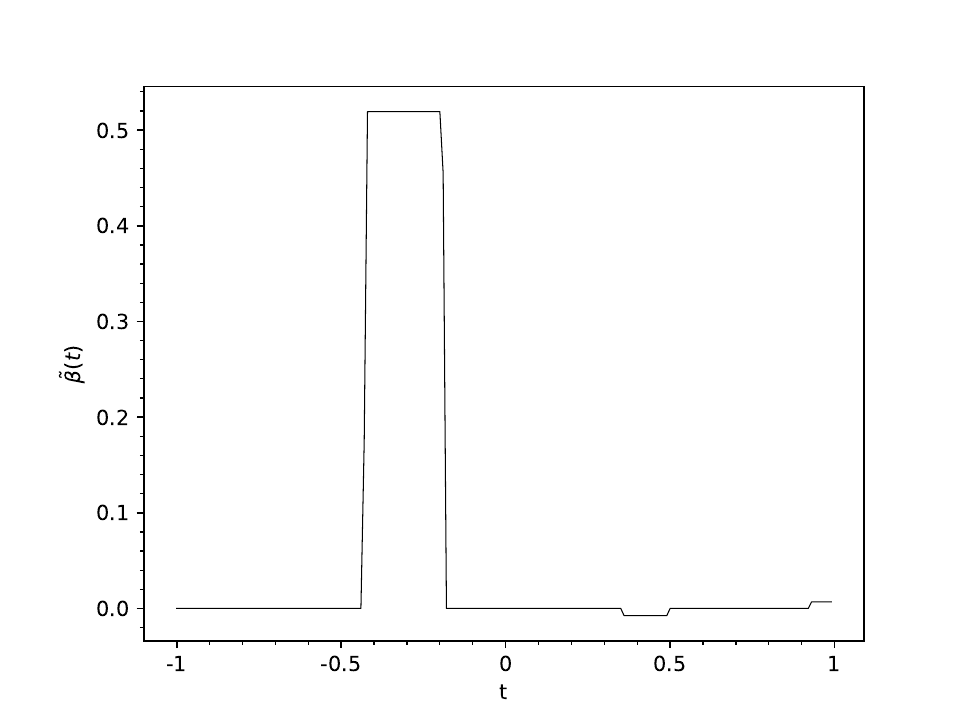}
      \caption{fused lasso}
    \end{subfigure}
  \caption{Independent spline coefficients: fitted coefficient functions $\tilde{\beta}$ by penalty type}
  \label{fig:comparison_1rect_ind}
\end{figure}

\begin{figure}[H]
  \centering
  	\begin{subfigure}[c]{0.22\textwidth}
      \includegraphics[width=\textwidth]{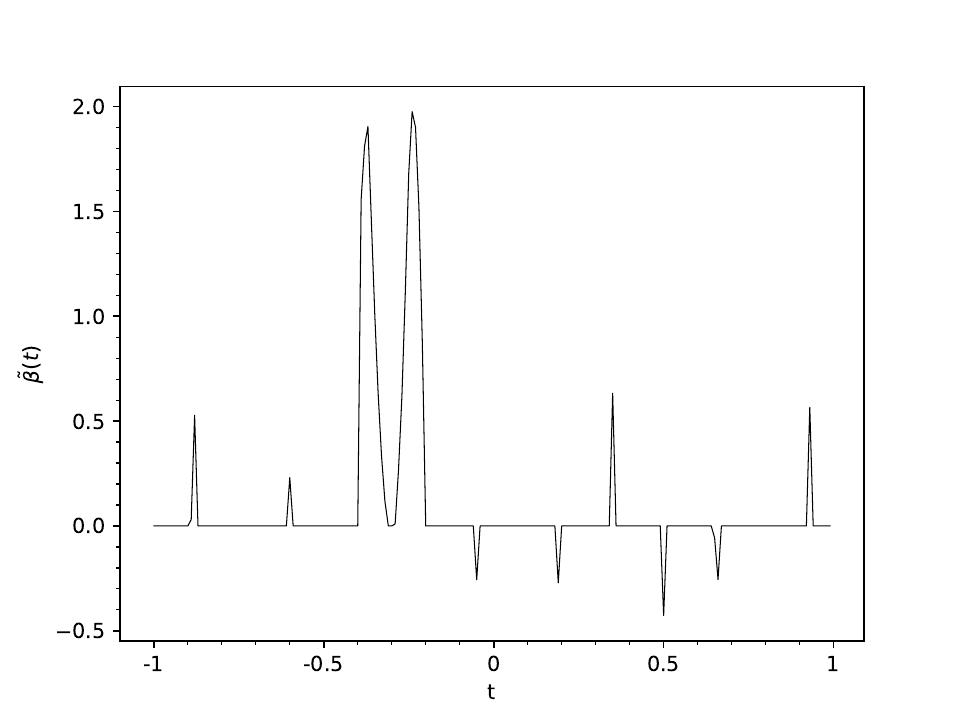}
      \caption{lasso}
    \end{subfigure}
    \begin{subfigure}[c]{0.22\textwidth}
      \includegraphics[width=\textwidth]{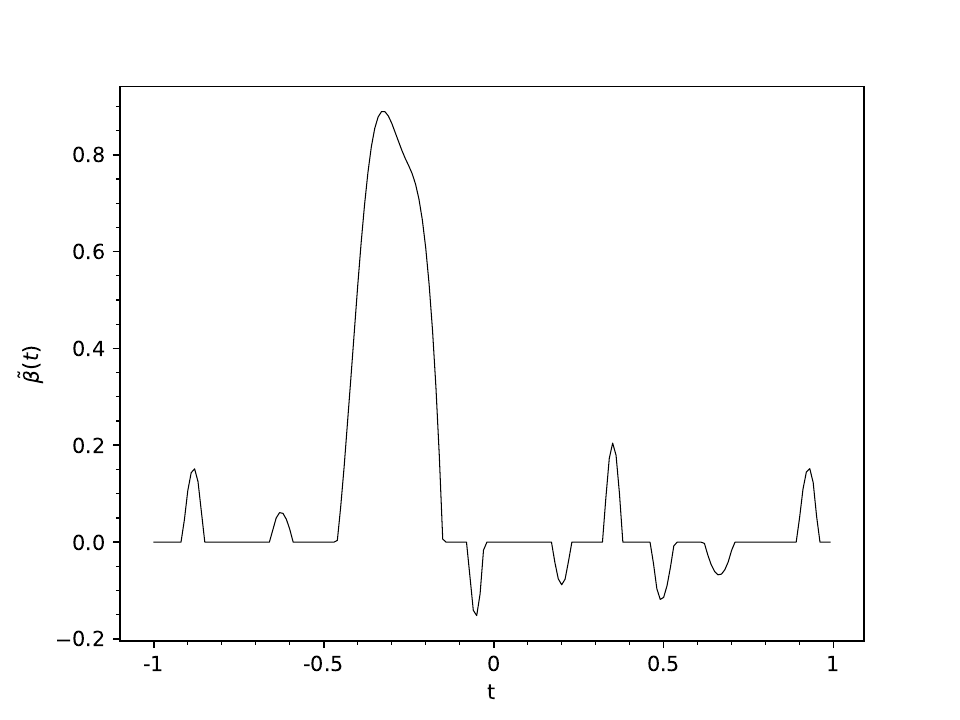}
      \caption{elastic net}
    \end{subfigure}
    \begin{subfigure}[c]{0.22\textwidth}
      \includegraphics[width=\textwidth]{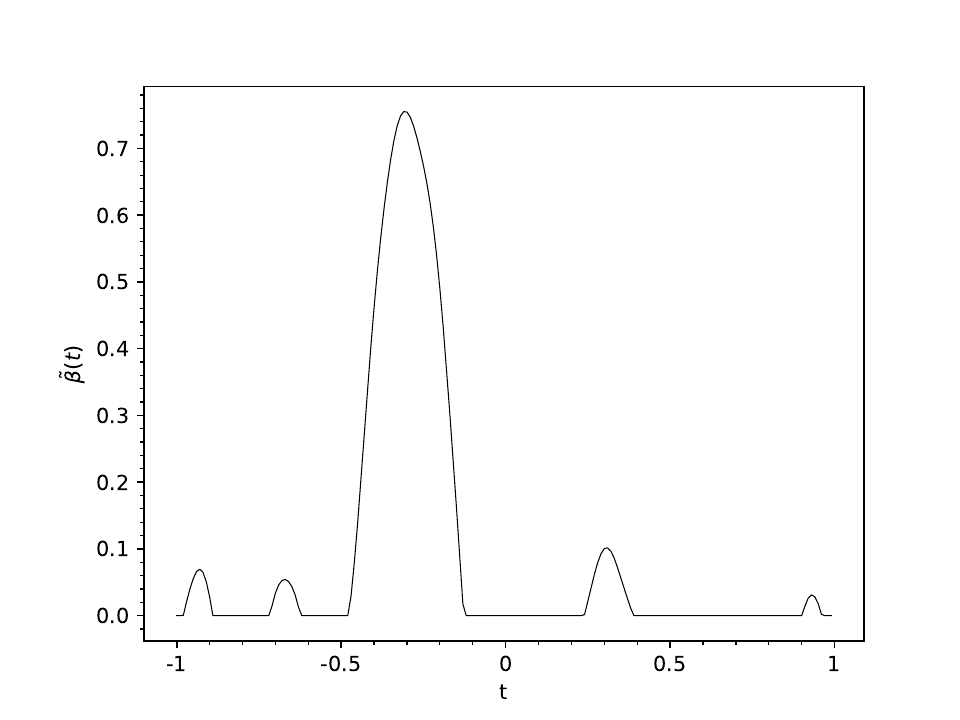}
      \caption{elastic SCAD}
    \end{subfigure}
    \begin{subfigure}[c]{0.22\textwidth}
      \includegraphics[width=\textwidth]{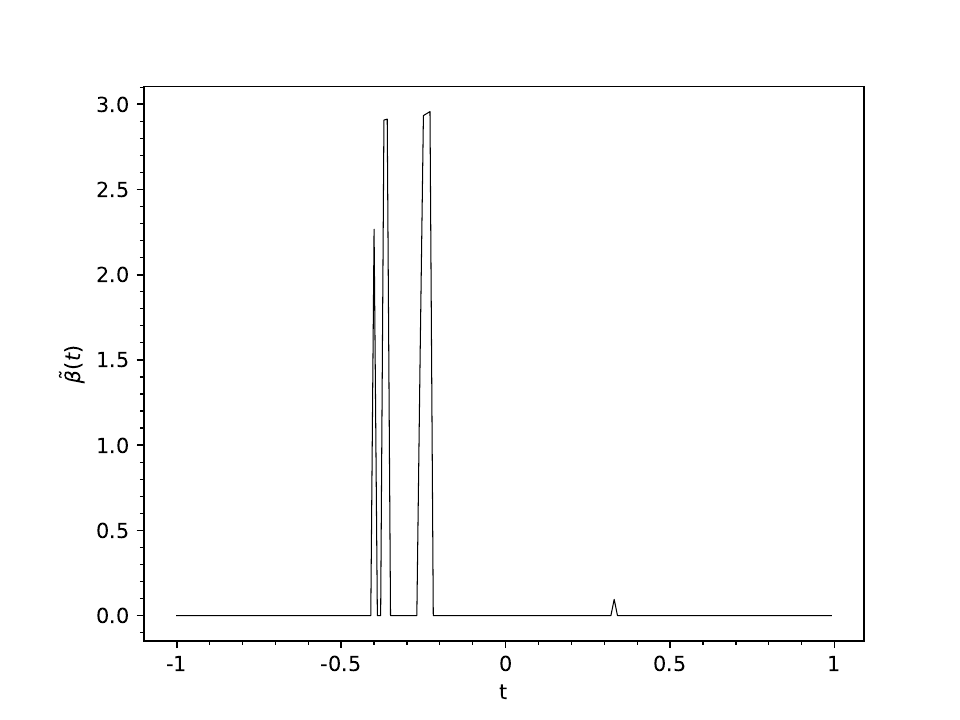}
      \caption{elastic MCP}
    \end{subfigure}

    \begin{subfigure}[c]{0.22\textwidth}
      \includegraphics[width=\textwidth]{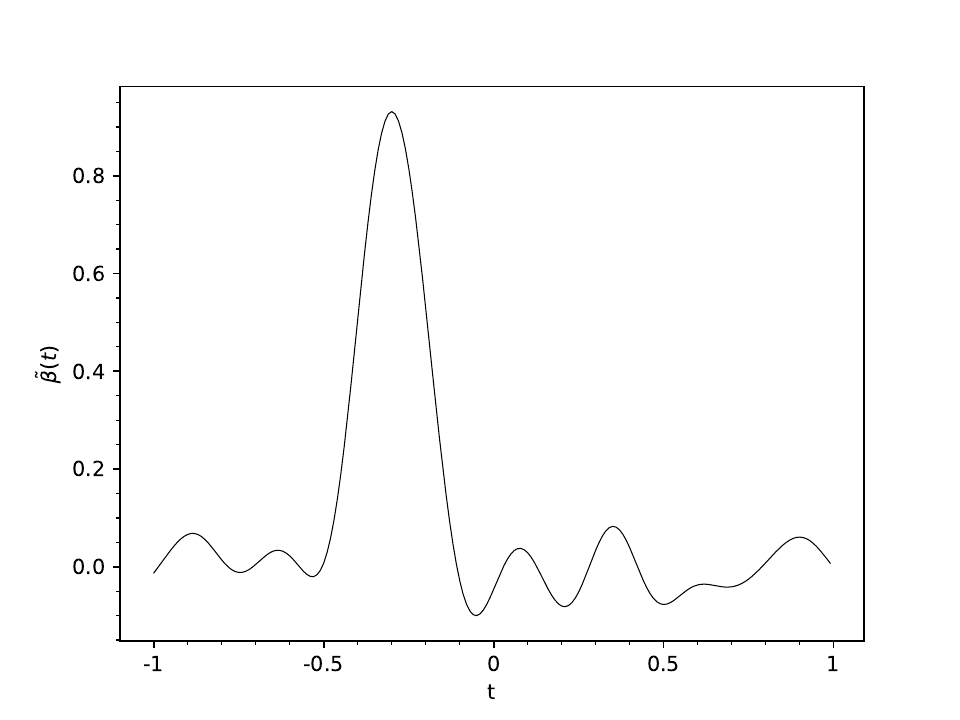}
      \caption{roughness}
    \end{subfigure}    
    \begin{subfigure}[c]{0.22\textwidth}
      \includegraphics[width=\textwidth]{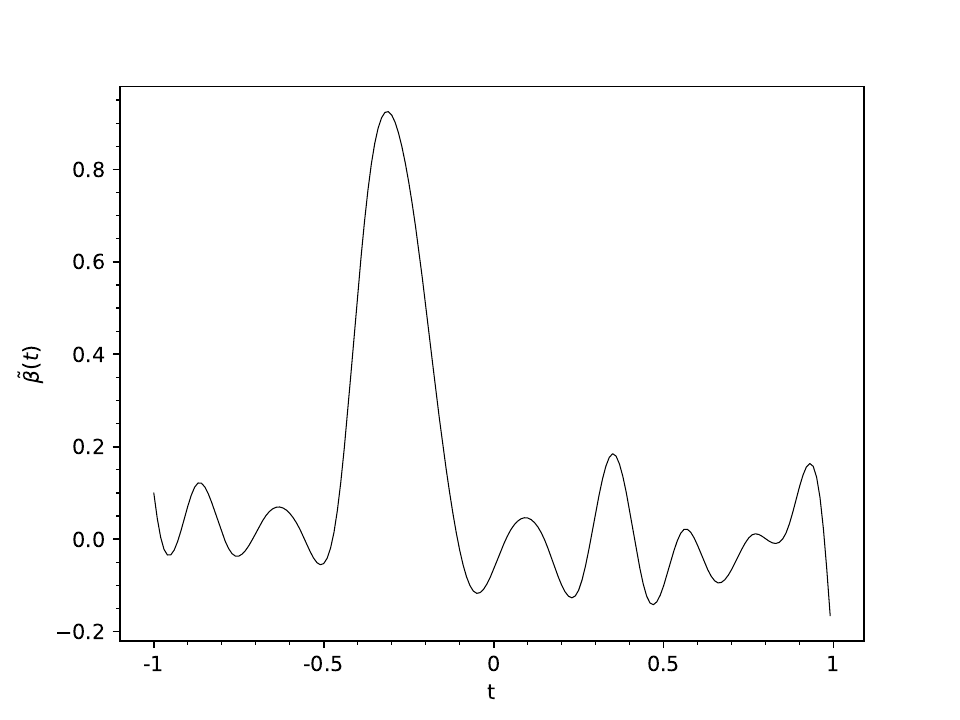}
      \caption{ridge}
    \end{subfigure}
    \begin{subfigure}[c]{0.22\textwidth}
      \includegraphics[width=\textwidth]{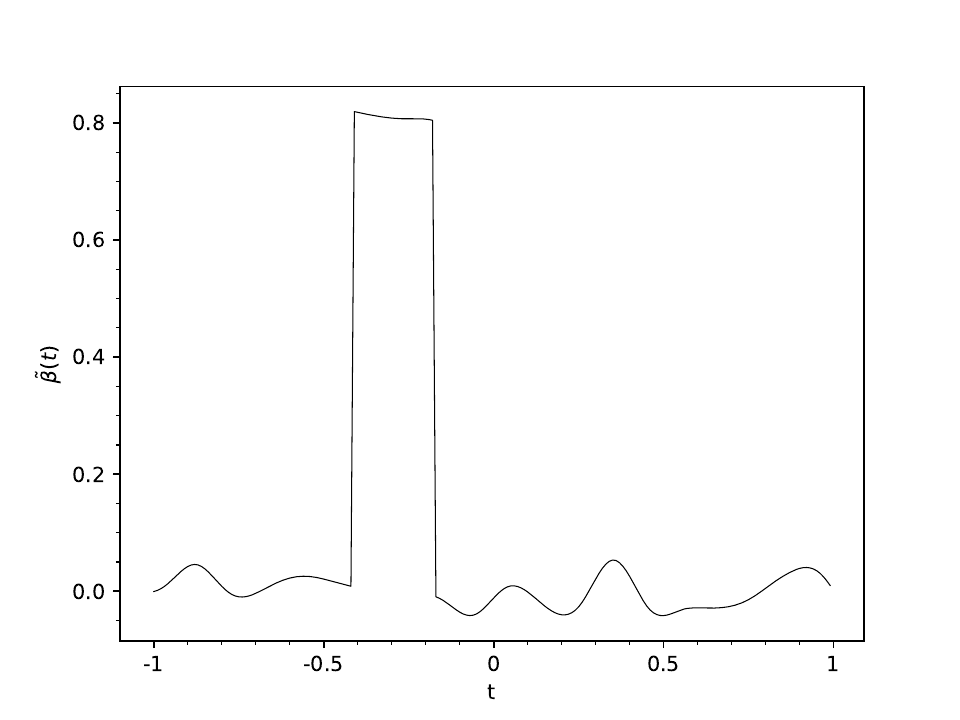}
      \caption{AATR}
    \end{subfigure}
    \begin{subfigure}[c]{0.22\textwidth}
      \includegraphics[width=\textwidth]{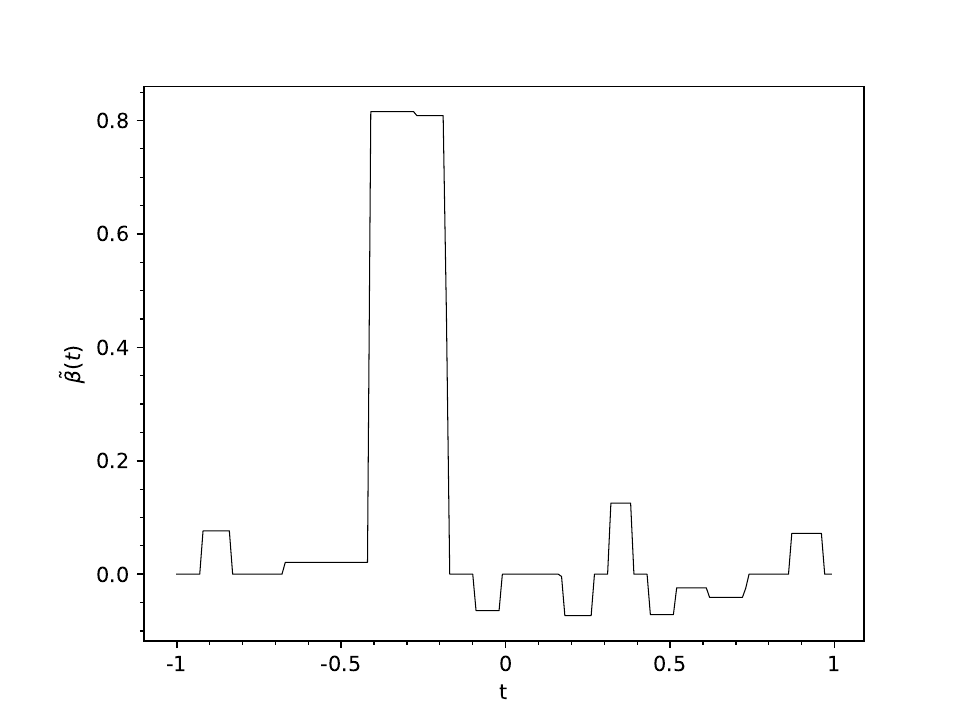}
      \caption{fused lasso}
    \end{subfigure}
  \caption{Highly dependent spline coefficients: fitted coefficient functions $\tilde{\beta}$ by penalty type}
  \label{fig:comparison_1rect_dep}
\end{figure}

\begin{figure}[H]
  \centering
    \begin{subfigure}[c]{0.27\textwidth}
      \includegraphics[width=\textwidth]{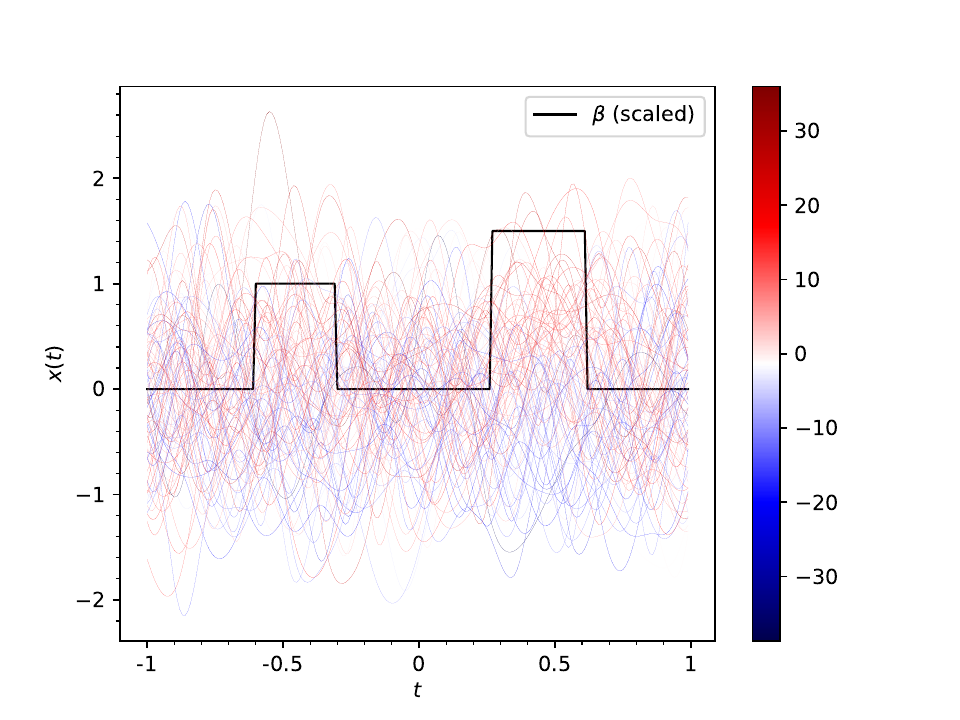}
      \caption{$x_{i}$ - independent}
    \end{subfigure}
    \begin{subfigure}[c]{0.27\textwidth}
      \includegraphics[width=\textwidth]{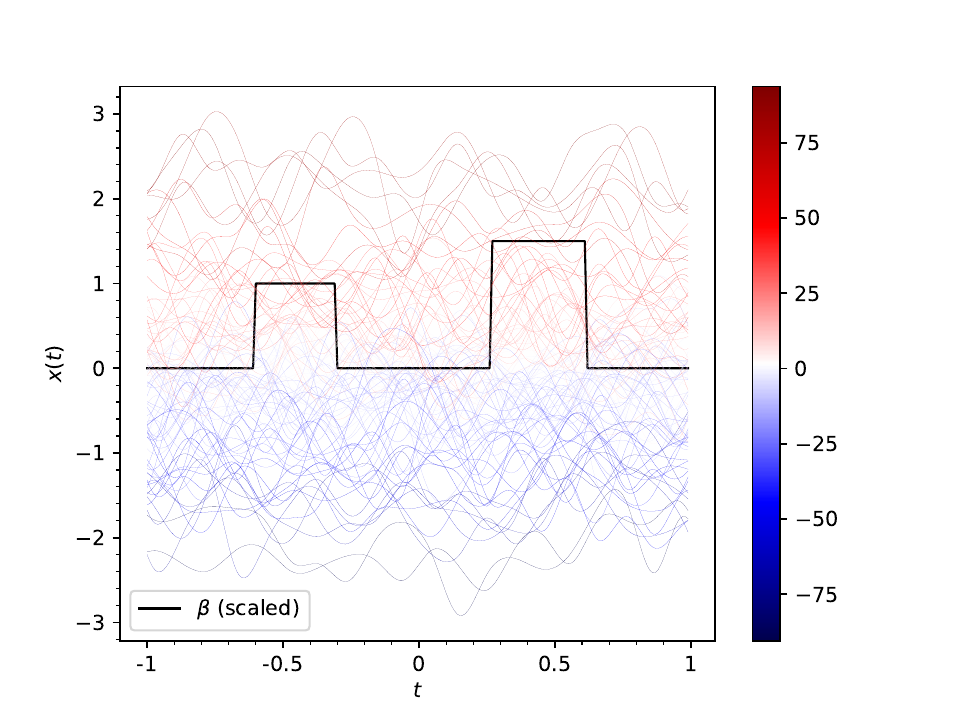}
      \caption{$x_{i}$ - dependent}
    \end{subfigure}
    \begin{subfigure}[c]{0.27\textwidth}
      \includegraphics[width=\textwidth]{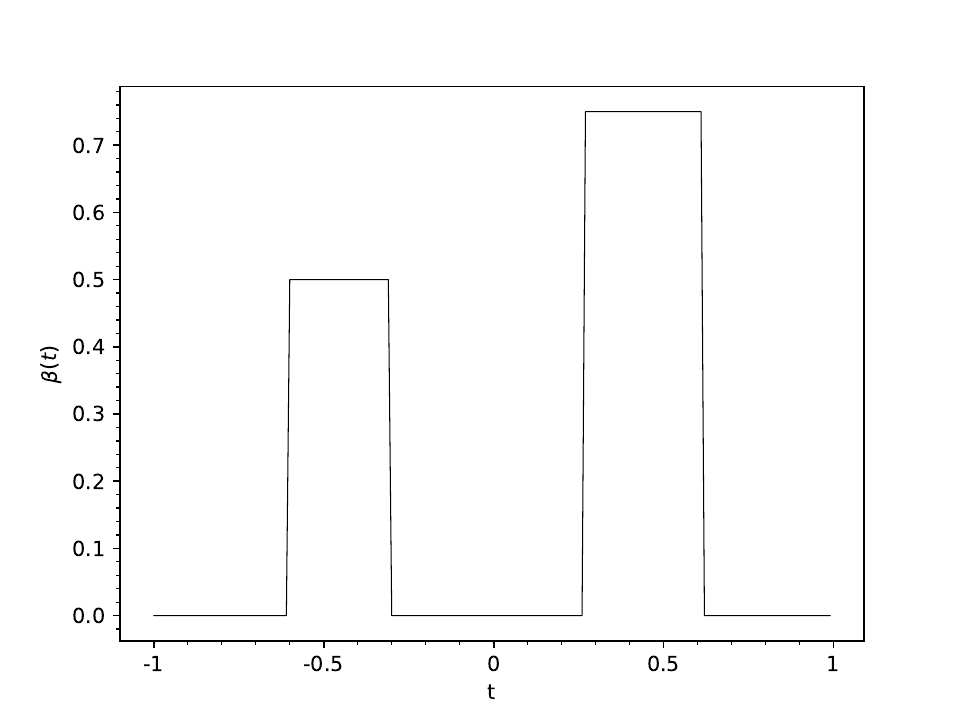}
      \caption{true $\beta$}
    \end{subfigure}
  \caption{Simulated data with independent and highly dependent spline coefficients, true $\beta$ with $q$=2}
  \label{fig:data_2rect}
\end{figure}

\begin{figure}[H]
  \centering
  	\begin{subfigure}[c]{0.22\textwidth}
      \includegraphics[width=\textwidth]{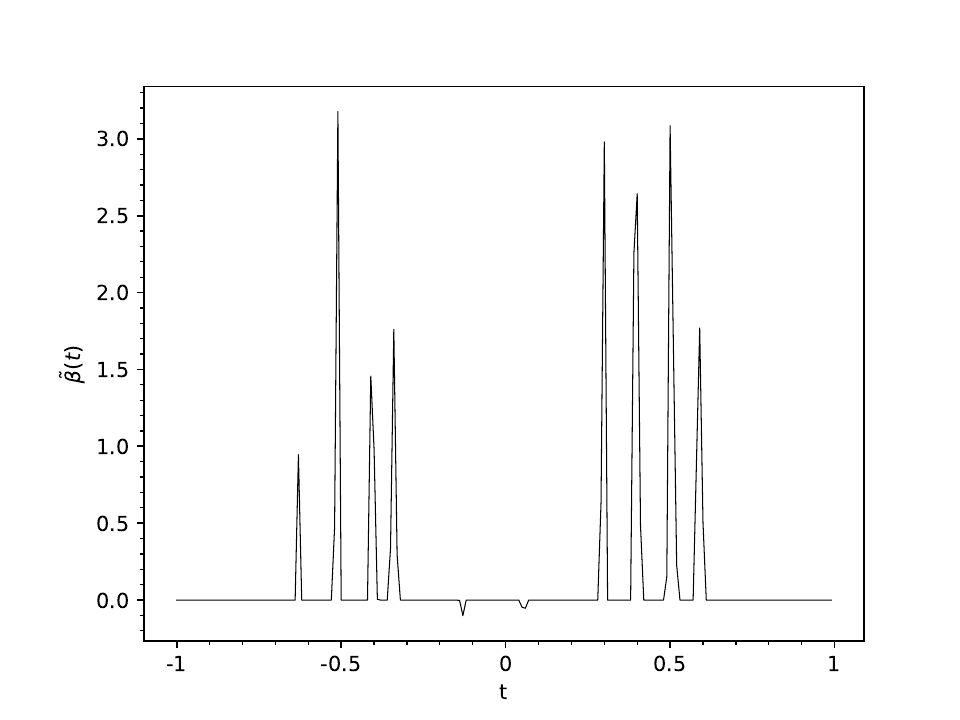}
      \caption{lasso}
    \end{subfigure}
    \begin{subfigure}[c]{0.22\textwidth}
      \includegraphics[width=\textwidth]{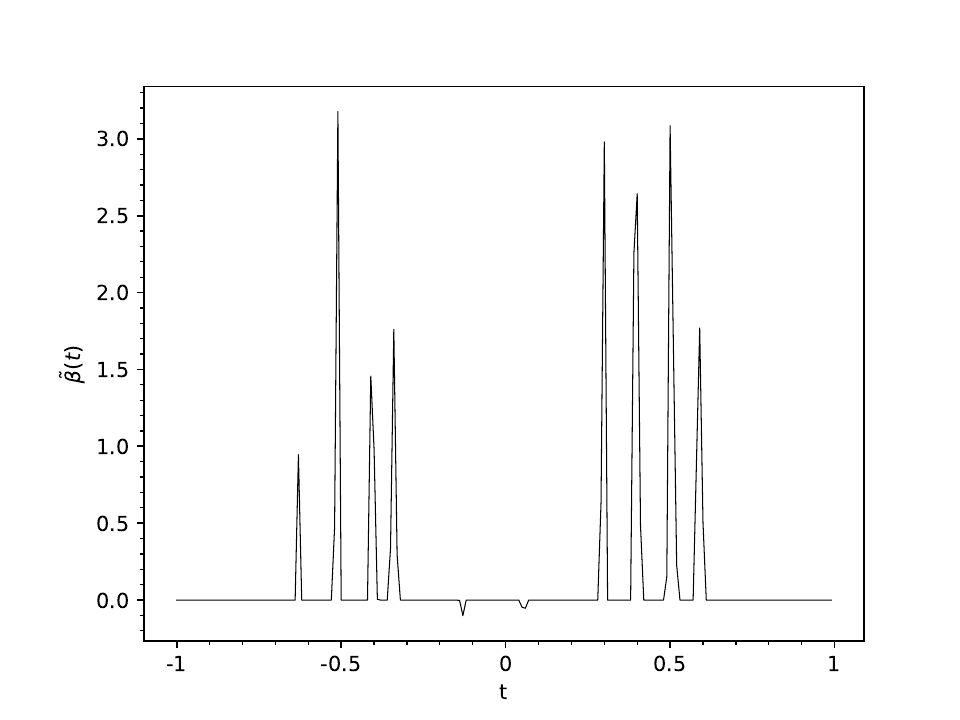}
      \caption{elastic net}
    \end{subfigure}
    \begin{subfigure}[c]{0.22\textwidth}
      \includegraphics[width=\textwidth]{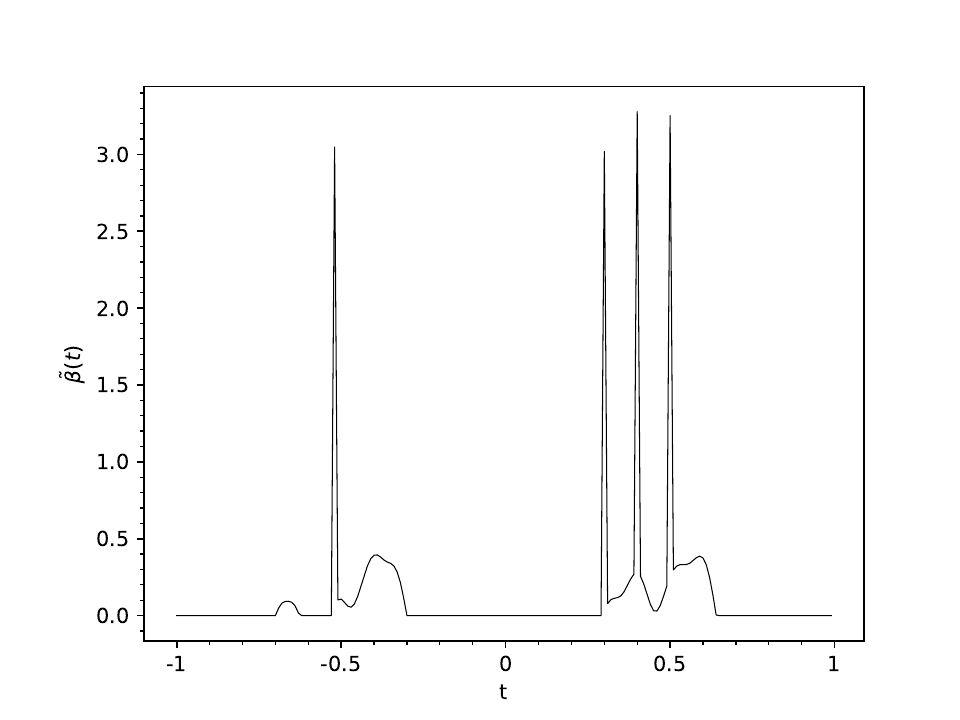}
      \caption{elastic SCAD}
    \end{subfigure}
    \begin{subfigure}[c]{0.22\textwidth}
      \includegraphics[width=\textwidth]{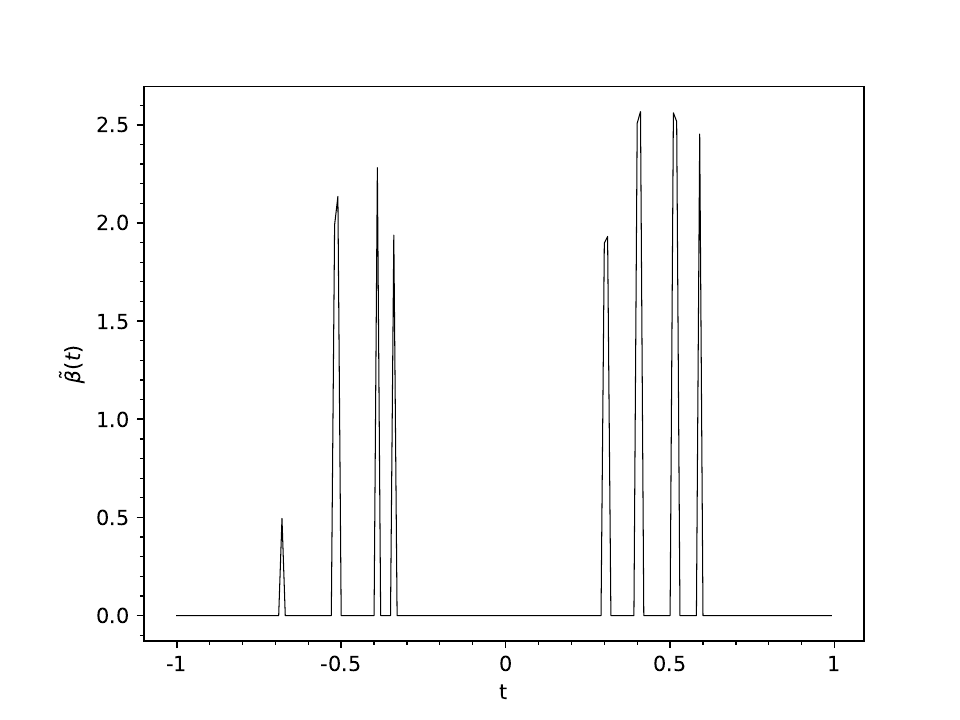}
      \caption{elastic MCP}
    \end{subfigure}

    \begin{subfigure}[c]{0.22\textwidth}
      \includegraphics[width=\textwidth]{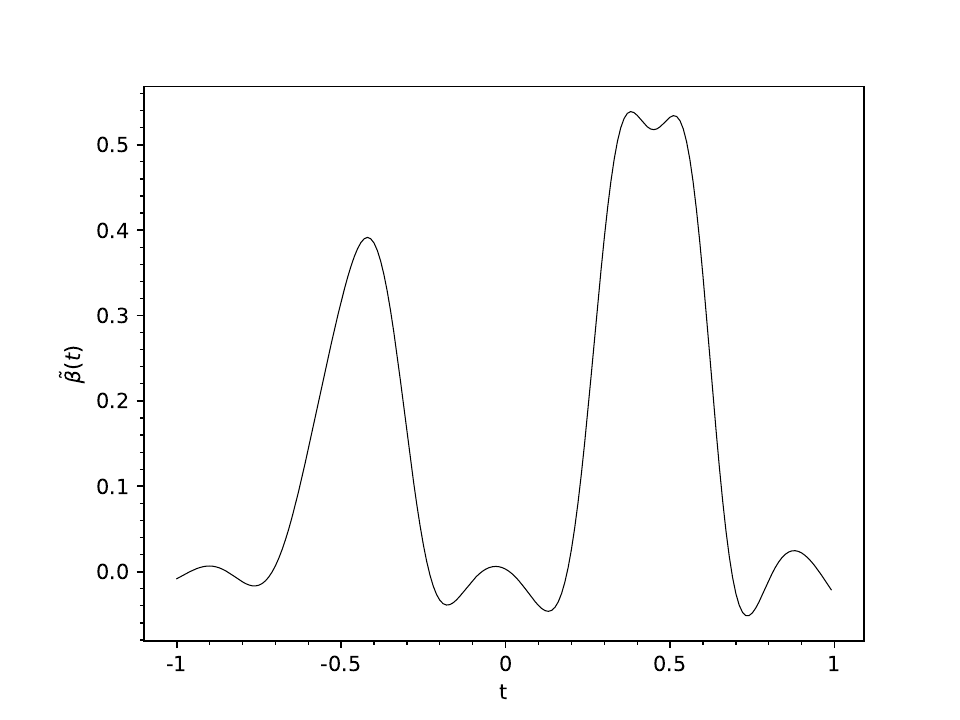}
      \caption{roughness}
    \end{subfigure}    
    \begin{subfigure}[c]{0.22\textwidth}
      \includegraphics[width=\textwidth]{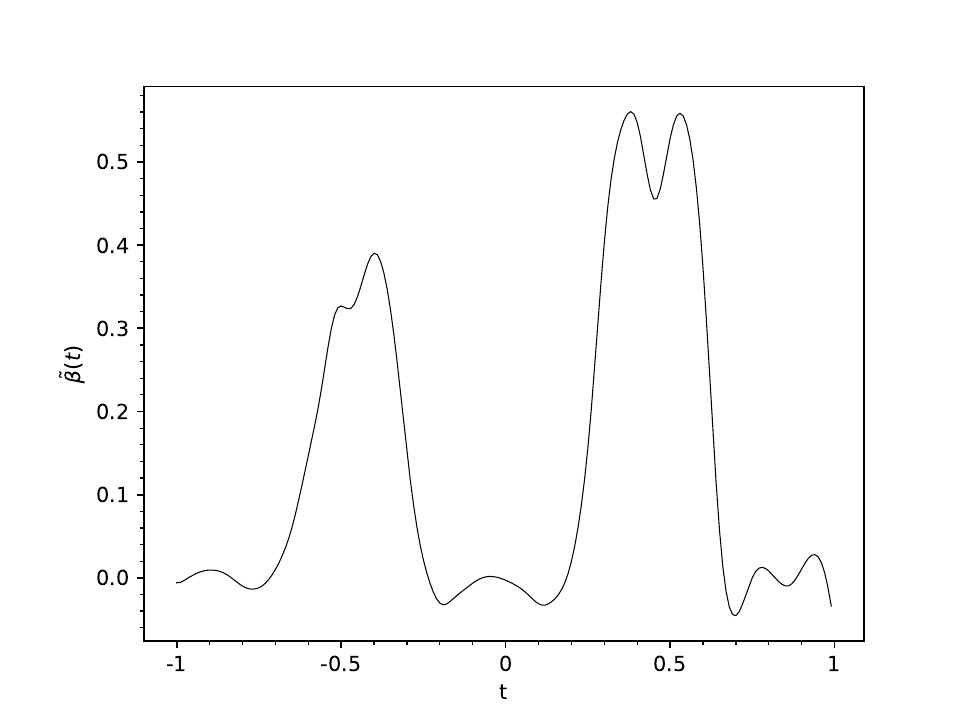}
      \caption{ridge}
    \end{subfigure}
    \begin{subfigure}[c]{0.22\textwidth}
      \includegraphics[width=\textwidth]{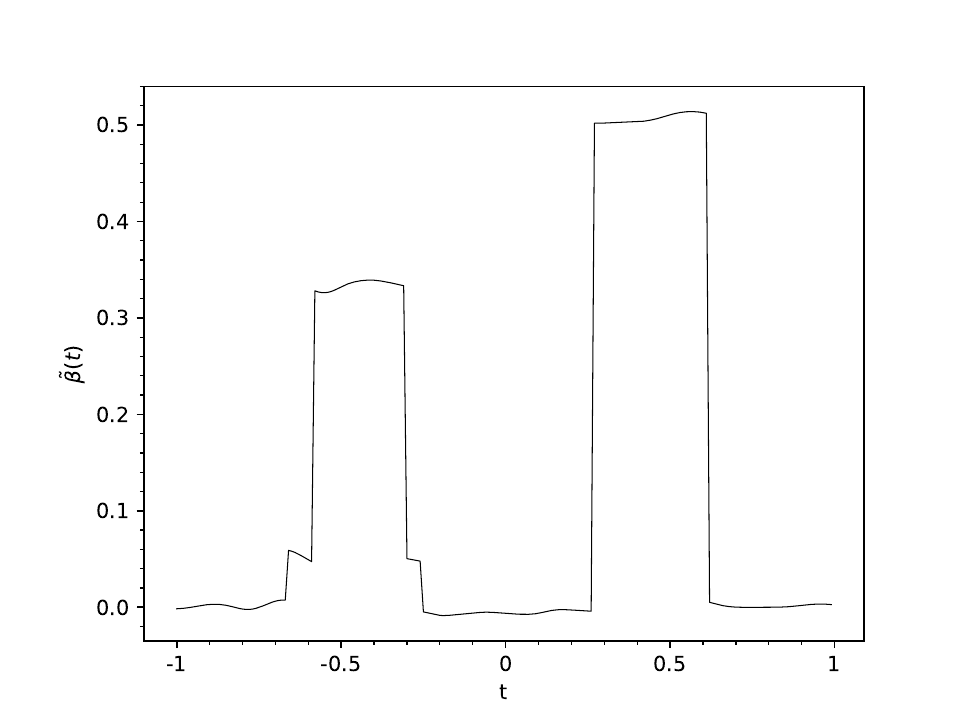}
      \caption{AATR}
    \end{subfigure}
    \begin{subfigure}[c]{0.22\textwidth}
      \includegraphics[width=\textwidth]{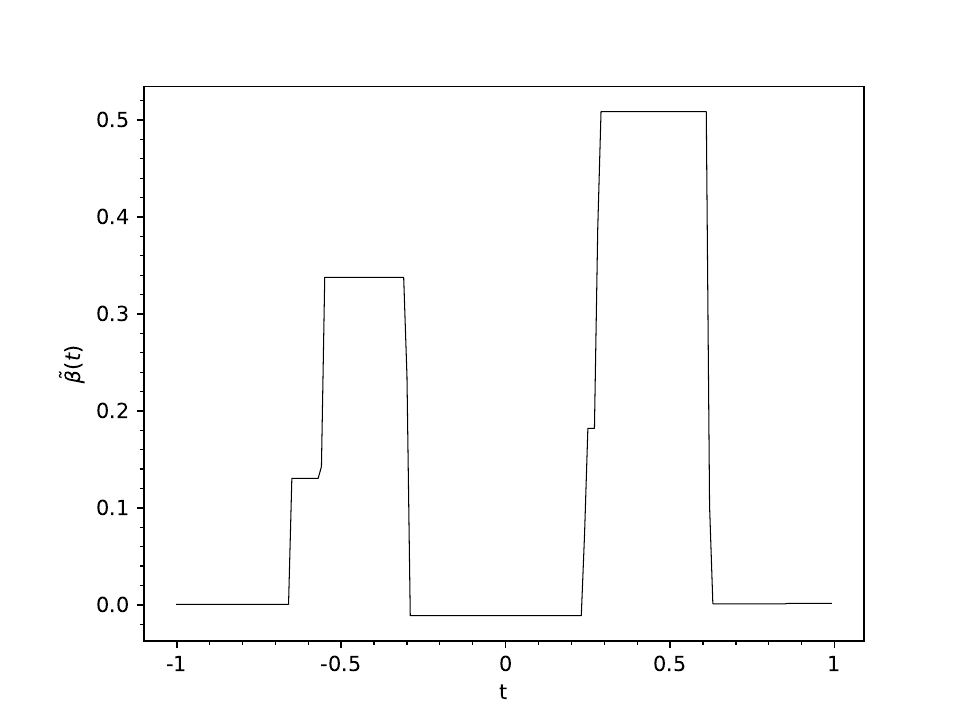}
      \caption{fused lasso}
    \end{subfigure}
  \caption{Independent spline coefficients: fitted coefficient functions $\tilde{\beta}$ by penalty type}
  \label{fig:comparison_2rect_ind}
\end{figure}

\begin{figure}[H]
  \centering
  	\begin{subfigure}[c]{0.22\textwidth}
      \includegraphics[width=\textwidth]{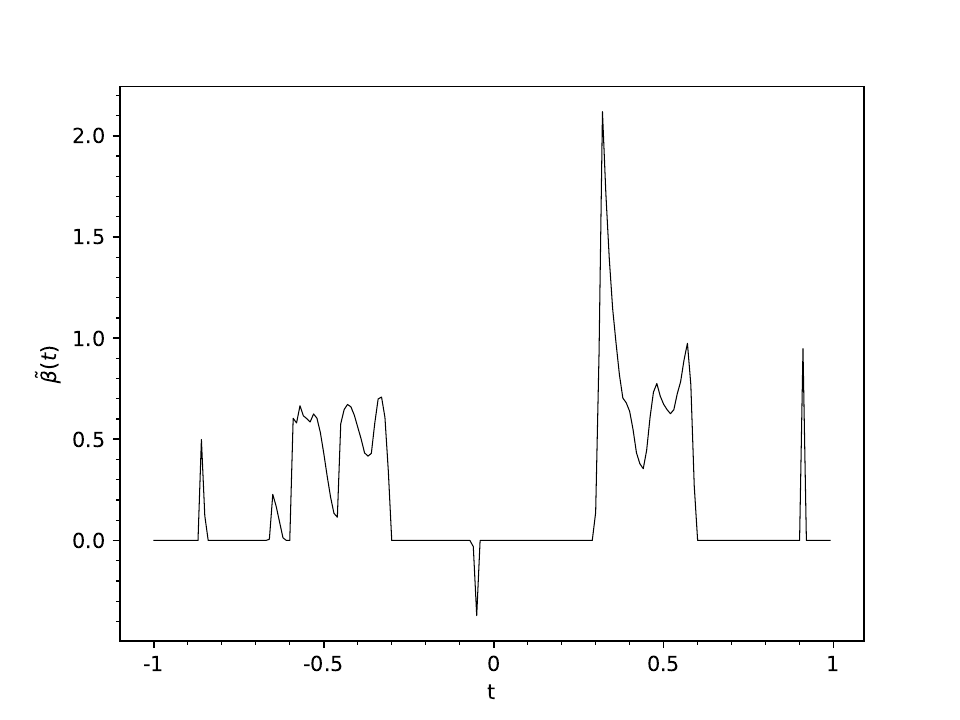}
      \caption{lasso}
    \end{subfigure}
    \begin{subfigure}[c]{0.22\textwidth}
      \includegraphics[width=\textwidth]{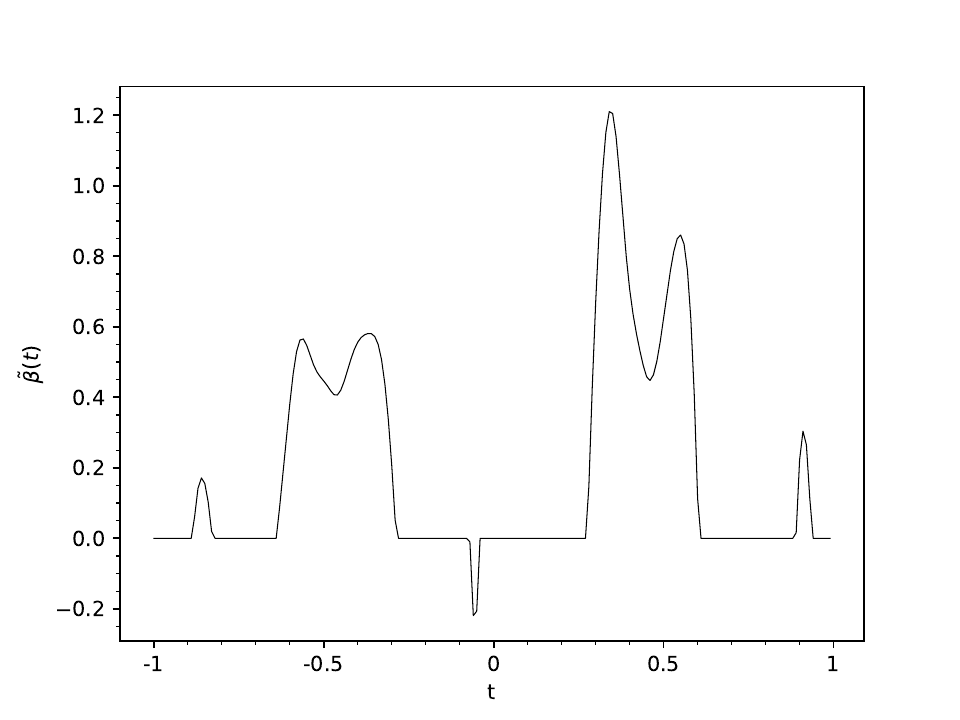}
      \caption{elastic net}
    \end{subfigure}
    \begin{subfigure}[c]{0.22\textwidth}
      \includegraphics[width=\textwidth]{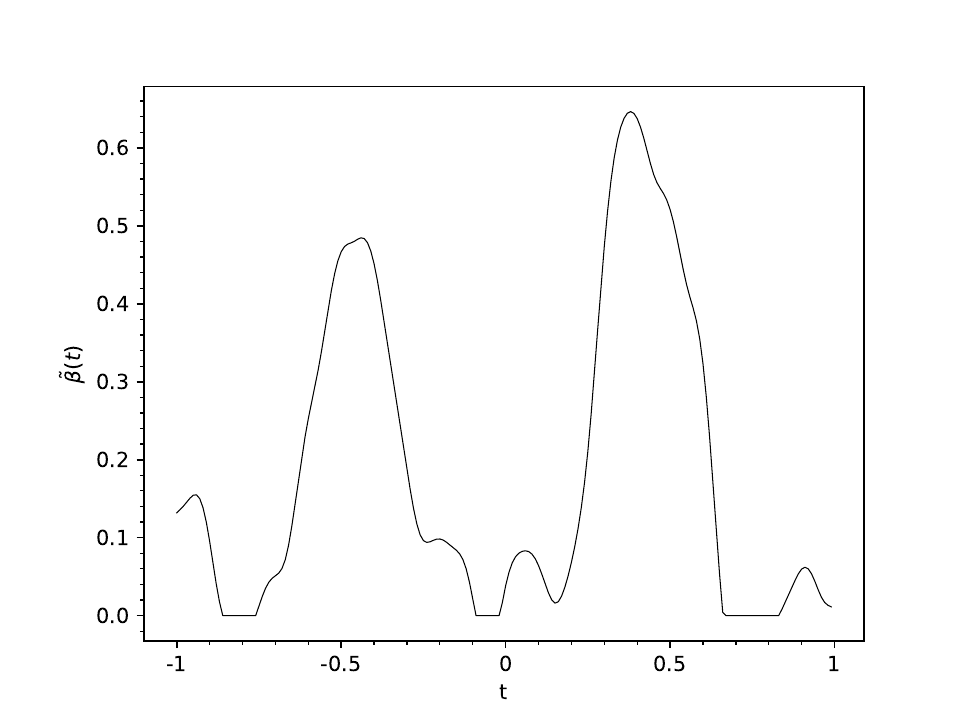}
      \caption{elastic SCAD}
    \end{subfigure}
    \begin{subfigure}[c]{0.22\textwidth}
      \includegraphics[width=\textwidth]{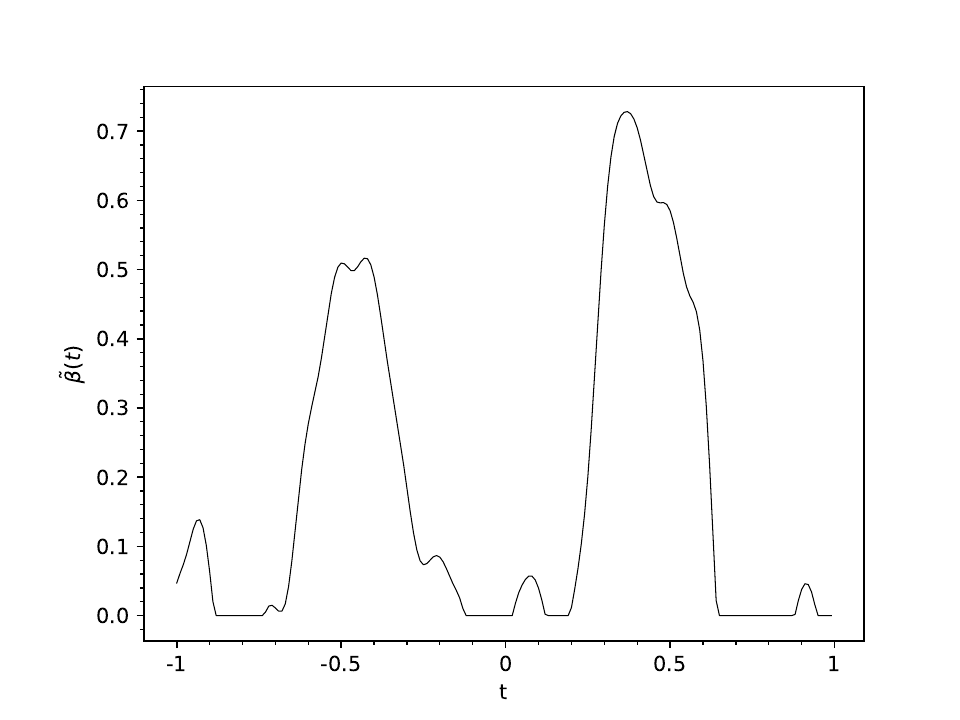}
      \caption{elastic MCP}
    \end{subfigure}

    \begin{subfigure}[c]{0.22\textwidth}
      \includegraphics[width=\textwidth]{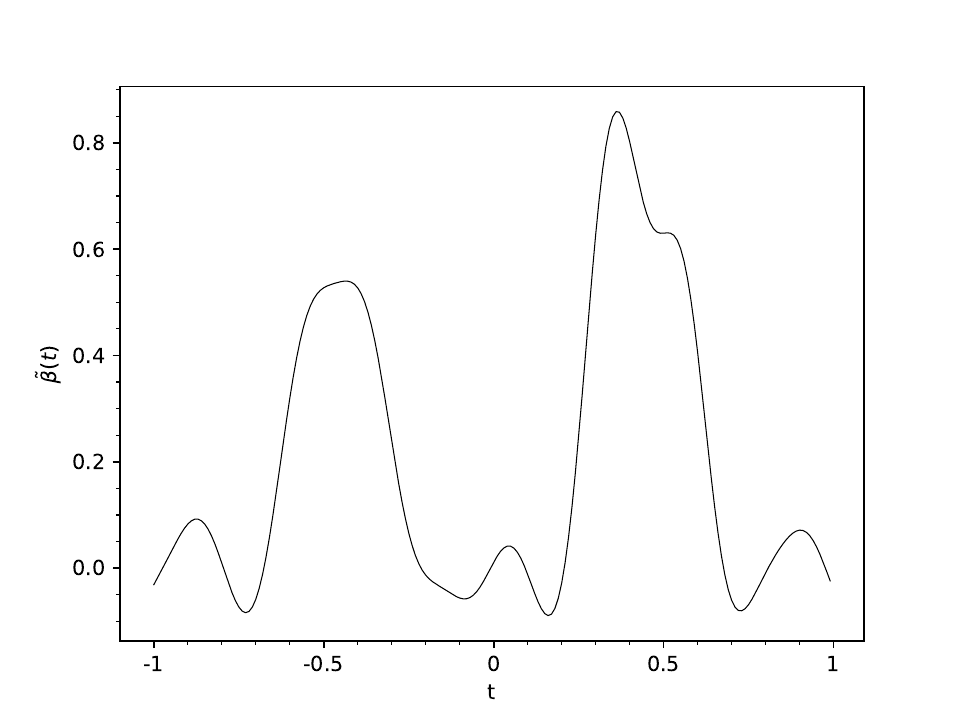}
      \caption{roughness}
    \end{subfigure}    
    \begin{subfigure}[c]{0.22\textwidth}
      \includegraphics[width=\textwidth]{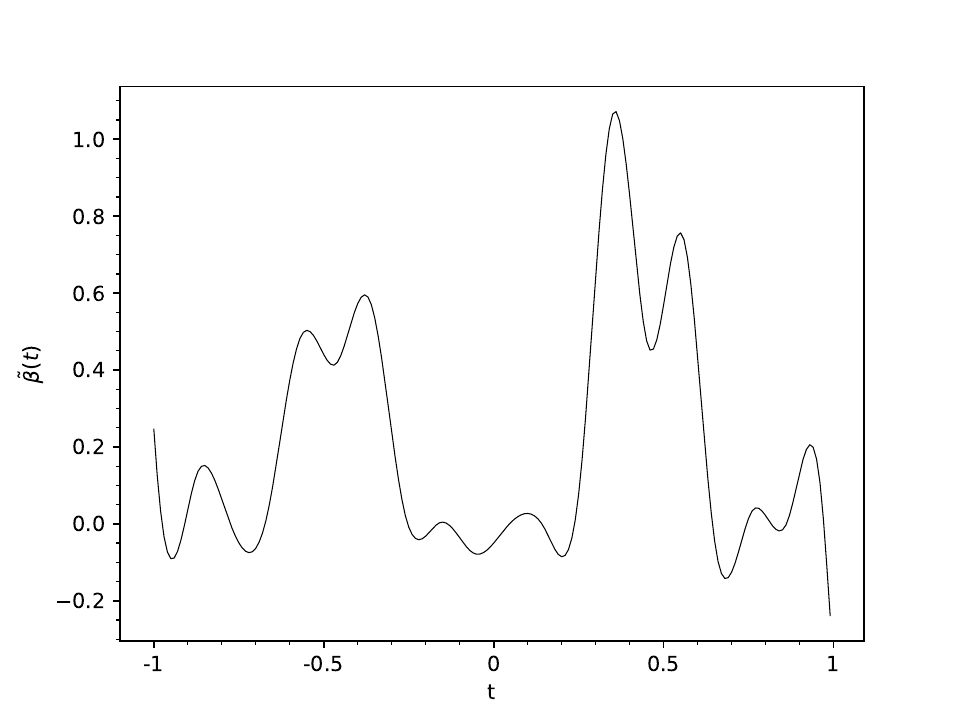}
      \caption{ridge}
    \end{subfigure}
    \begin{subfigure}[c]{0.22\textwidth}
      \includegraphics[width=\textwidth]{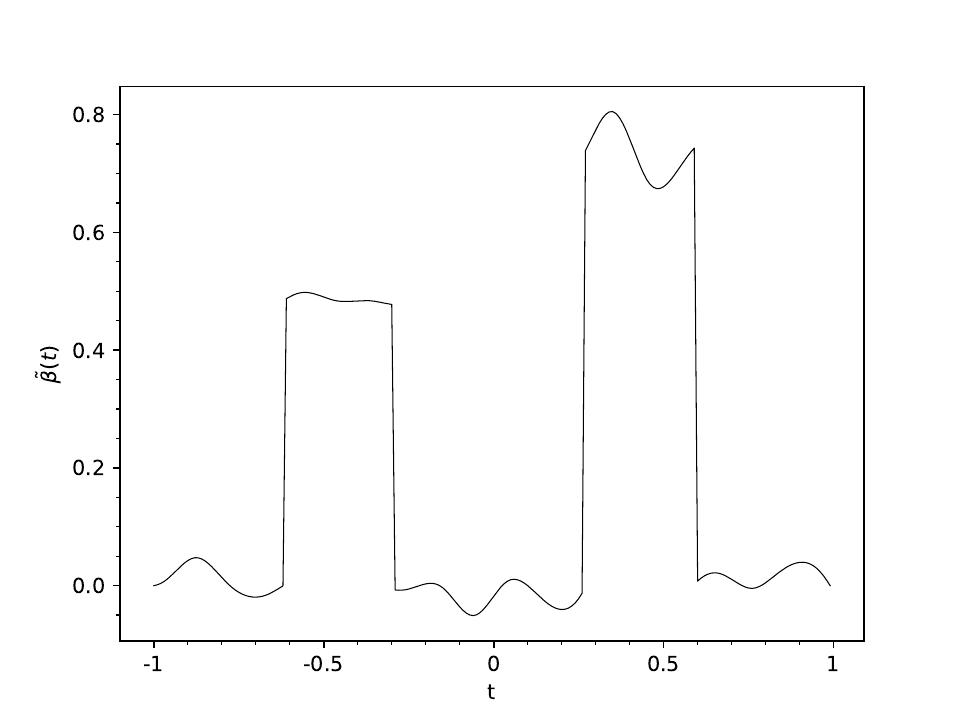}
      \caption{AATR}
    \end{subfigure}
    \begin{subfigure}[c]{0.22\textwidth}
      \includegraphics[width=\textwidth]{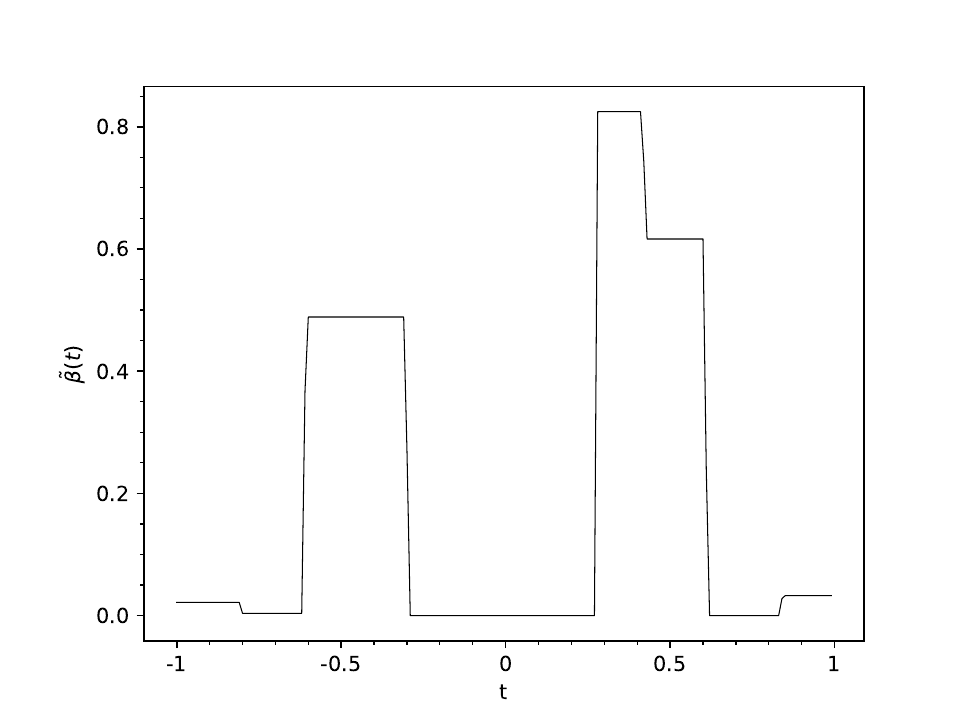}
      \caption{fused lasso}
    \end{subfigure}
  \caption{Highly dependent spline coefficients: fitted coefficient functions $\tilde{\beta}$ by penalty type}
  \label{fig:comparison_2rect_dep}
\end{figure}

\begin{figure}[H]
  \centering
    \begin{subfigure}[c]{0.27\textwidth}
      \includegraphics[width=\textwidth]{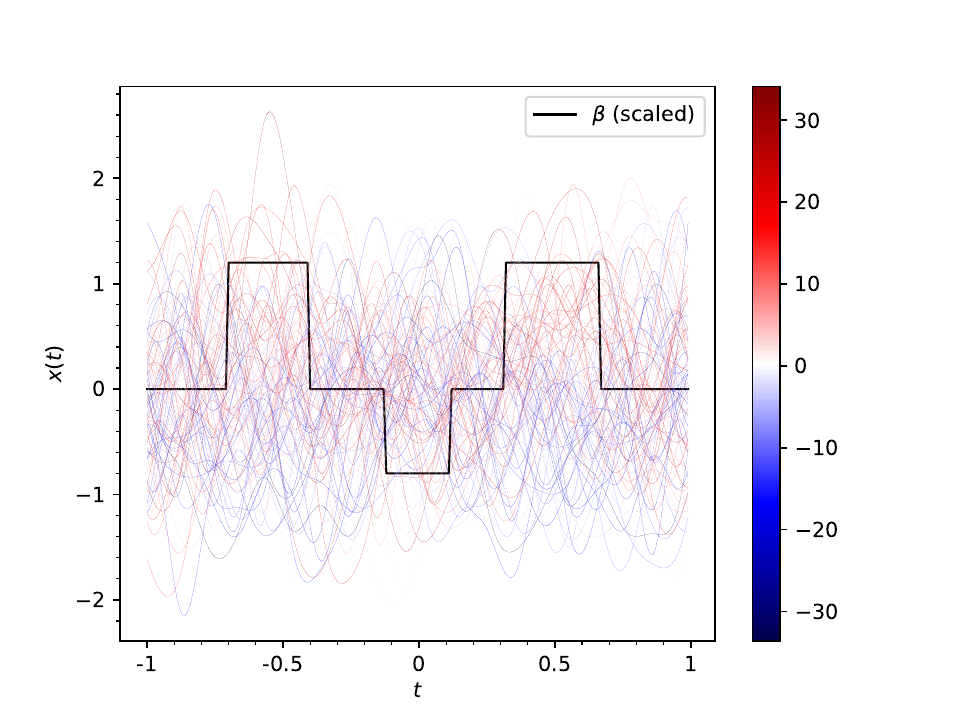}
      \caption{$x_{i}$ - independent}
    \end{subfigure}
    \begin{subfigure}[c]{0.27\textwidth}
      \includegraphics[width=\textwidth]{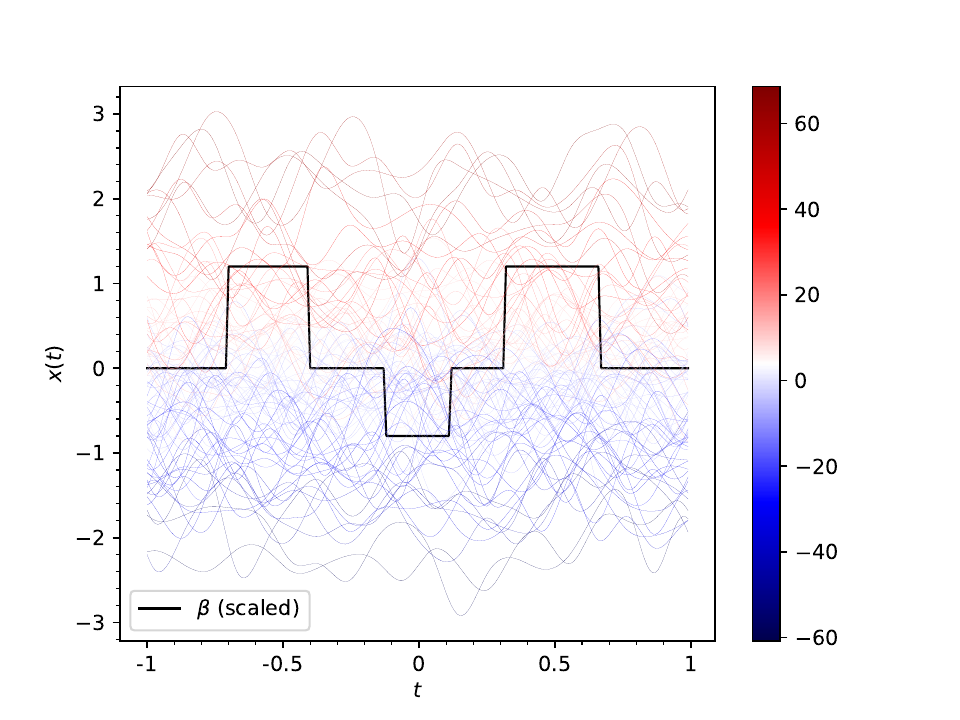}
      \caption{$x_{i}$ - dependent}
    \end{subfigure}
    \begin{subfigure}[c]{0.27\textwidth}
      \includegraphics[width=\textwidth]{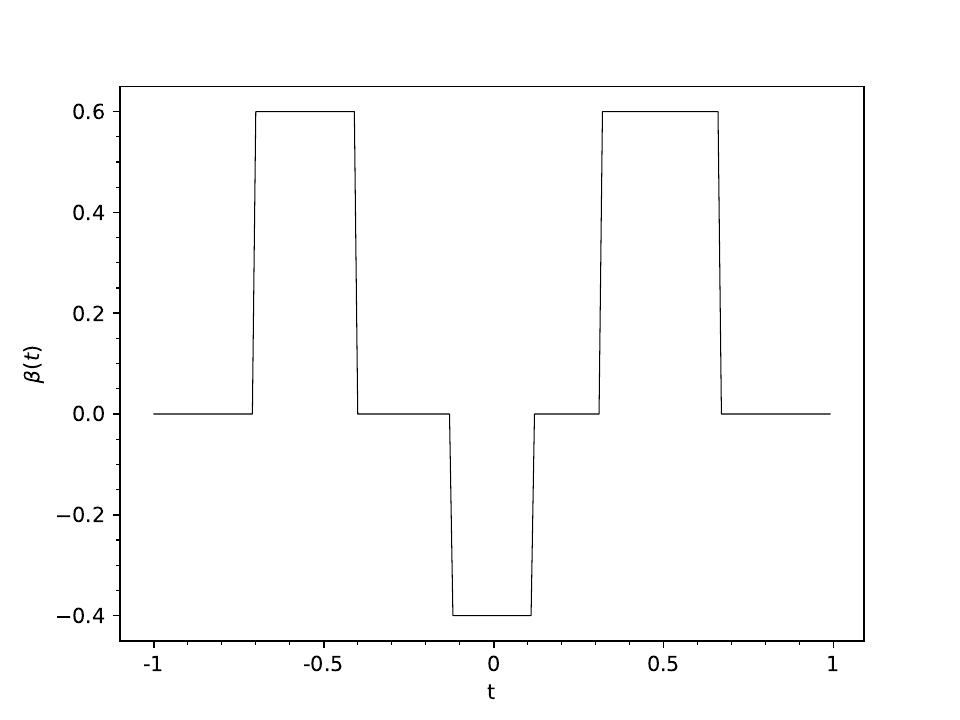}
      \caption{true $\beta$}
    \end{subfigure}
  \caption{Simulated data with independent and highly dependent spline coefficients, true $\beta$ with $q$=3}
  \label{fig:data_3rect}
\end{figure}

\begin{figure}[H]
  \centering
  	\begin{subfigure}[c]{0.22\textwidth}
      \includegraphics[width=\textwidth]{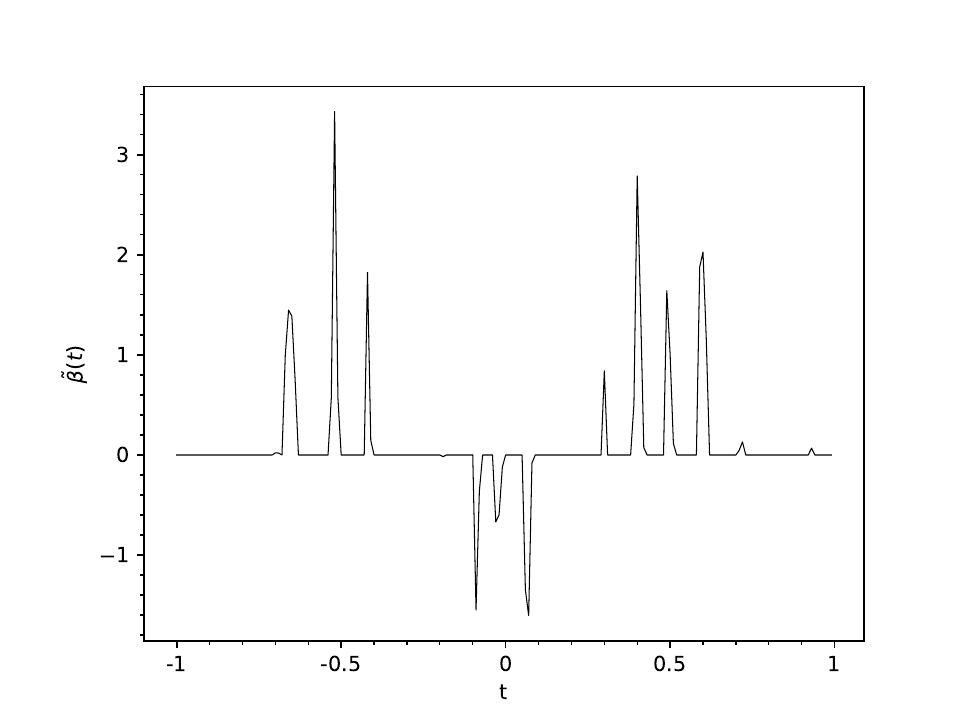}
      \caption{lasso}
    \end{subfigure}
    \begin{subfigure}[c]{0.22\textwidth}
      \includegraphics[width=\textwidth]{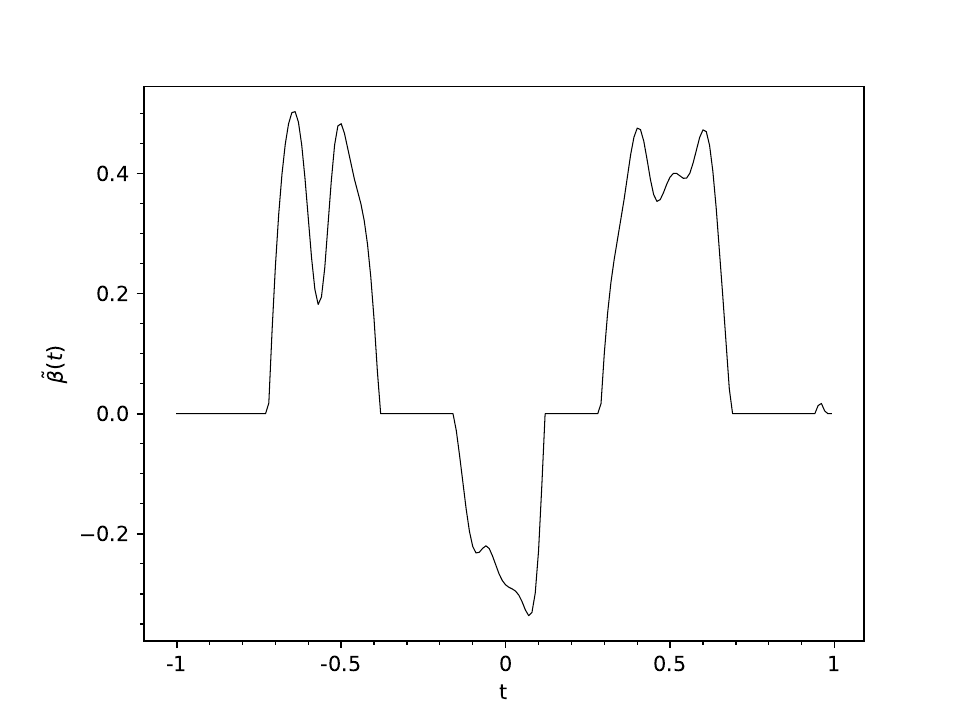}
      \caption{elastic net}
    \end{subfigure}
    \begin{subfigure}[c]{0.22\textwidth}
      \includegraphics[width=\textwidth]{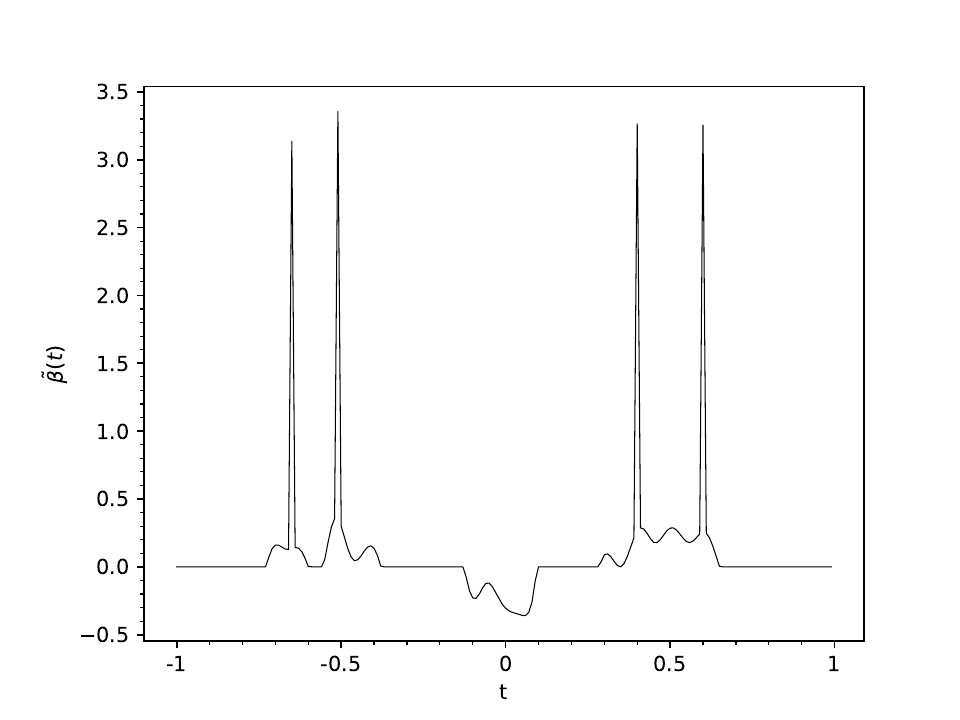}
      \caption{elastic SCAD}
    \end{subfigure}
    \begin{subfigure}[c]{0.22\textwidth}
      \includegraphics[width=\textwidth]{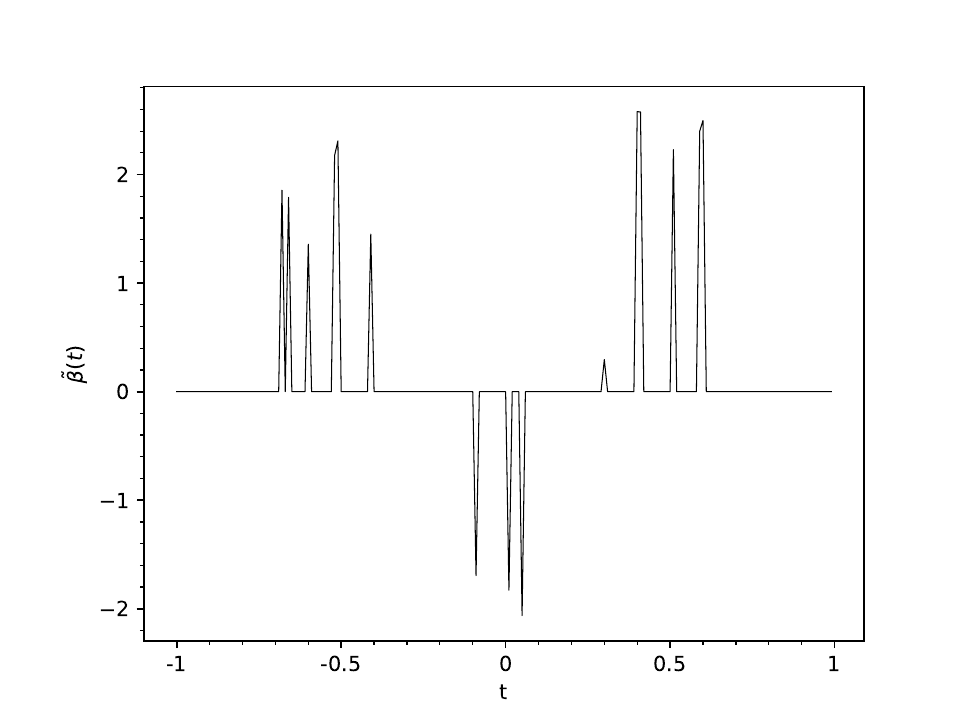}
      \caption{elastic MCP}
    \end{subfigure}

    \begin{subfigure}[c]{0.22\textwidth}
      \includegraphics[width=\textwidth]{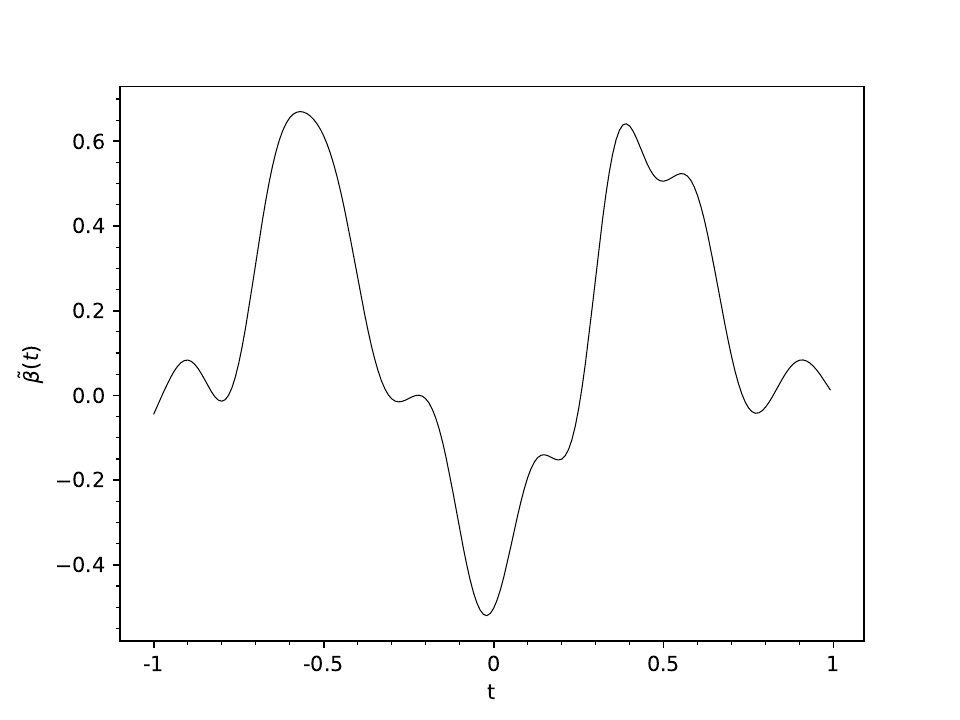}
      \caption{roughness}
    \end{subfigure}    
    \begin{subfigure}[c]{0.22\textwidth}
      \includegraphics[width=\textwidth]{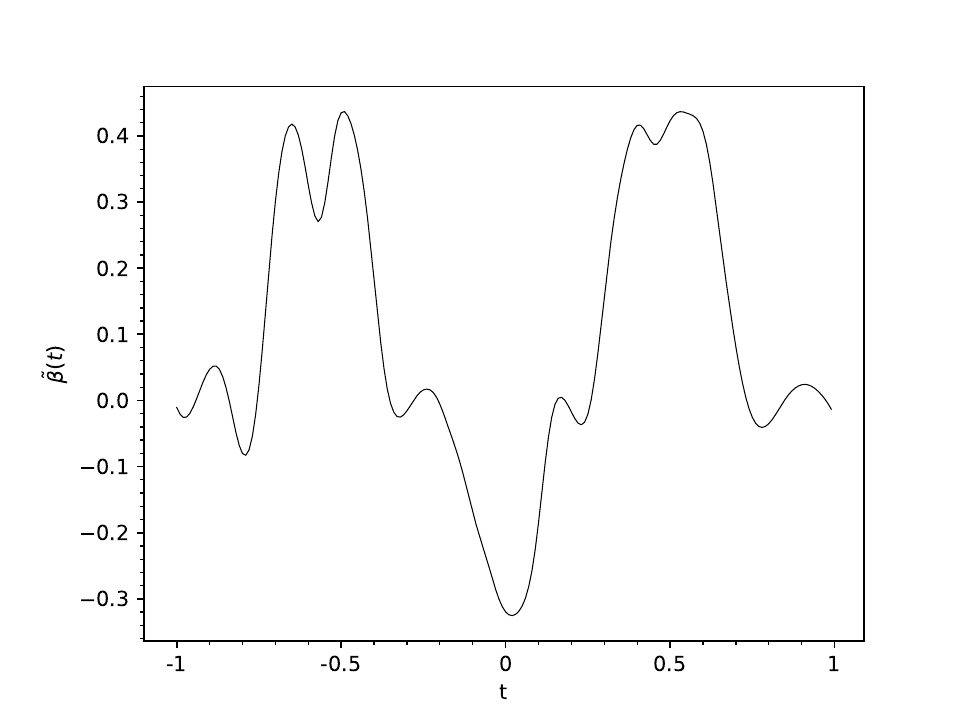}
      \caption{ridge}
    \end{subfigure}
    \begin{subfigure}[c]{0.22\textwidth}
      \includegraphics[width=\textwidth]{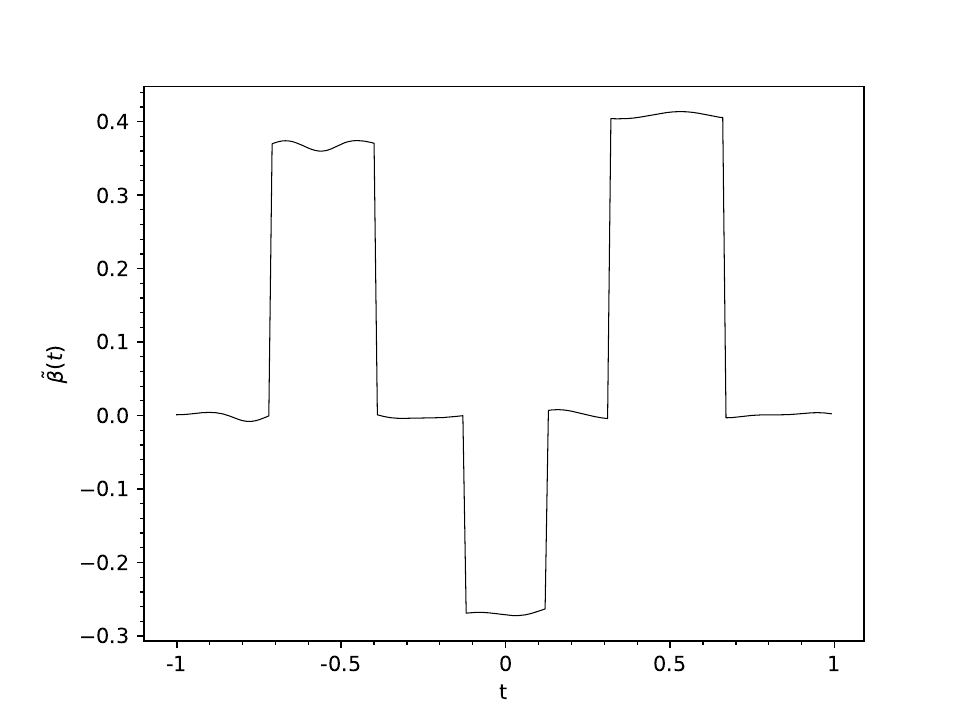}
      \caption{AATR}
    \end{subfigure}
    \begin{subfigure}[c]{0.22\textwidth}
      \includegraphics[width=\textwidth]{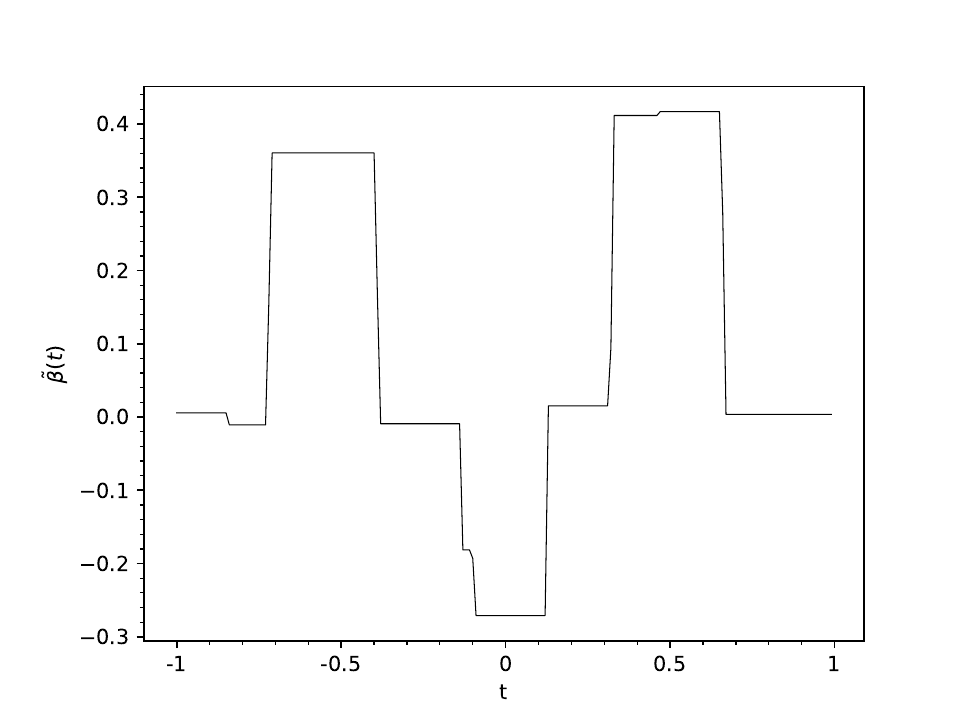}
      \caption{fused lasso}
    \end{subfigure}
  \caption{Independent spline coefficients: fitted coefficient functions $\tilde{\beta}$ by penalty type}
  \label{fig:comparison_3rect_ind}
\end{figure}

\begin{figure}[H]
  \centering
  	\begin{subfigure}[c]{0.22\textwidth}
      \includegraphics[width=\textwidth]{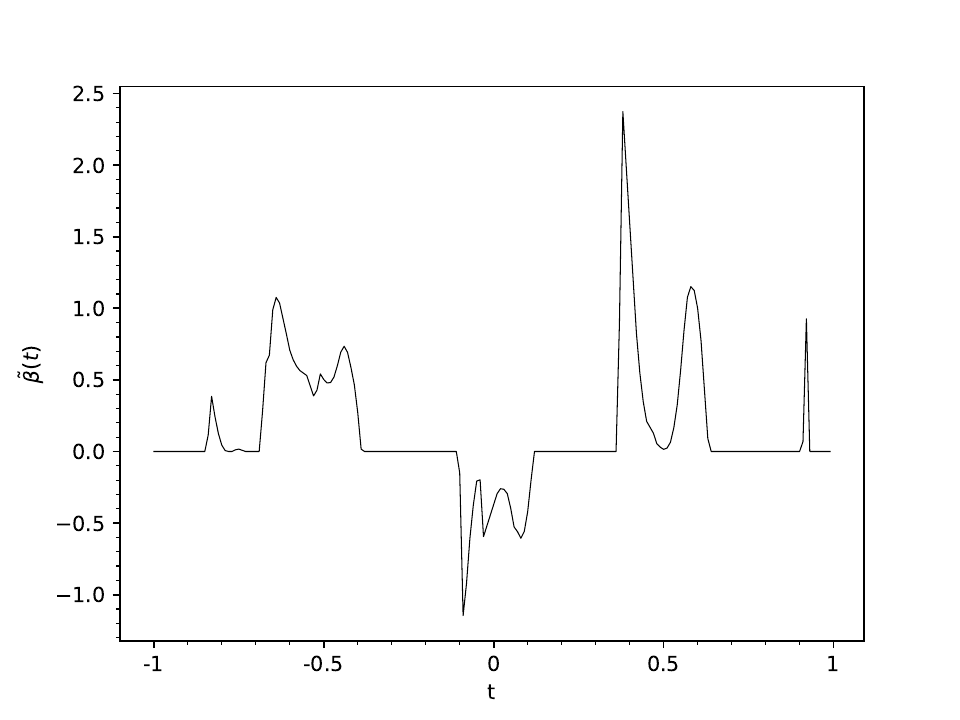}
      \caption{lasso}
    \end{subfigure}
    \begin{subfigure}[c]{0.22\textwidth}
      \includegraphics[width=\textwidth]{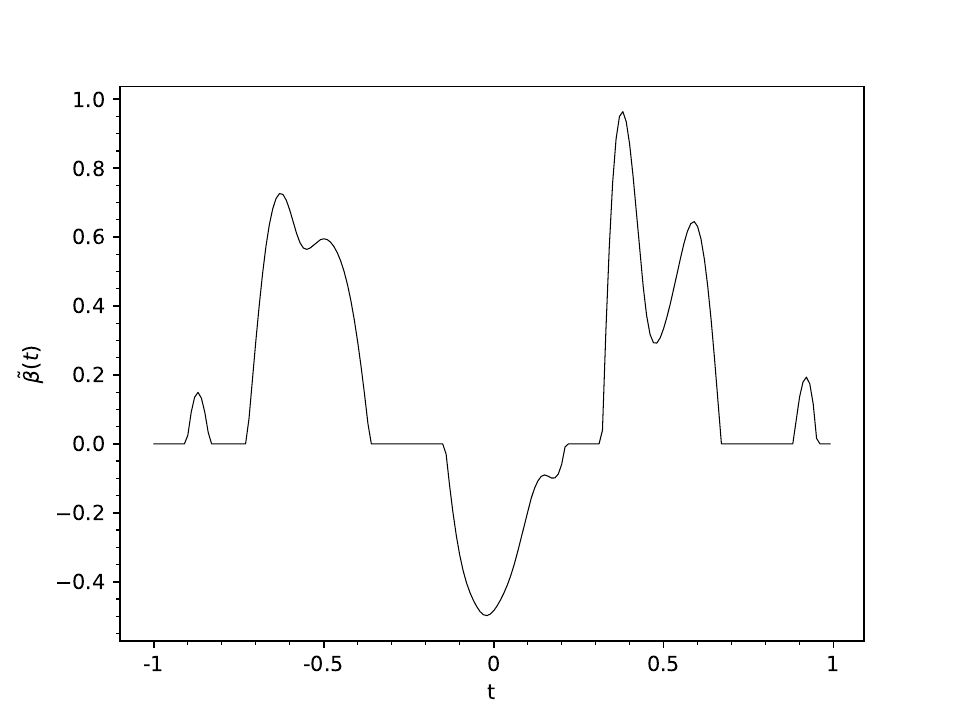}
      \caption{elastic net}
    \end{subfigure}
    \begin{subfigure}[c]{0.22\textwidth}
      \includegraphics[width=\textwidth]{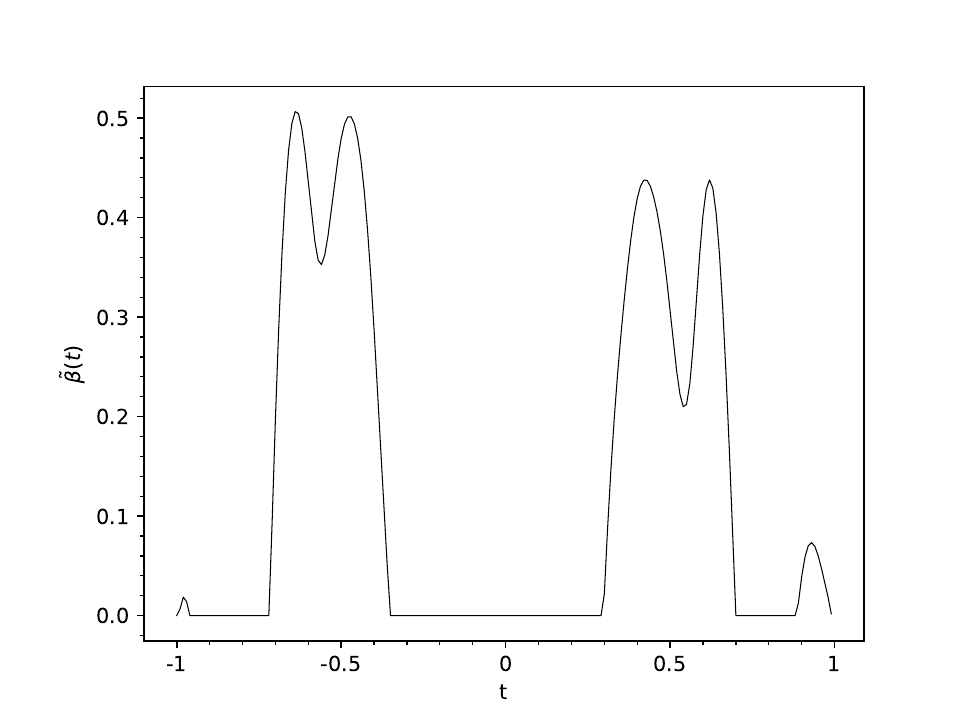}
      \caption{elastic SCAD}
    \end{subfigure}
    \begin{subfigure}[c]{0.22\textwidth}
      \includegraphics[width=\textwidth]{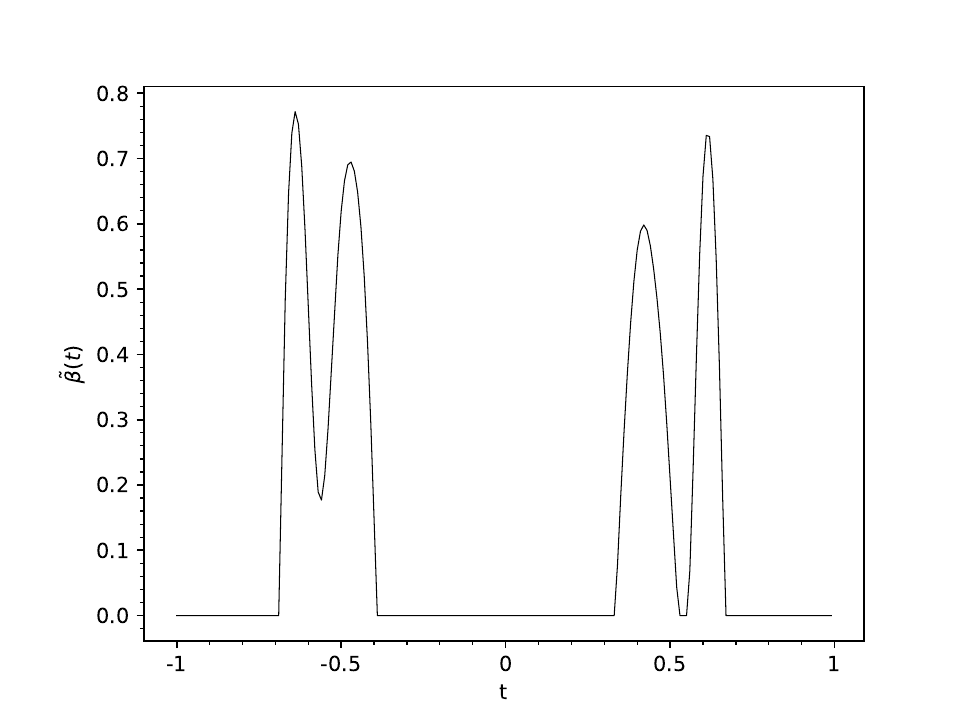}
      \caption{elastic MCP}
    \end{subfigure}

    \begin{subfigure}[c]{0.22\textwidth}
      \includegraphics[width=\textwidth]{figs/3rect_dep/roughness.pdf}
      \caption{roughness}
    \end{subfigure}    
    \begin{subfigure}[c]{0.22\textwidth}
      \includegraphics[width=\textwidth]{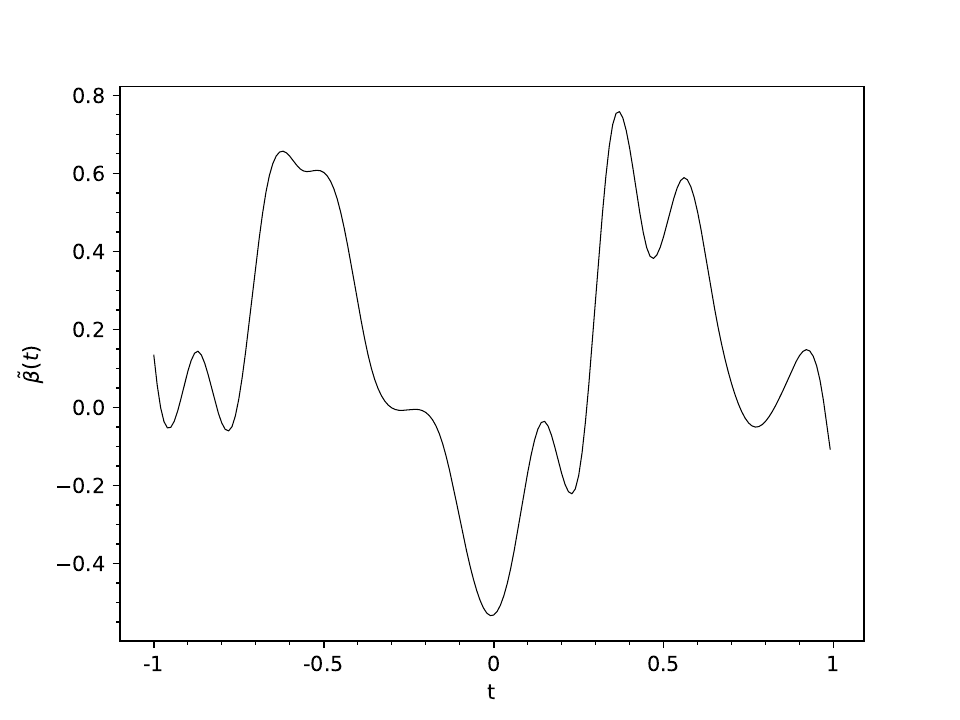}
      \caption{ridge}
    \end{subfigure}
    \begin{subfigure}[c]{0.22\textwidth}
      \includegraphics[width=\textwidth]{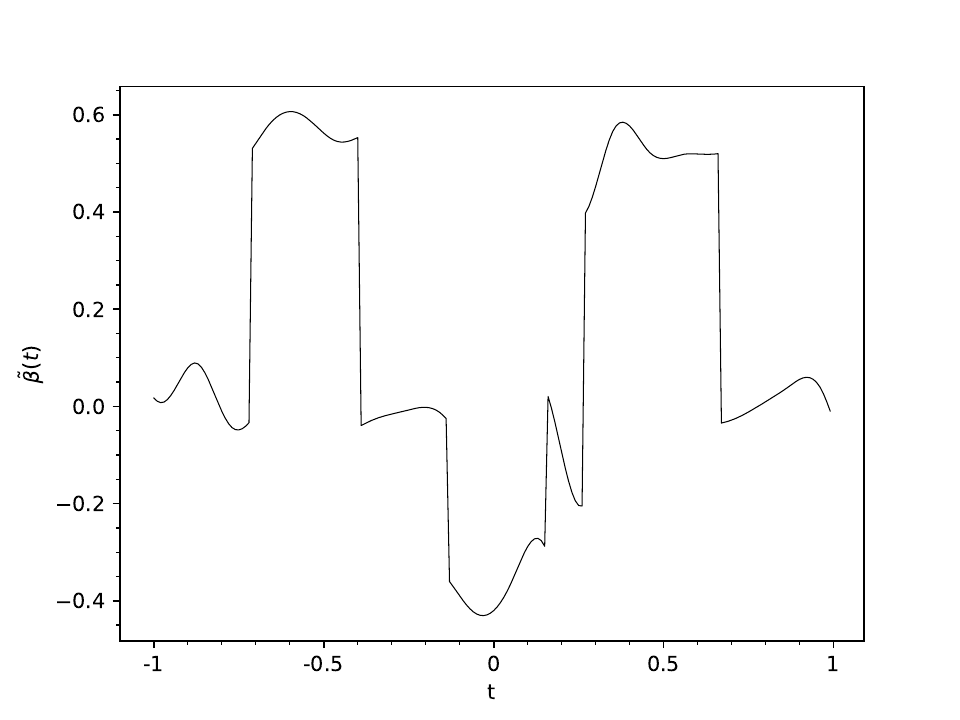}
      \caption{AATR}
    \end{subfigure}
    \begin{subfigure}[c]{0.22\textwidth}
      \includegraphics[width=\textwidth]{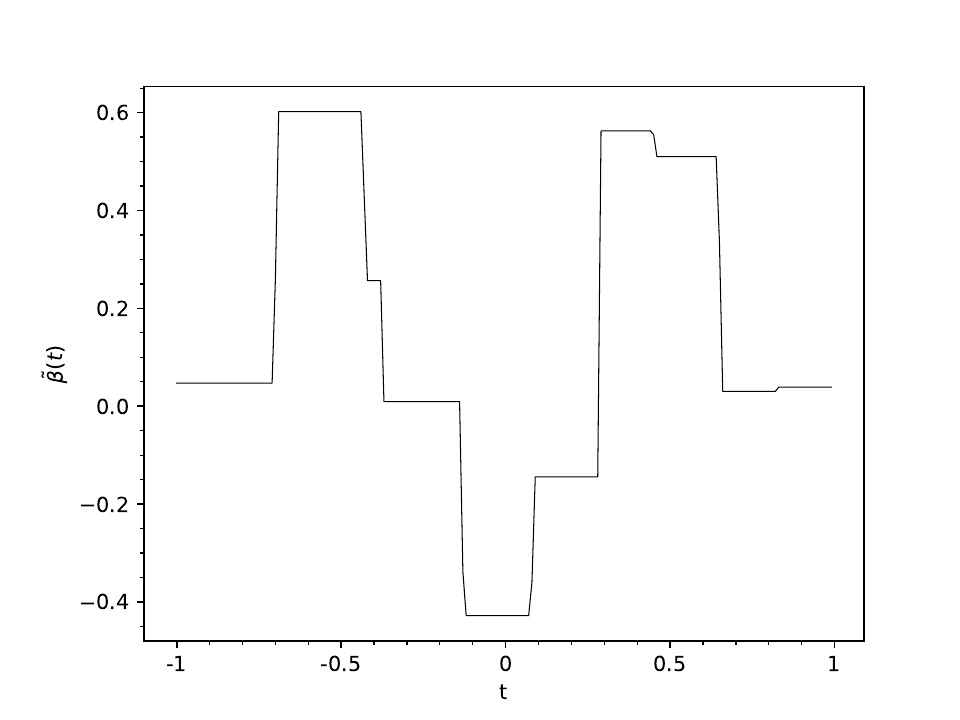}
      \caption{fused lasso}
    \end{subfigure}
  \caption{Highly dependent spline coefficients: fitted coefficient functions $\tilde{\beta}$ by penalty type}
  \label{fig:comparison_3rect_dep}
\end{figure}

\begin{figure}[H]
  \centering
    \begin{subfigure}[c]{0.27\textwidth}
      \includegraphics[width=\textwidth]{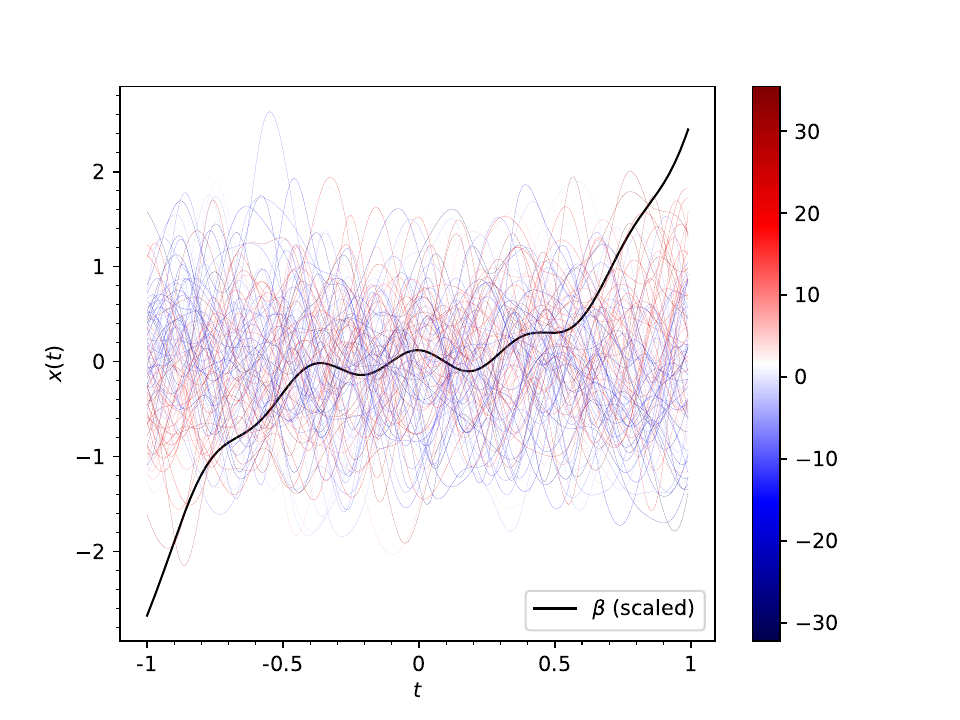}
      \caption{$x_{i}$ - independent}
    \end{subfigure}
    \begin{subfigure}[c]{0.27\textwidth}
      \includegraphics[width=\textwidth]{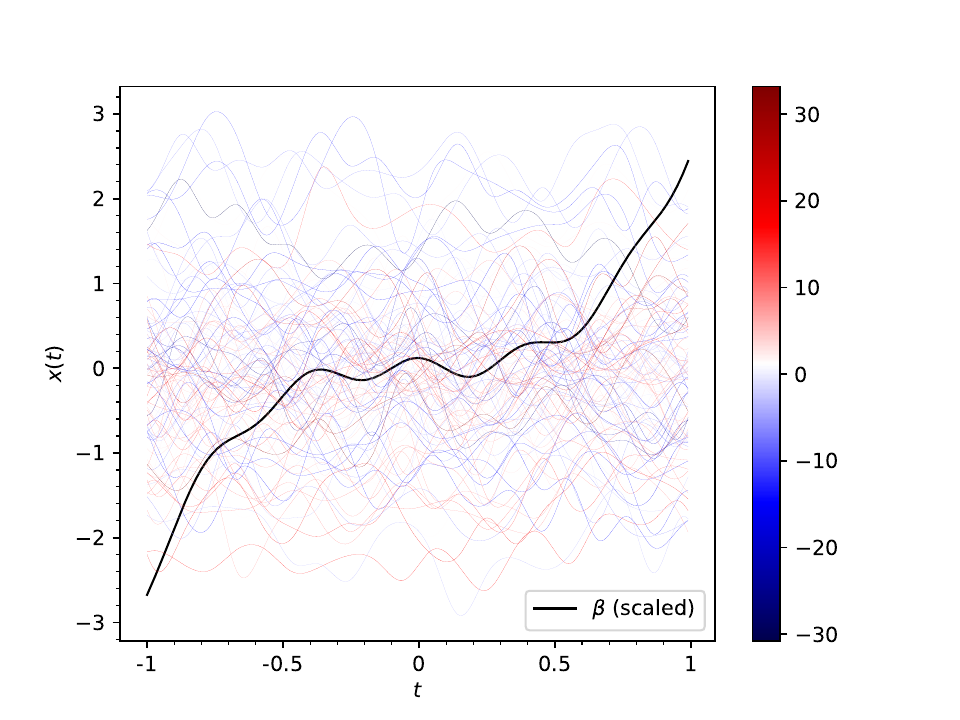}
      \caption{$x_{i}$ - dependent}
    \end{subfigure}
    \begin{subfigure}[c]{0.27\textwidth}
      \includegraphics[width=\textwidth]{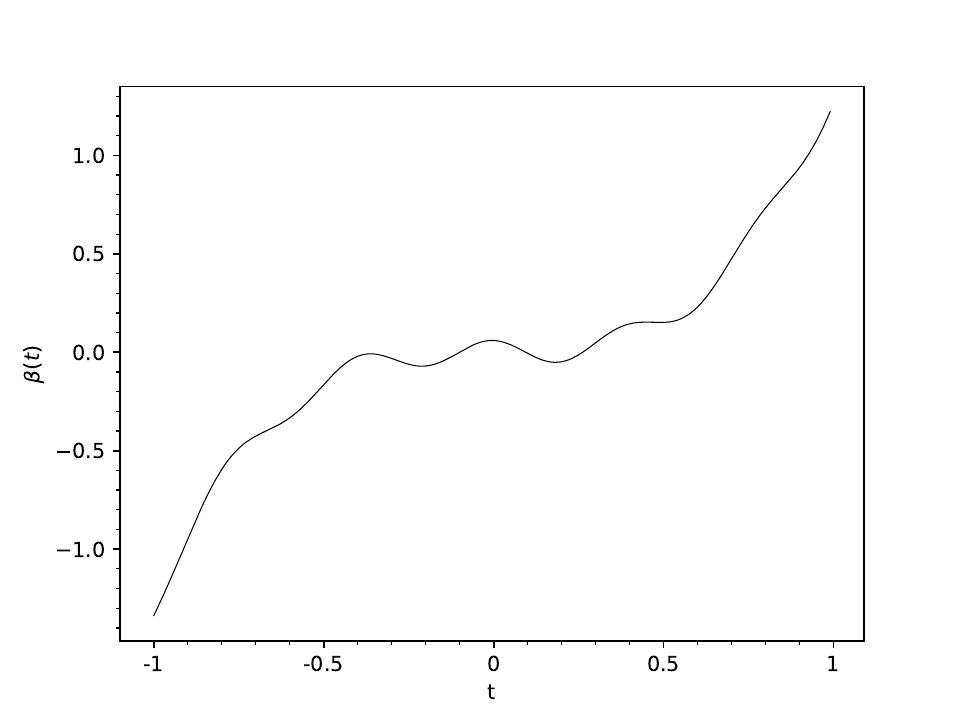}
      \caption{true $\beta$}
    \end{subfigure}
  \caption{Simulated data with independent and highly dependent spline coefficients, true $\beta$ as a smooth shape}
  \label{fig:data_smooth}
\end{figure}

\begin{figure}[H]
  \centering
  	\begin{subfigure}[c]{0.22\textwidth}
      \includegraphics[width=\textwidth]{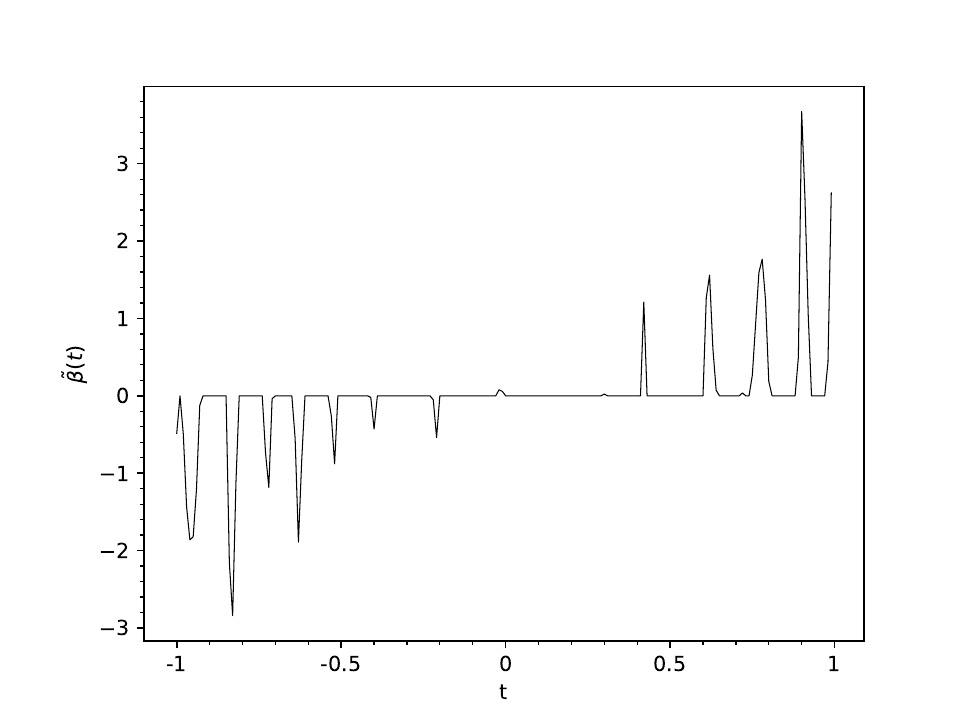}
      \caption{lasso}
    \end{subfigure}
    \begin{subfigure}[c]{0.22\textwidth}
      \includegraphics[width=\textwidth]{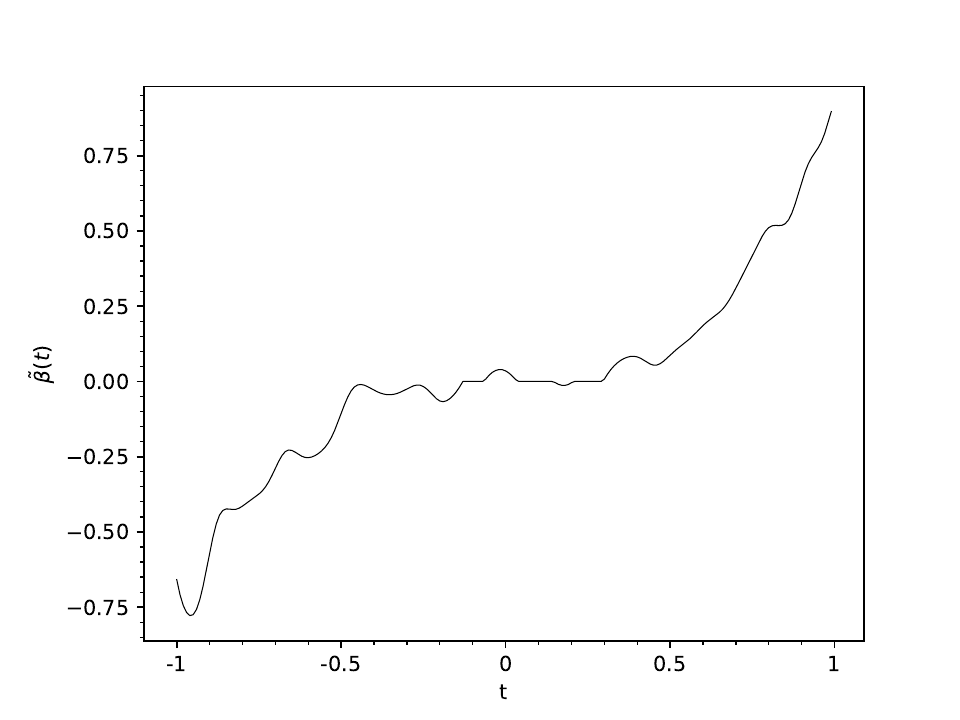}
      \caption{elastic net}
    \end{subfigure}
    \begin{subfigure}[c]{0.22\textwidth}
      \includegraphics[width=\textwidth]{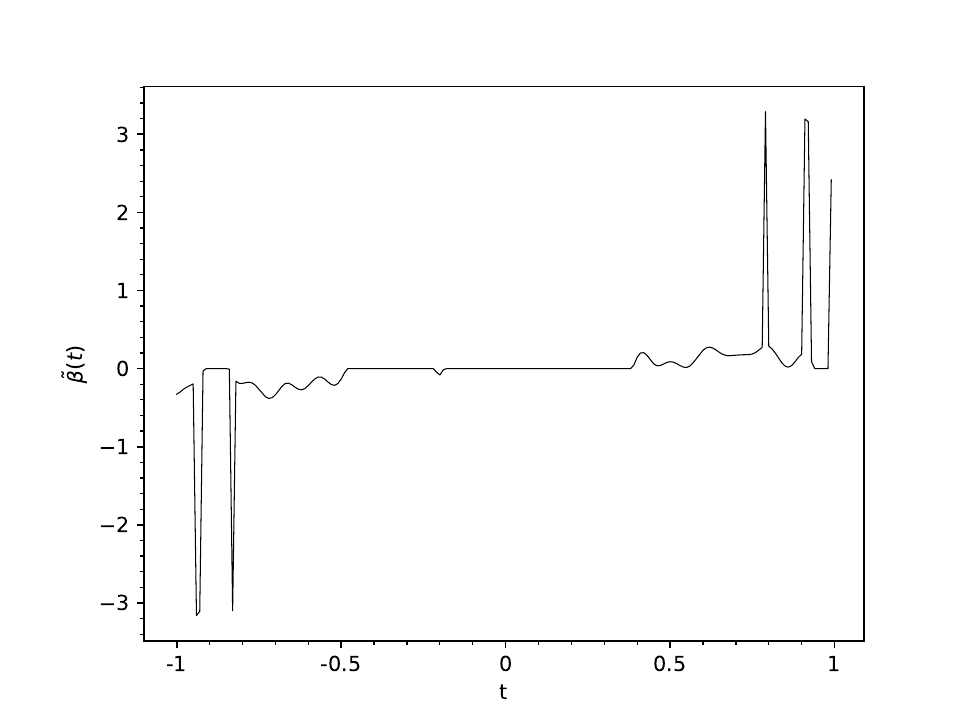}
      \caption{elastic SCAD}
    \end{subfigure}
    \begin{subfigure}[c]{0.22\textwidth}
      \includegraphics[width=\textwidth]{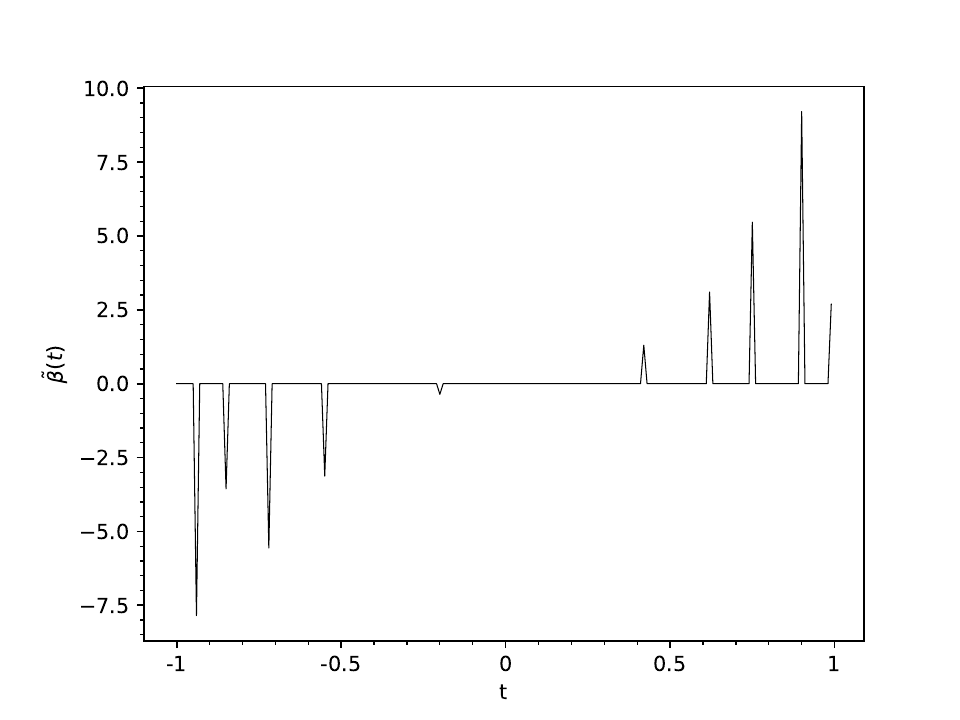}
      \caption{elastic MCP}
    \end{subfigure}

    \begin{subfigure}[c]{0.22\textwidth}
      \includegraphics[width=\textwidth]{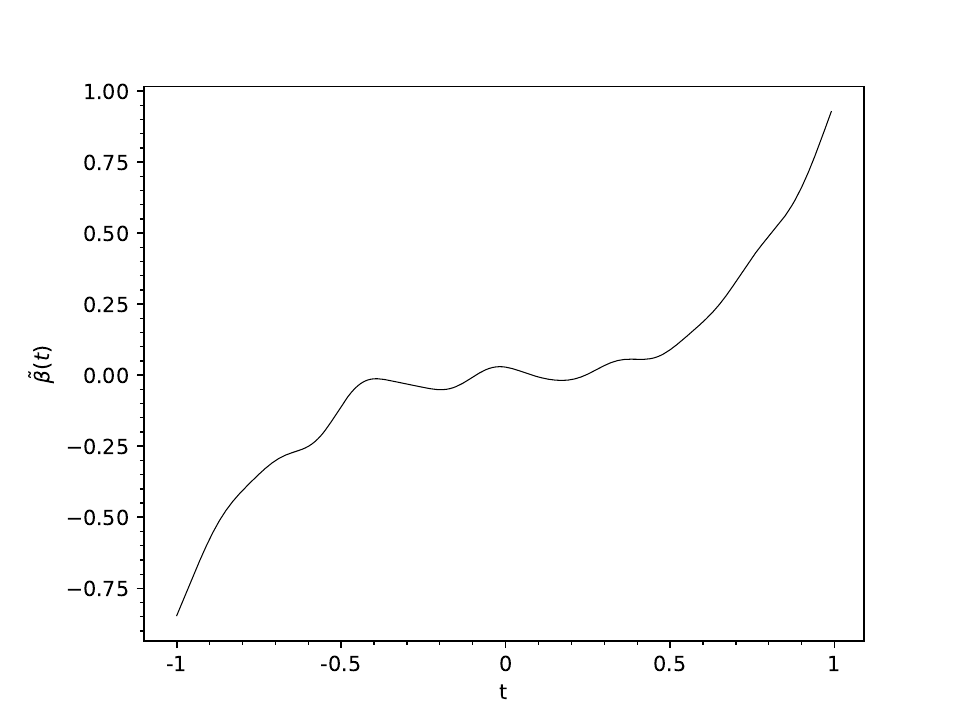}
      \caption{roughness}
    \end{subfigure}    
    \begin{subfigure}[c]{0.22\textwidth}
      \includegraphics[width=\textwidth]{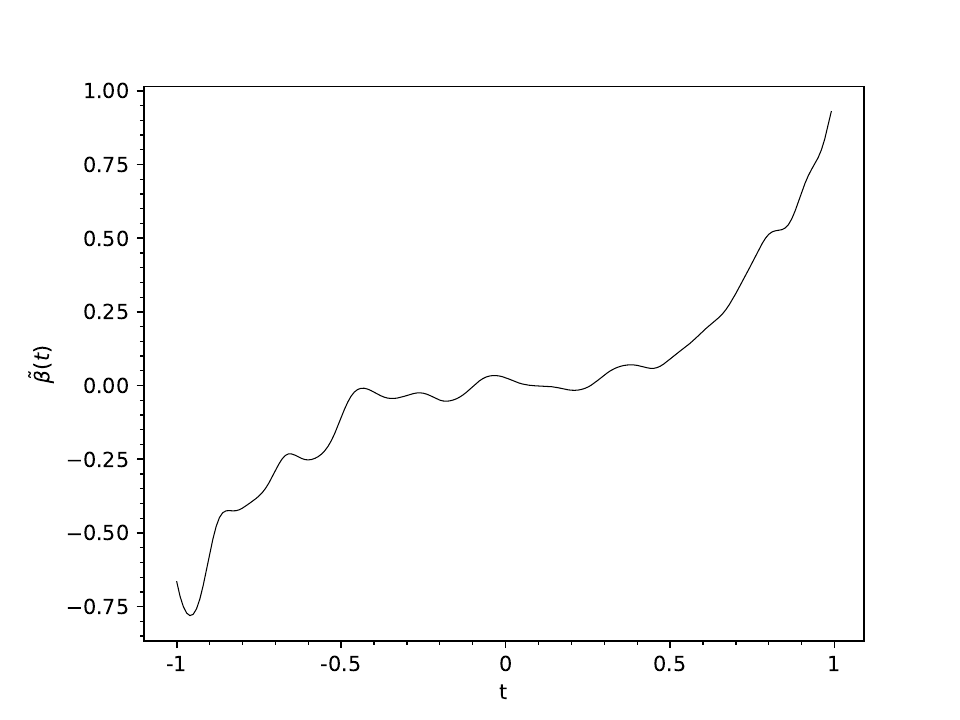}
      \caption{ridge}
    \end{subfigure}
    \begin{subfigure}[c]{0.22\textwidth}
      \includegraphics[width=\textwidth]{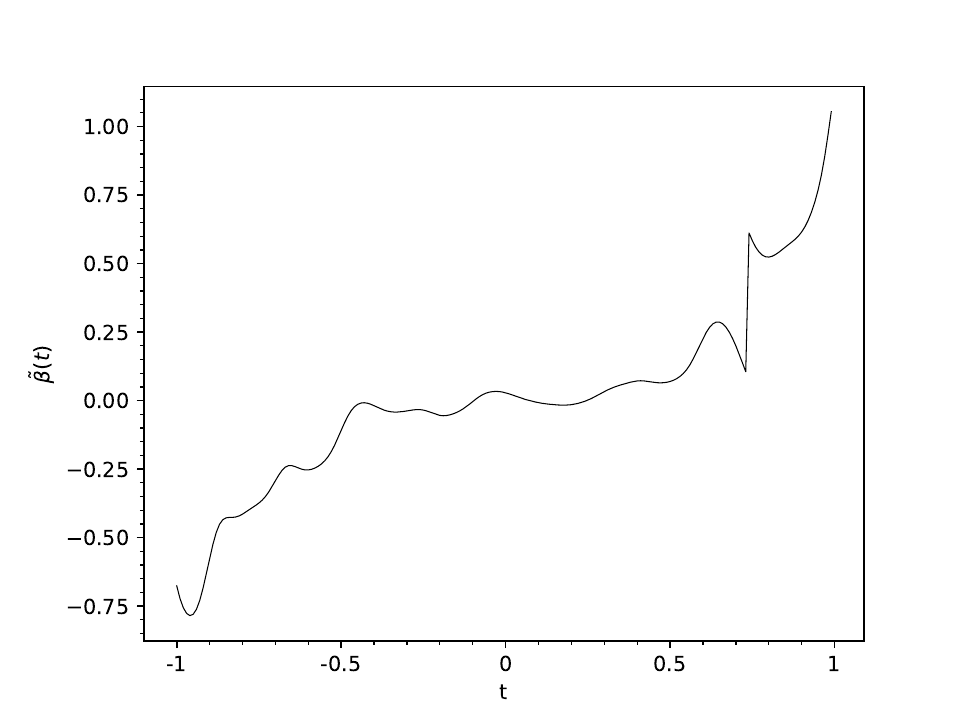}
      \caption{AATR}
    \end{subfigure}
    \begin{subfigure}[c]{0.22\textwidth}
      \includegraphics[width=\textwidth]{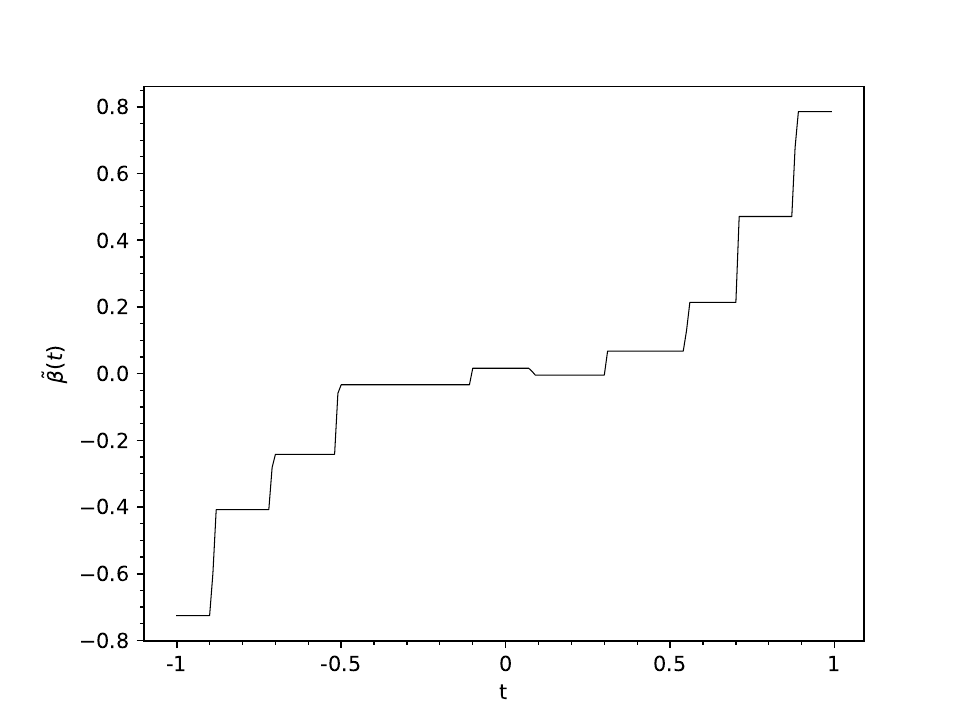}
      \caption{fused lasso}
    \end{subfigure}
  \caption{Independent spline coefficients: fitted coefficient functions $\tilde{\beta}$ by penalty type}
  \label{fig:comparison_smooth_ind}
\end{figure}

\begin{figure}[H]
  \centering
  	\begin{subfigure}[c]{0.22\textwidth}
      \includegraphics[width=\textwidth]{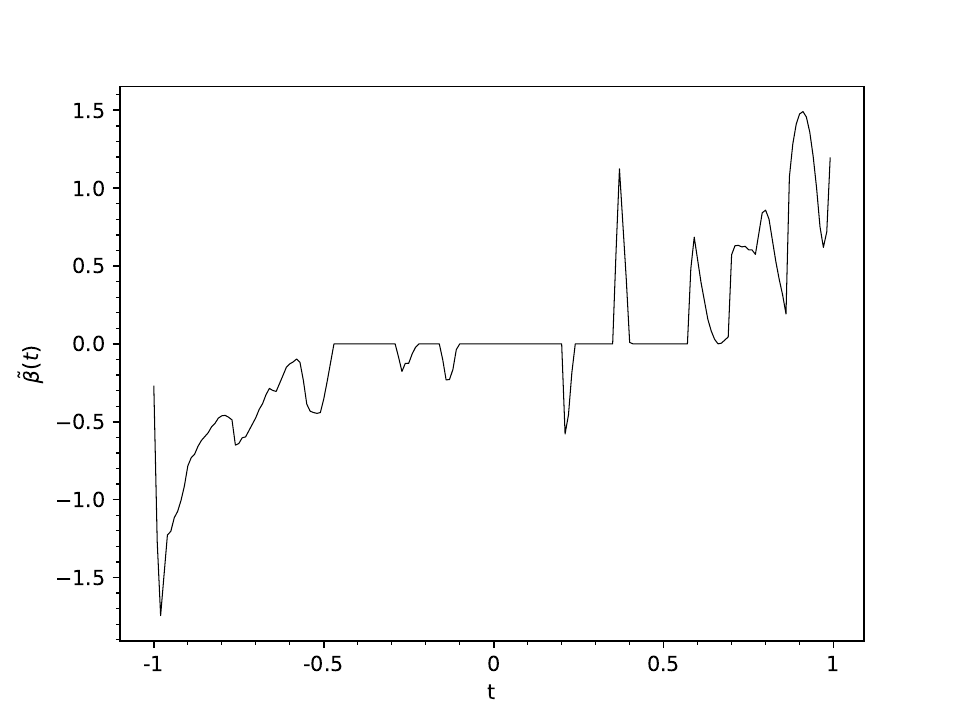}
      \caption{lasso}
    \end{subfigure}
    \begin{subfigure}[c]{0.22\textwidth}
      \includegraphics[width=\textwidth]{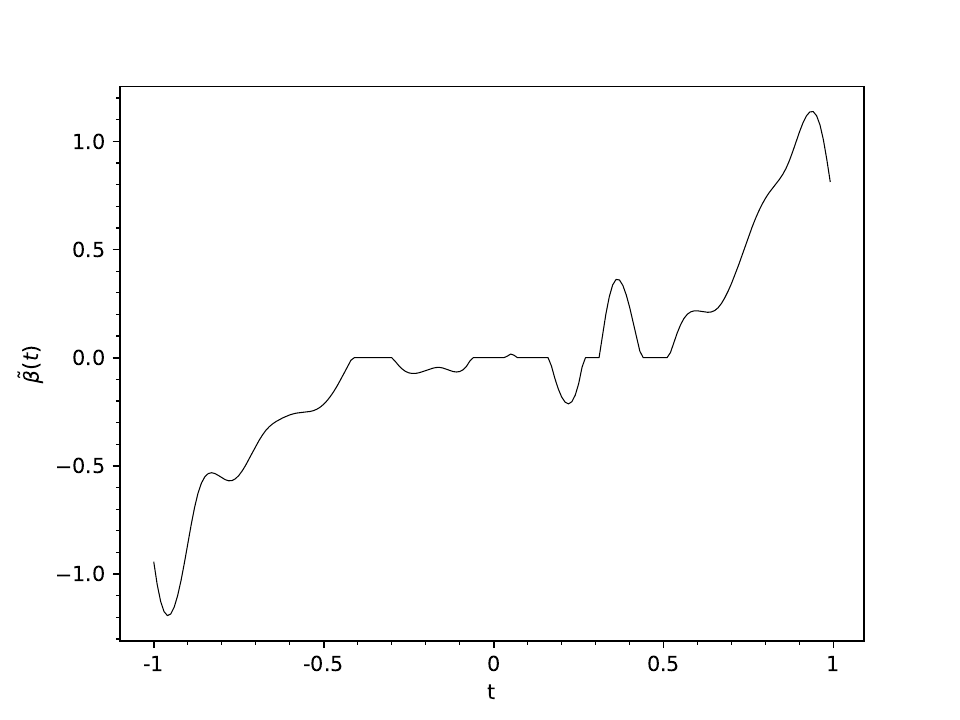}
      \caption{elastic net}
    \end{subfigure}
    \begin{subfigure}[c]{0.22\textwidth}
      \includegraphics[width=\textwidth]{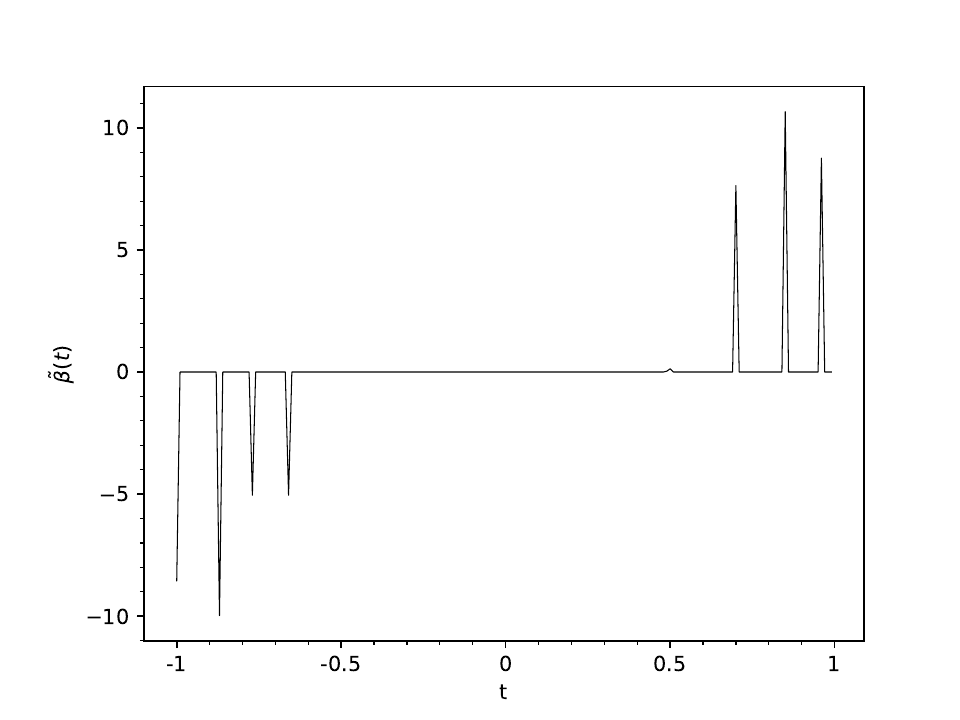}
      \caption{elastic SCAD}
    \end{subfigure}
    \begin{subfigure}[c]{0.22\textwidth}
      \includegraphics[width=\textwidth]{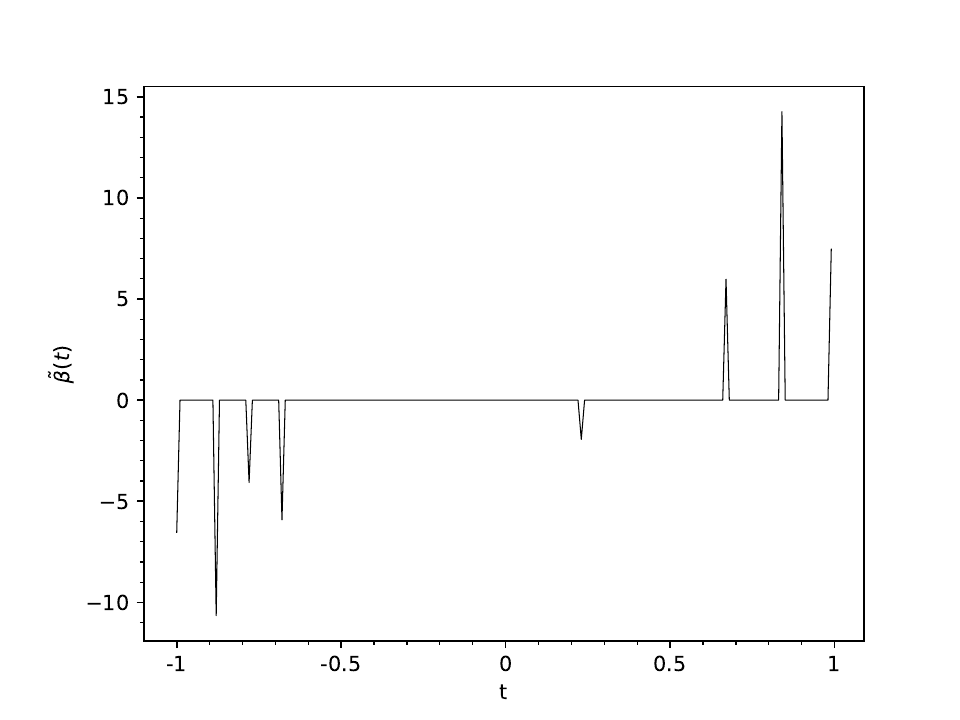}
      \caption{elastic MCP}
    \end{subfigure}

    \begin{subfigure}[c]{0.22\textwidth}
      \includegraphics[width=\textwidth]{figs/smooth_ind/roughness.pdf}
      \caption{roughness}
    \end{subfigure}        
    \begin{subfigure}[c]{0.22\textwidth}
      \includegraphics[width=\textwidth]{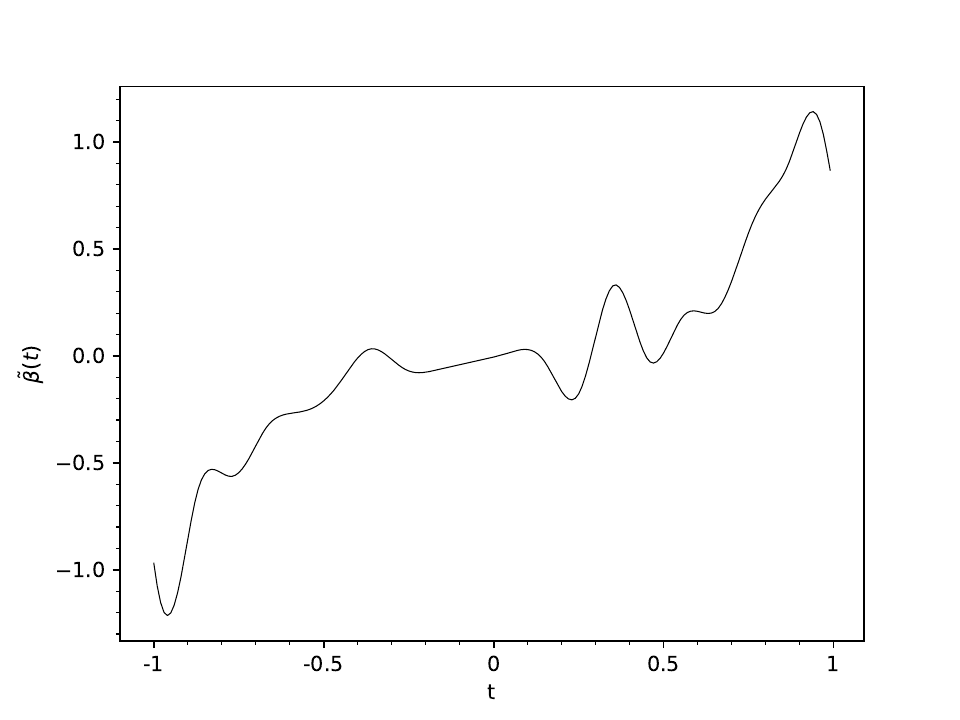}
      \caption{ridge}
    \end{subfigure}
    \begin{subfigure}[c]{0.22\textwidth}
      \includegraphics[width=\textwidth]{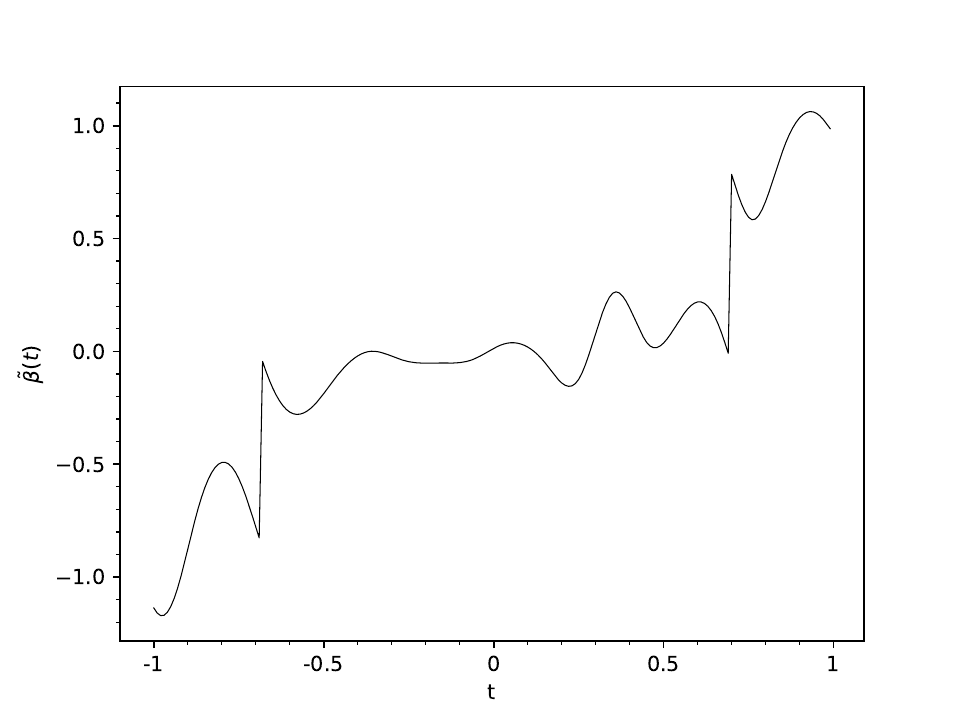}
      \caption{AATR}
    \end{subfigure}
    \begin{subfigure}[c]{0.22\textwidth}
      \includegraphics[width=\textwidth]{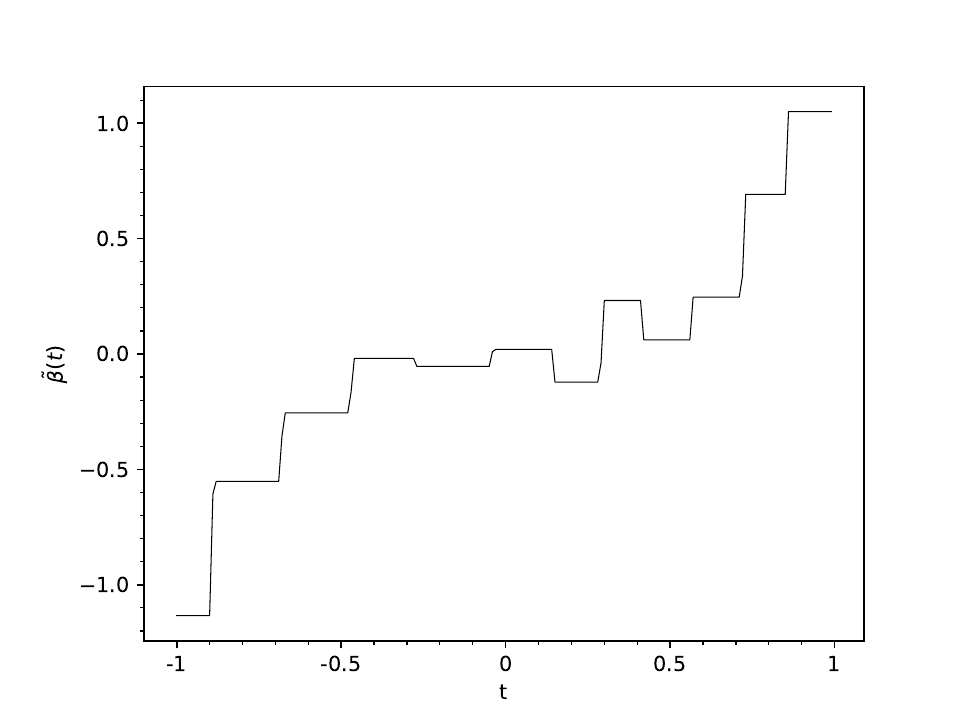}
      \caption{fused lasso}
    \end{subfigure}
  \caption{Highly dependent spline coefficients: fitted coefficient functions $\tilde{\beta}$ by penalty type}
  \label{fig:comparison_smooth_dep}
\end{figure}

The objective of the simulations is to show the behaviour of our method in different scenarios, where the true coefficient function $\beta$ is either a mixture of $q$=$1/2/3$ rectangles or a smooth shape that in principle is not compatible with a rectangle-based penalization, resulting in four different shapes tested. All the simulations share the same base model for the input data, which is a cubic B-spline with 40 inner knots equispaced between $[-2,2]$, while the spline coefficients are sampled from a multivariate normal for each of the $N$ observations. In particular, we repeat the analysis with two different settings, one with independent spline coefficients, and the other with highly dependent ones, for a total of eight configurations. Once the coefficients are generated and the functional model is determined for the single experiment, the regressors $x_{i}$ are obtained by evaluating the functions in $p$=$200$ equispaced points in $[-1,1]$, while the responses $y_{i}$ are computed according to Equation \ref{eq:linearmodel} with $\beta_{0}$=$0$ and $\epsilon_{i} \sim \mathcal{N}(0,1)$. The sample size is fixed at $N$=$100$ for all the experiments. Figures \ref{fig:data_1rect}, \ref{fig:comparison_1rect_ind} and  \ref{fig:comparison_1rect_dep} portray the input curves and the fitted coefficient functions for the configurations with $q$=$1$, while Figures \ref{fig:data_2rect}, \ref{fig:comparison_2rect_ind} and  \ref{fig:comparison_2rect_dep} show the case $q$=$2$, Figures \ref{fig:data_3rect}, \ref{fig:comparison_3rect_ind} and \ref{fig:comparison_3rect_dep} show the case $q$=$3$ and finally Figures \ref{fig:data_smooth}, \ref{fig:comparison_smooth_ind} and \ref{fig:comparison_smooth_dep} illustrate the smooth scenario. The results for all the simulations are presented together in Table \ref{tab:simulations}, which reports the mean regression error on the test set obtained by 3-fold cross-validation of the $N$ samples. The hyperparameters are selected with grid search by 3-fold cross-validation on the current split. It is worth noting that for a given configuration, the method with the lowest error is not always the one that is closer to recovering the correct shape, and overall the fused lasso and AATR are the ones that are able to better approximate the true pattern. While the $L_{1}$ norm allows the fused lasso to find nonsmooth coefficient functions, with the fusion penalty in particular that imposes the piecewise constant behaviour, AATR achieves similar results by smooth $L_{2}$ shrinkage towards a nonsmooth shape. This is clear by looking at Figure \ref{fig:comparison_1rect_dep} for instance, where AATR recovers a shape that is a rectified version of the one resulting from the roughness penalized model or the ridge, keeping a smooth behaviour in the artifacts outside of the rectangular part, where instead the fused lasso produces the same artifacts, but in a piecewise constant fashion. A similar argument can be held for the case in which the true $\beta$ is smooth and not sparse, where as expected the roughness penalty performs best. Moreover, similarly to the "elastic" methods that are able to produce both smooth and sparse solutions, AATR will produce sparse solutions only if the shape $\gamma$ itself is sparse. On the downside, the main limitation of AATR is its high reliance on the heuristic approximation of $\gamma$, for both the initialization and the subsequent iterations. While finding a suitable initial value can drastically reduce the number of alternating iterations, often to a single one as we empirically observed in our experiments, it is also true that the number of variables in the global optimization problem grows (linearly) with the number of basis $q$, making our approach viable only for small values of $q$.

\begin{table}[H]
\centering
\caption{\label{tab:simulations} Regression results for all the simulations: mean-square error}
\scalebox{.6}{
\begin{tabular}{l D{,}{\, \pm \,}{-1} D{,}{\, \pm \,}{-1} D{,}{\, \pm \,}{-1} D{,}{\, \pm \,}{-1} D{,}{\, \pm \,}{-1} D{,}{\, \pm \,}{-1} D{,}{\, \pm \,}{-1} D{,}{\, \pm \,}{-1}}
\toprule
\midrule
           
 & \multicolumn{4}{c}{independent spline coefficients} & \multicolumn{4}{c}{dependent spline coefficients} \\
\midrule

 & \multicolumn{1}{c}{$q$=1} & \multicolumn{1}{c}{$q$=2} & \multicolumn{1}{c}{$q$=3} & \multicolumn{1}{c}{\textit{smooth}} & \multicolumn{1}{c}{$q$=1} & \multicolumn{1}{c}{$q$=2} & \multicolumn{1}{c}{$q$=3} & \multicolumn{1}{c}{\textit{smooth}}\\
           
\midrule

mnlstsq     & 1.857,.81 & 1.833,.78  & 1.567,.54 & 1.848,.80  & 1.483,.62 & 1.483,.62  & 1.483,.62 & 1.483,.62  \\

ridge       & 1.343,.45 & 1.309,.38  & 1.338,.46 & 1.308,.34  & 1.063,.11 & 1.132,.20  & 1.141,.23 & 1.083,.17  \\

roughness   & 1.235,.35 & 1.399,.21  & 1.335,.27 & 1.183,.33  & 1.036,.03 & 1.145,.14  & 1.115,.16 & 0.948,.01  \\
  
lasso       & 0.999,.31 & 1.225,.49  & 1.435,.50 & 1.307,.36  & 1.029,.02 & 1.093,.16  & 1.196,.18 & 1.140,.21  \\
  
elastic net & 0.999,.31 & 1.222,.49  & 1.327,.48 & 1.244,.30  & 1.028,.02 & 1.087,.15  & 1.119,.11 & 1.070,.14  \\

elastic SCAD & 1.317,.14 & 2.867,.72  & 4.678,1.4 & 6.598,1.4  & 2.552,.59 & 7.938,.38  & 15.51,3.8 & 5.295,1.1  \\

elastic MCP & 1.282,.51 & 1.993,.67  & 3.721,1.3 & 3.662,1.2  & 1.923,.43 & 5.957,.42  & 14.27,3.4 & 5.634,1.7  \\
  
fused lasso & 0.967,.28 & 1.041,.19  & 1.278,.23 & 1.275,.33  & 1.017,.05 & 0.940,.04  & 1.048,.08 & 1.040,.05  \\ 

AATR        & 1.019,.24 & 1.052,.29  & 1.253,.28 & 1.344,.43  & 0.975,.05 & 0.960,.08  & 1.012,.01 & 1.011,.06  \\ 
\midrule
\bottomrule  
\end{tabular}}
\end{table}

\subsection{Solar radiation}
\begin{figure}[H]
  \centering
    \begin{subfigure}[c]{0.33\textwidth}
      \includegraphics[width=\textwidth]{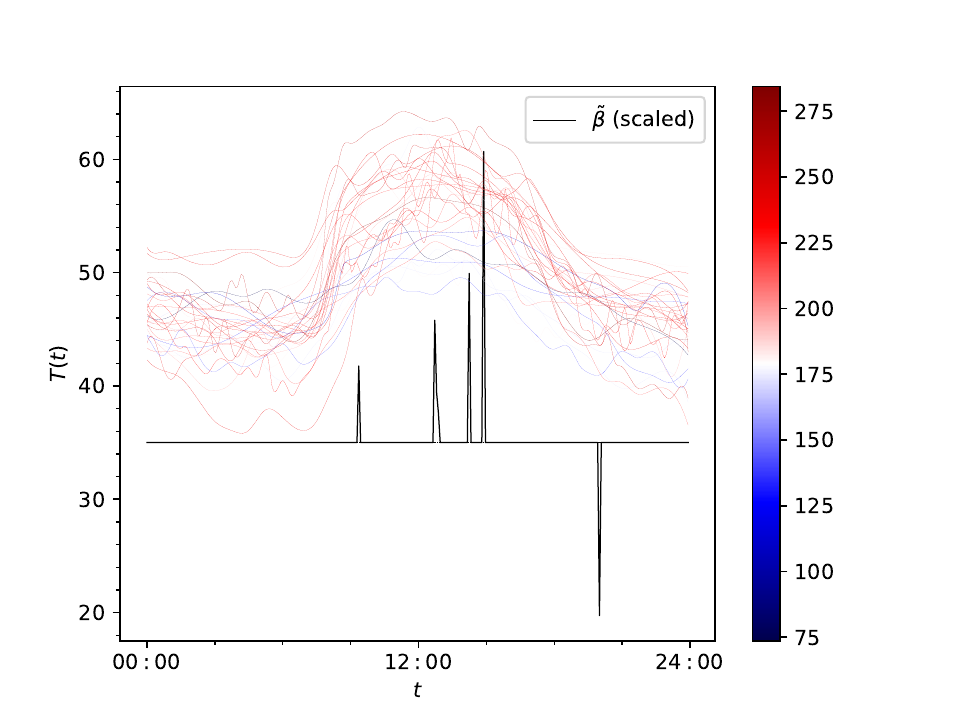}
      \caption{lasso}
    \end{subfigure}
    \begin{subfigure}[c]{0.33\textwidth}
      \includegraphics[width=\textwidth]{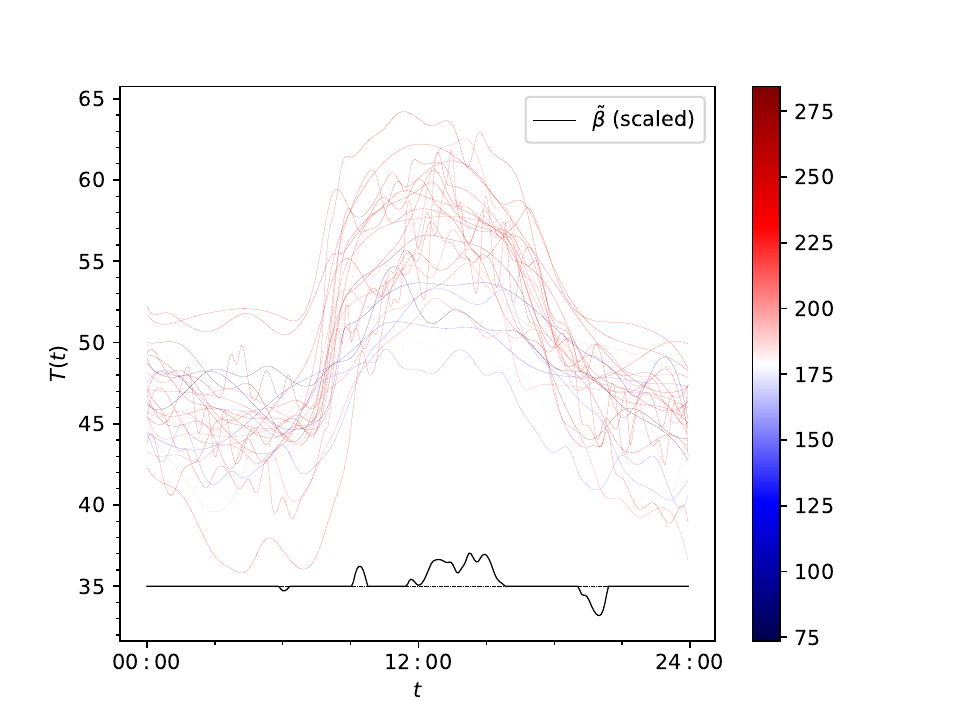}
      \caption{elastic net}
    \end{subfigure}
    \begin{subfigure}[c]{0.33\textwidth}
      \includegraphics[width=\textwidth]{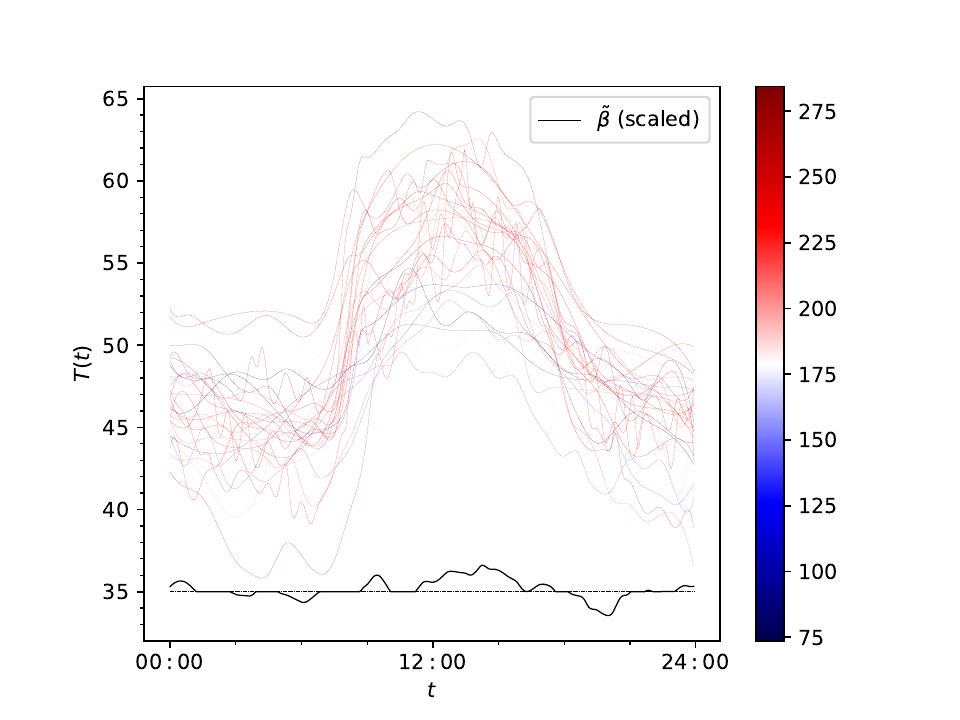}
      \caption{elastic SCAD}
    \end{subfigure}
    \begin{subfigure}[c]{0.33\textwidth}
      \includegraphics[width=\textwidth]{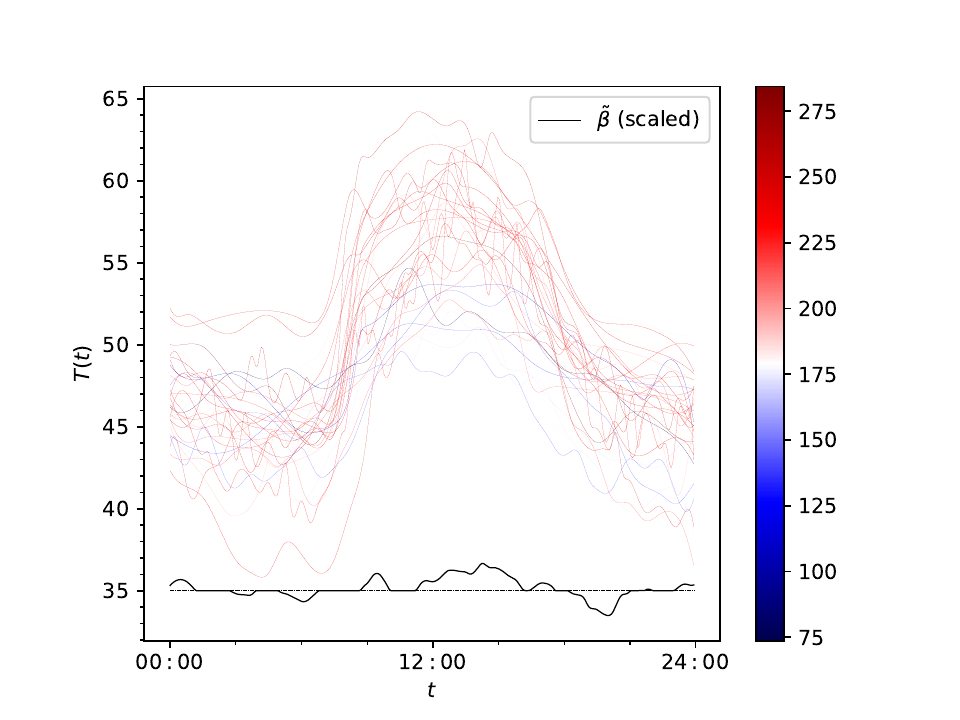}
      \caption{elastic MCP}
    \end{subfigure}
    \begin{subfigure}[c]{0.33\textwidth}
      \includegraphics[width=\textwidth]{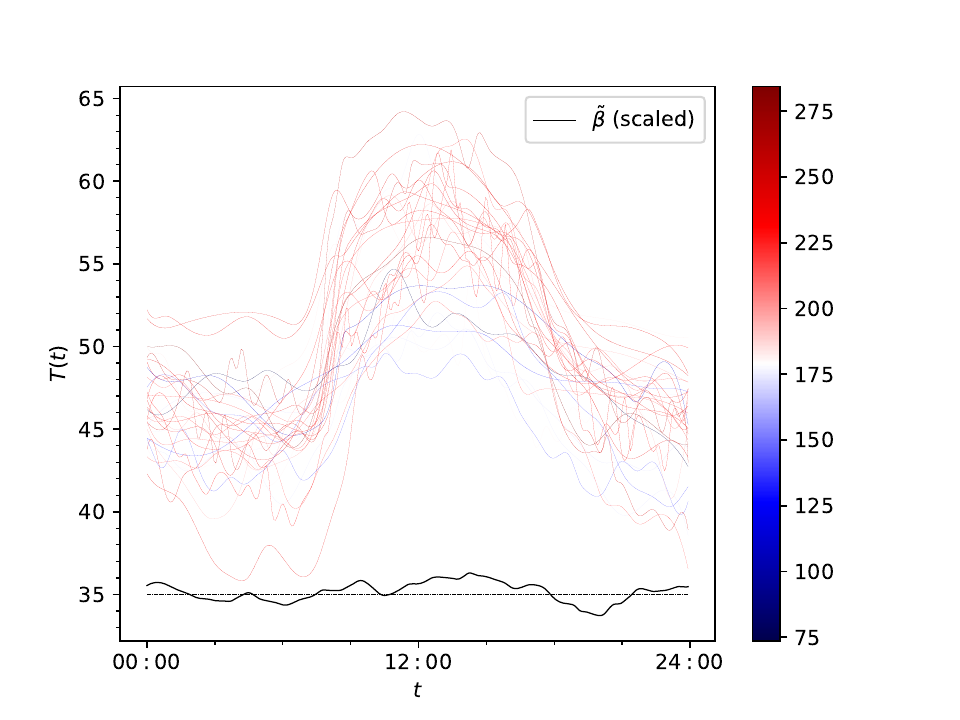}
      \caption{ridge}
    \end{subfigure}
    \begin{subfigure}[c]{0.33\textwidth}
      \includegraphics[width=\textwidth]{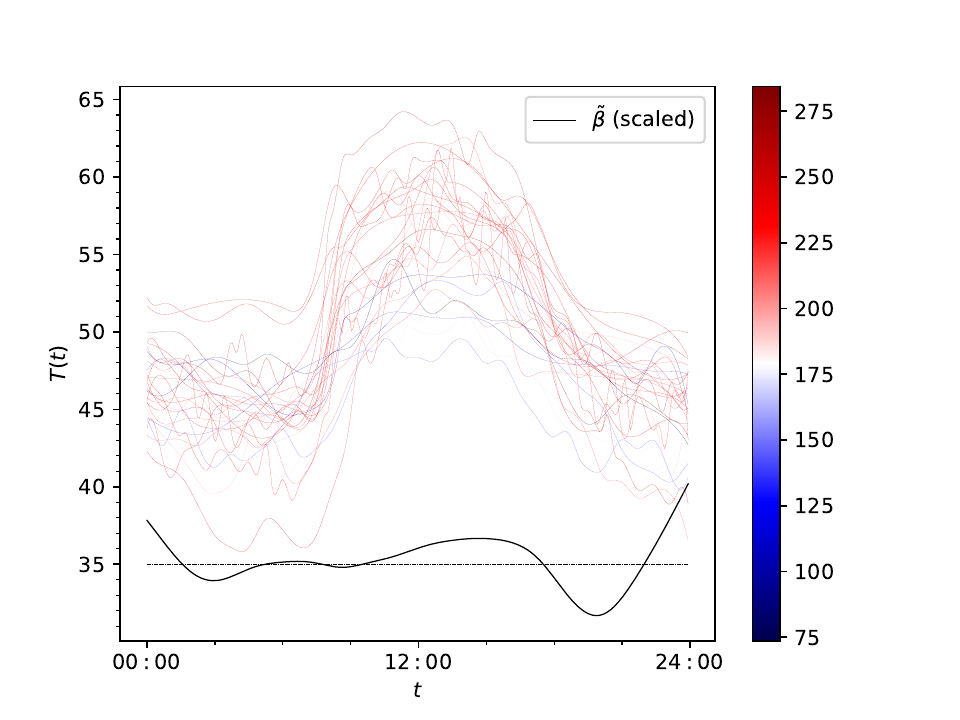}
      \caption{roughness}
    \end{subfigure}
    \begin{subfigure}[c]{0.33\textwidth}
      \includegraphics[width=\textwidth]{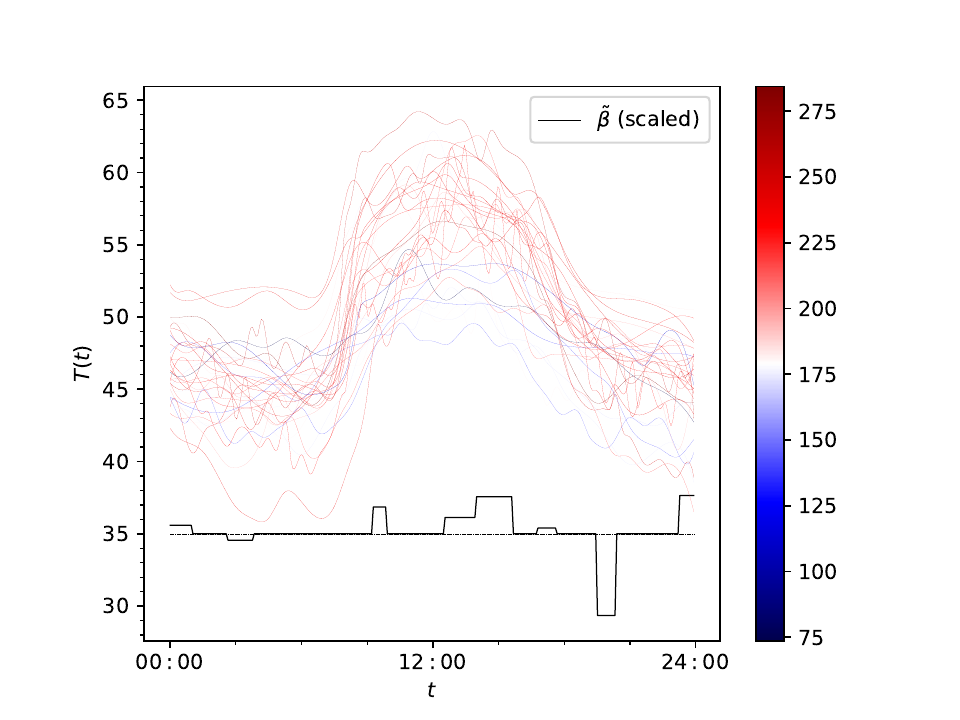}
      \caption{fused lasso}
    \end{subfigure}
    \begin{subfigure}[c]{0.33\textwidth}
      \includegraphics[width=\textwidth]{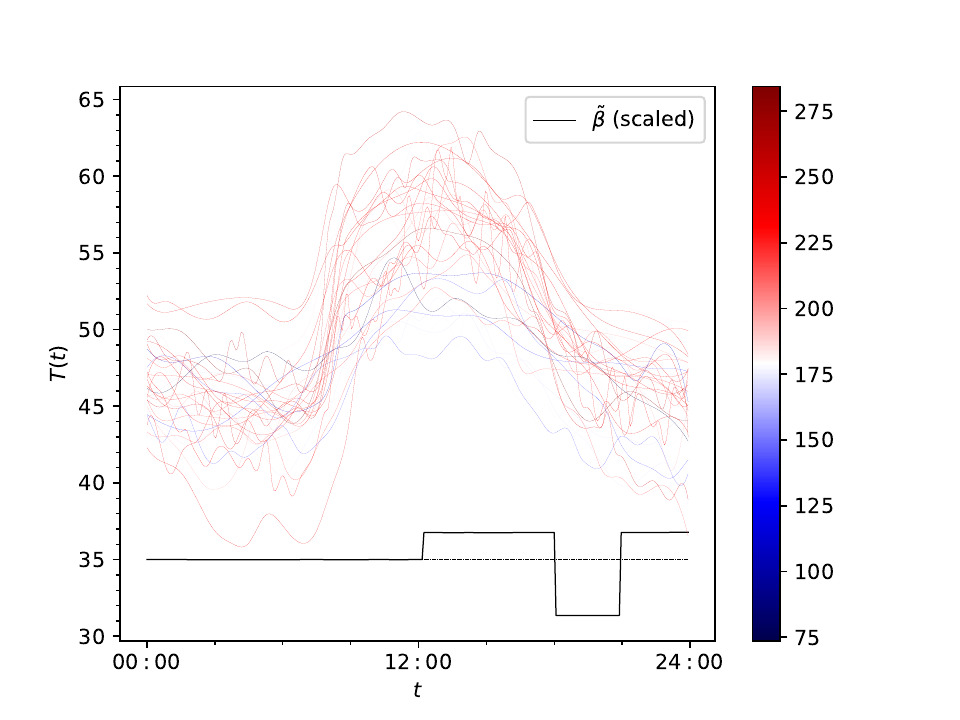}
      \caption{AATR}
    \end{subfigure}
  \caption{Solar radiation: comparison of the fitted coefficient functions $\tilde{\beta}$ scaled with respect to the temperature curves, the black dashed line represents the zero level for $\tilde{\beta}$}
  \label{fig:comparison_solar}
\end{figure}

\begin{figure}[H]
\begin{center}
\includegraphics[scale=0.5]{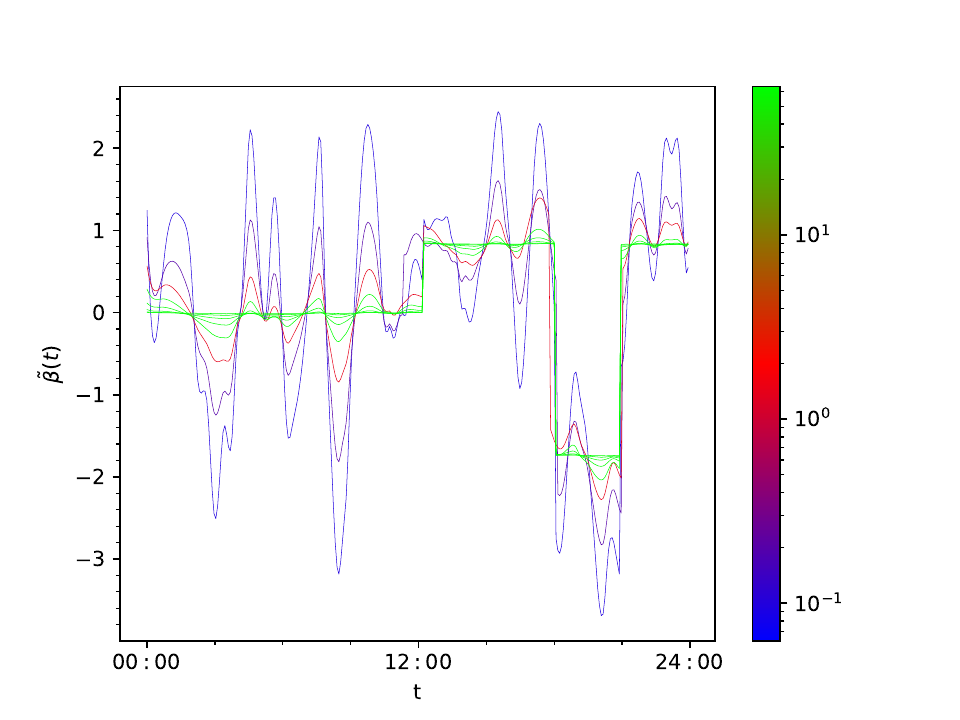}
\caption{Solar radiation: fitted $\tilde{\beta}$ obtained by solving Problem \ref{eq:L2} for increasing values of $\lambda$}
\label{fig:solar_by_lambda}
\end{center}
\end{figure}

\begin{figure}[H]
\begin{center}
\includegraphics[scale=0.5]{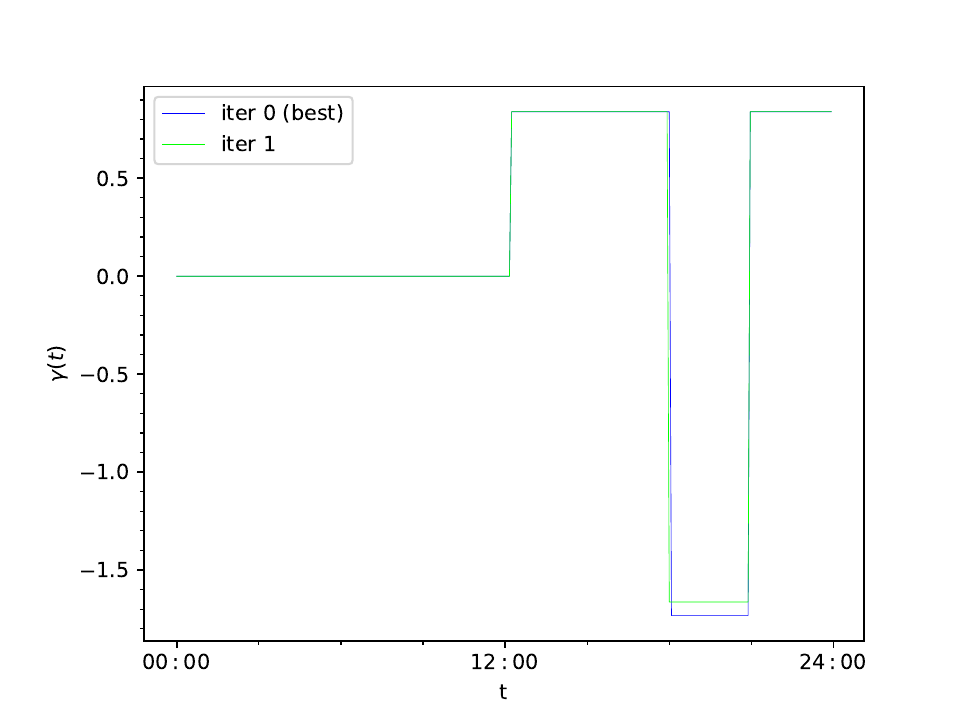}
\caption{Solar radiation: initial $\gamma$ resulting from Problem \ref{eq:hardinit} and subsequent values obtained by solving Problem \ref{eq:reshape}}
\label{fig:solar_by_gamma}
\end{center}
\end{figure}

The first real world application that we present belongs to the renewable energy/smart grid field, and is about predicting the mean daily solar radiation by looking at the daily temperature curves. The full data set is available on kaggle at \url{https://www.kaggle.com/dronio/SolarEnergy} and was originally provided by NASA through the Space Apps challenge. As the raw measurements are irregularly sampled during the day, with roughly one measurement every five minutes, we estimate the underlying daily temperature functions by using penalized free-knot cubic splines \citep{book_fittingsplines}, which are then evaluated on an equispaced grid of length $p$=$300$, while the raw solar radiation measurements are instead averaged over the day. The sample size is $N$=$39$ days and is obtained after removing some outliers that had constant temperature values over most of the day. The daily temperature profiles are shown with the fitted coefficient functions in Figure \ref{fig:comparison_solar}, where all methods seem to highlight some type of contrast between the temperature during the afternoon and the evening hours. Figure \ref{fig:solar_by_lambda} depicts the behaviour of $\beta$ in Problem \ref{eq:L2} for increasing values of $\lambda$, where as expected, higher values of $\lambda$ shrink $\beta$ towards the current $\gamma$, while Figure \ref{fig:solar_by_gamma} instead shows the progression of the different shapes $\gamma$ resulting from Problem \ref{eq:reshape} with $\lambda$ and $q$ fixed, where the algorithm stops after one iteration, as the initial value obtained from Problem \ref{eq:hardinit} is already the one with the lowest train error. Table \ref{tab:solar} reports the mean regression error on the test set obtained by 5-fold cross-validation, with nested grid search by 3-fold cross-validation on the current split for hyperparameter selection.

\begin{table}
\caption{Solar radiation: regression results, mean-square error}
\label{tab:solar}
\begin {center}
\begin{tabular}{l D{,}{\, \pm \,}{-1}}
\toprule
\midrule
mnlstsq     & 3701,1987  \\
  
ridge       & 1445,625  \\

roughness   & 1521,389  \\
  
lasso       & 1626,670  \\
  
elastic net & 1513,568  \\

elastic SCAD & 1383,578  \\

elastic MCP & 1378,567  \\
  
fused lasso & 1326,383  \\

AATR        & 958,231  \\
\midrule
\bottomrule
\end{tabular}
\end {center}
\end{table}

\subsection{London bike sharing}
\begin{figure}[H]
  \centering
    \begin{subfigure}[c]{0.33\textwidth}
      \includegraphics[width=\textwidth]{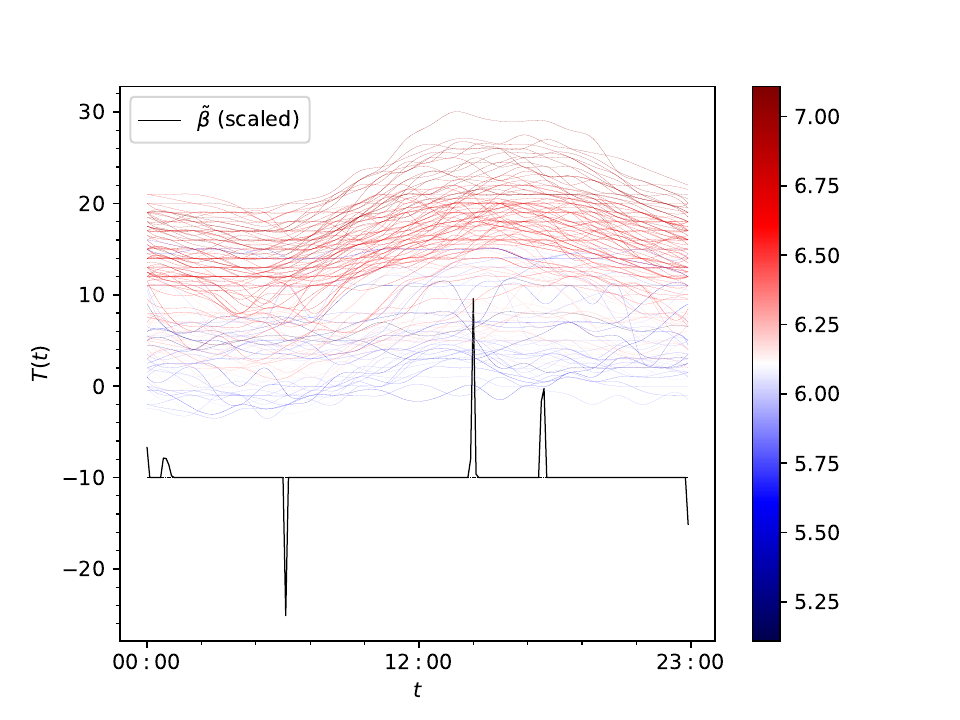}
      \caption{lasso}
    \end{subfigure}
    \begin{subfigure}[c]{0.33\textwidth}
      \includegraphics[width=\textwidth]{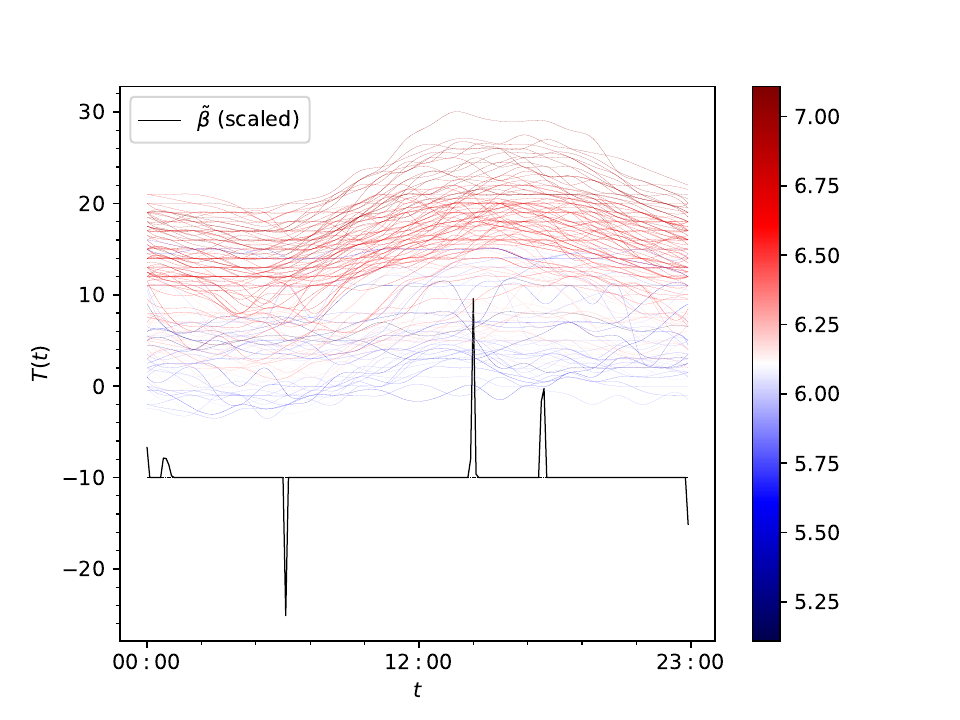}
      \caption{elastic net}
    \end{subfigure}
    \begin{subfigure}[c]{0.33\textwidth}
      \includegraphics[width=\textwidth]{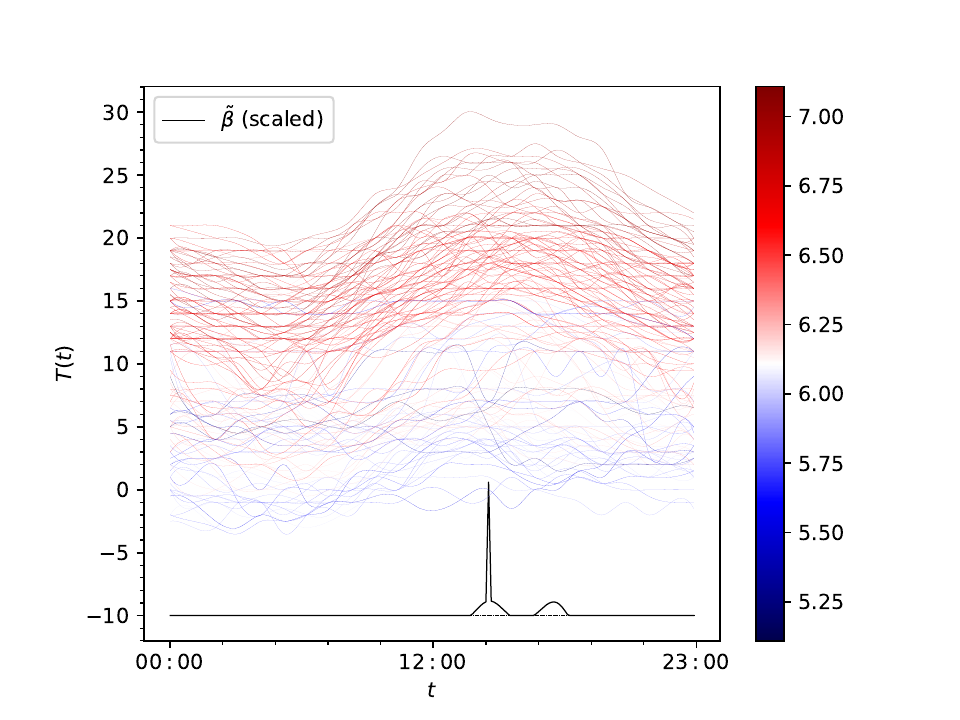}
      \caption{elastic SCAD}
    \end{subfigure}
    \begin{subfigure}[c]{0.33\textwidth}
      \includegraphics[width=\textwidth]{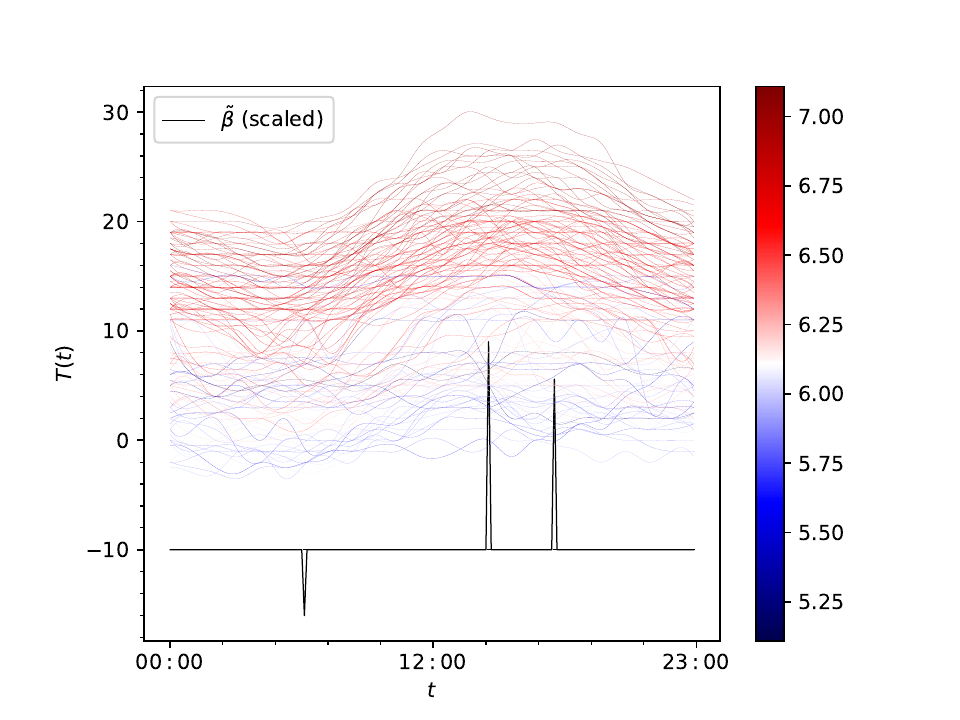}
      \caption{elastic MCP}
    \end{subfigure}
    \begin{subfigure}[c]{0.33\textwidth}
      \includegraphics[width=\textwidth]{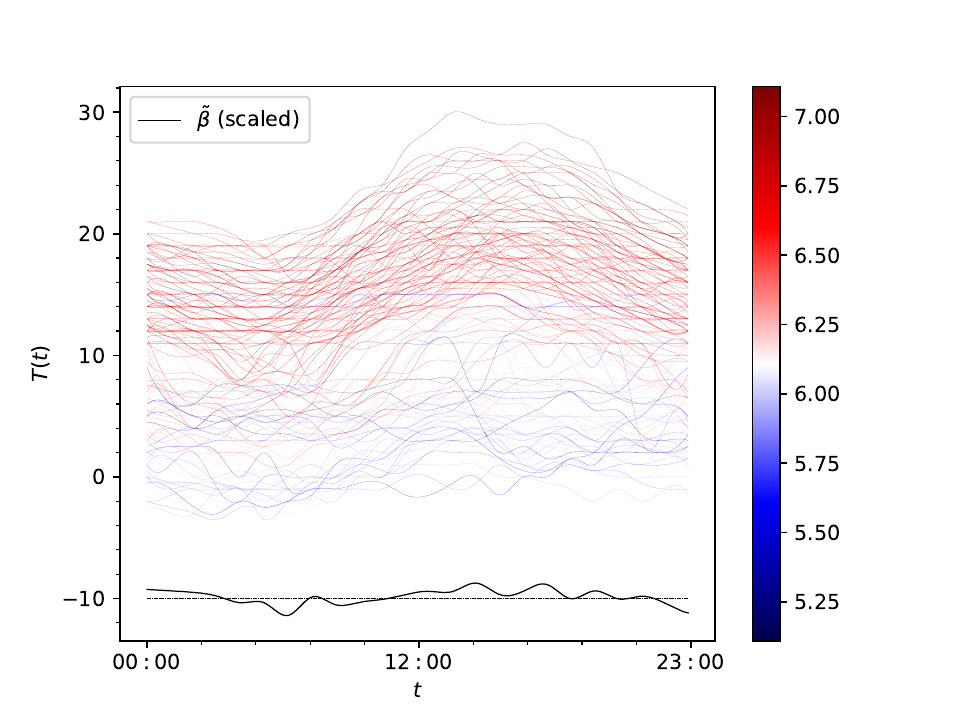}
      \caption{ridge}
    \end{subfigure}
    \begin{subfigure}[c]{0.33\textwidth}
      \includegraphics[width=\textwidth]{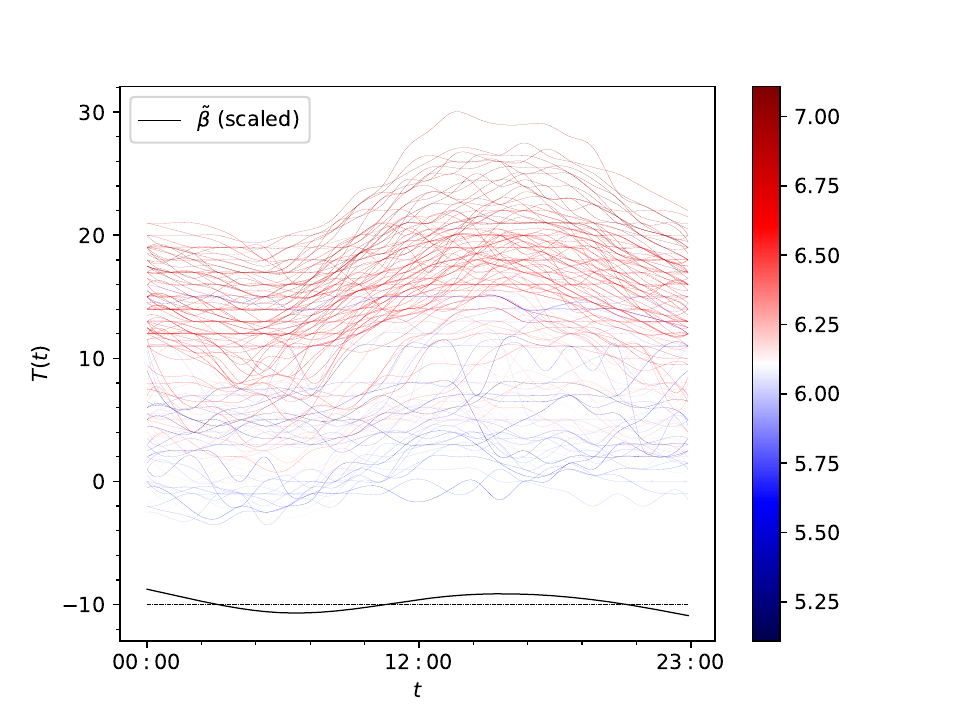}
      \caption{roughness}
    \end{subfigure}
    \begin{subfigure}[c]{0.33\textwidth}
      \includegraphics[width=\textwidth]{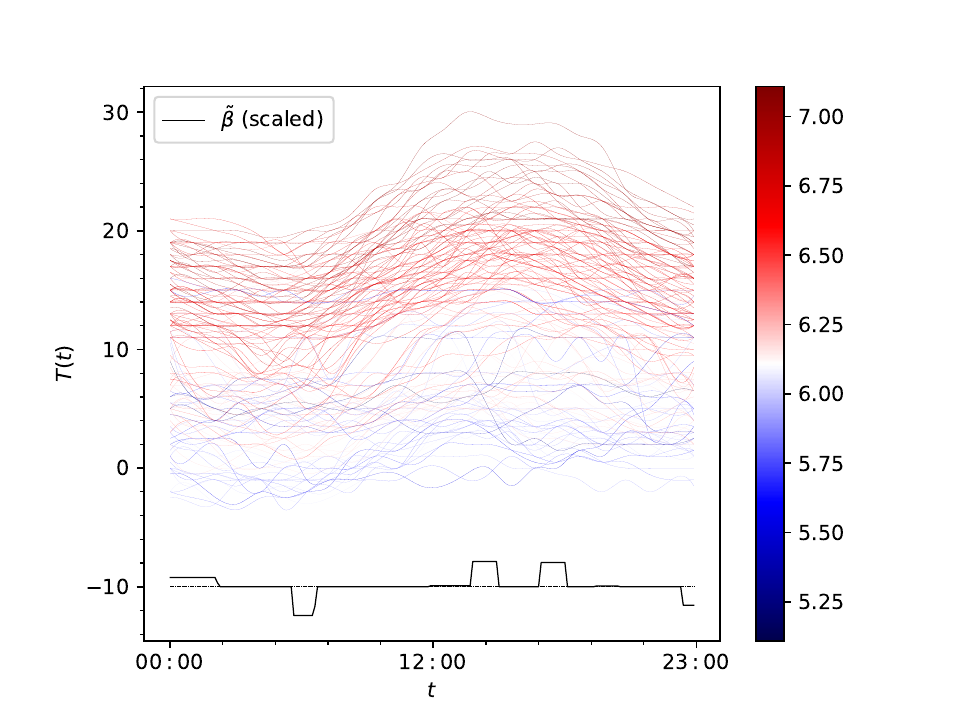}
      \caption{fused lasso}
    \end{subfigure}
    \begin{subfigure}[c]{0.33\textwidth}
      \includegraphics[width=\textwidth]{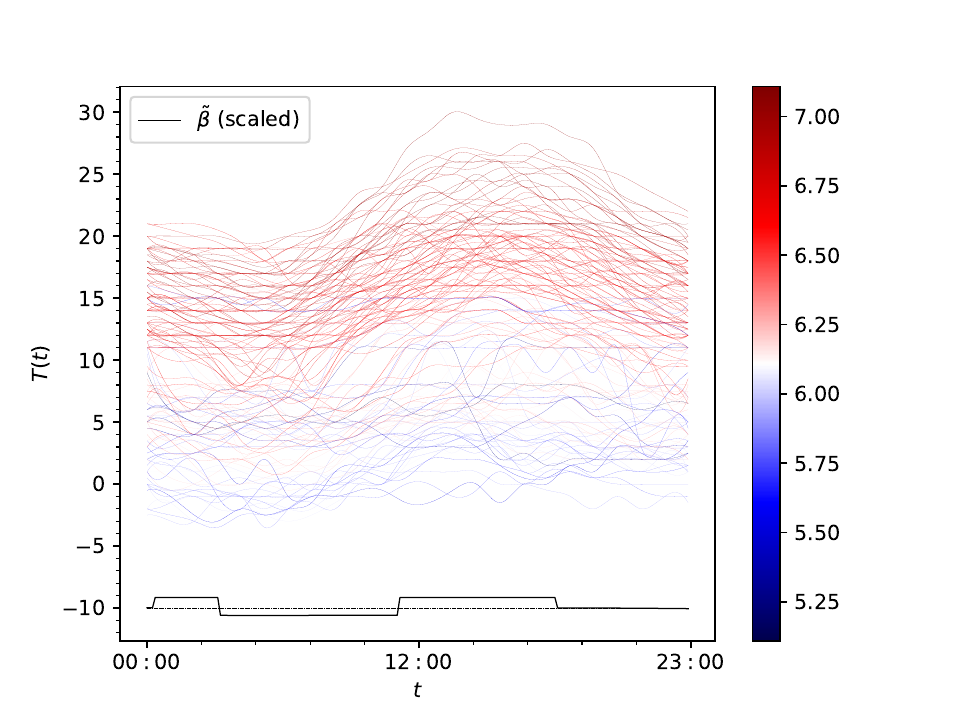}
      \caption{AATR}
    \end{subfigure}
  \caption{London bike sharing: comparison of the fitted coefficient functions $\tilde{\beta}$ scaled with respect to the temperature curves, the black dashed line represents the zero level for $\tilde{\beta}$}
  \label{fig:comparison_london}
\end{figure}

\begin{figure}[H]
\begin{center}
\includegraphics[scale=0.5]{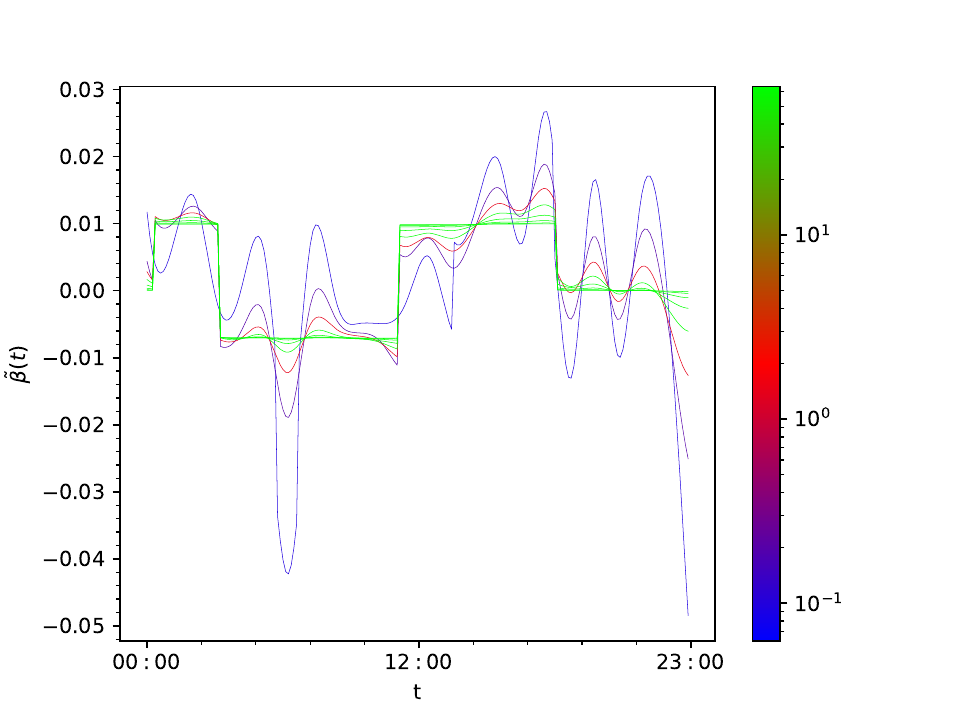}
\caption{London bike sharing: fitted $\tilde{\beta}$ obtained by solving Problem \ref{eq:L2} for increasing values of $\lambda$}
\label{fig:london_by_lambda}
\end{center}
\end{figure}

\begin{figure}[H]
\begin{center}
\includegraphics[scale=0.5]{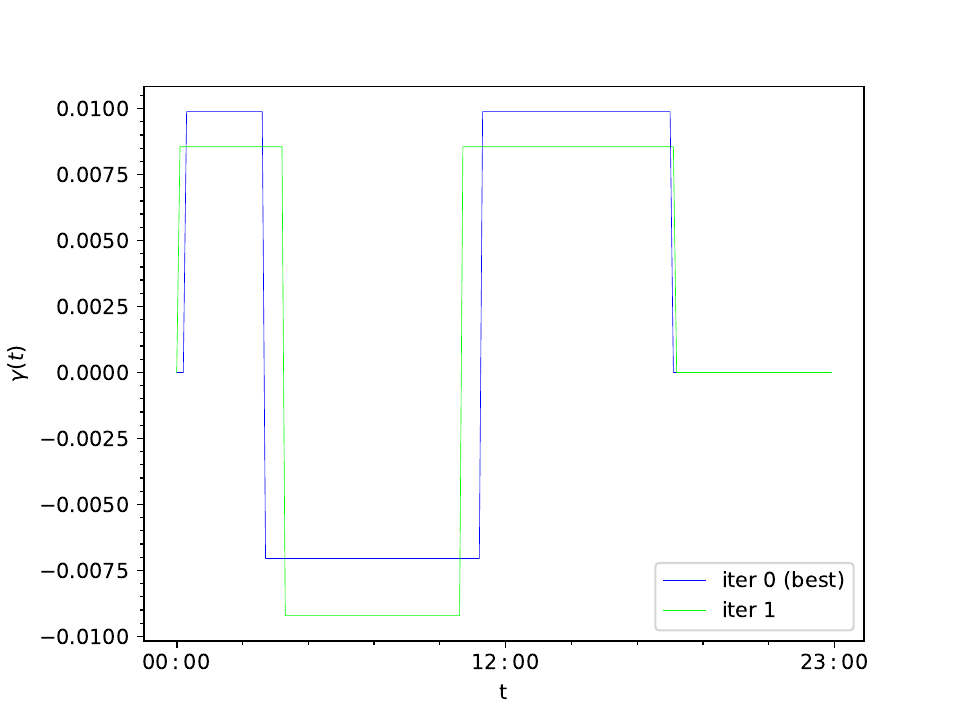}
\caption{London bike sharing: initial $\gamma$ resulting from Problem \ref{eq:hardinit} and subsequent values obtained by solving Problem \ref{eq:reshape}}
\label{fig:london_by_gamma}
\end{center}
\end{figure}

The second real world application is about one of London's bike sharing systems, and the full dataset is also available on kaggle at 
\url{https://www.kaggle.com/hmavrodiev/london-bike-sharing-dataset}. We focus on predicting the average daily (log) count of bikes by looking at the daily (feels like) temperature curves. The data is measured every hour and therefore each day has at most 24 points. As the sample size of the original dataset is large, we select only the weekend days where all the measurements are available, which results in $N$=$195$ days. Note that weekend and midweek days typically have very distinct renting patterns and therefore it makes sense to separate the data for this kind of analysis. The underlying temperature functions are estimated by cubic interpolating B-splines, which are then evaluated on the same equispaced grid of length $p$=$200$. As in the previous case study, the daily temperature profiles are shown with the scaled coefficient functions in Figure \ref{fig:comparison_london}, while Figure \ref{fig:london_by_lambda} illustrates the behaviour of $\beta$ in Problem \ref{eq:L2} for increasing values of $\lambda$, and Figure \ref{fig:london_by_gamma} depicts the progression of $\gamma$ with $\lambda$ and $q$ fixed. The results are also obtained by 5-fold cross-validation with the same grid search scheme for hyperparameter selection and are reported in Table \ref{tab:london}. It is worth to note that the two best performing methods, AATR and the roughness penalized model, are the ones that also show a coherent behaviour in their respective coefficient functions, despite the obvious difference in smoothness.

\begin{table}[H]
\caption{London bike sharing: regression results, mean-square error}
\label{tab:london}
\begin {center}
\begin{tabular}{l D{,}{\, \pm \,}{-1}}
\toprule
\midrule
mnlstsq     & .0931,.0397  \\
  
ridge       & .0506,.0213  \\

roughness   & .0488,.0194  \\
  
lasso       & .0523,.0236  \\
  
elastic net & .0514,.0218  \\

elastic SCAD & .0543,.0222  \\

elastic MCP & .0529,.0228  \\
  
fused lasso & .0496,.0207  \\

AATR        & .0486,.0187  \\
\midrule
\bottomrule
\end{tabular}
\end {center}
\end{table}

\section{Conclusions}
Shrinkage methods are widely used in high dimensional linear models in order to induce sparsity and perform variable selection between the the many regressors. In this work we proposed an $L_{2}$-based penalization algorithm for scalar on function linear regression models, where the coefficient function is shrunk towards a data-driven shape template $\gamma$. In particular, we focused on the case where $\gamma$ is the sum of $q$ rectangles, which results in a sparse and nonsmooth piecewise defined shape that is interpretable and well suited in the presence of dense and highly correlated variables. Finding the optimal knot placement of a piecewise function is a nonconvex problem, and to mitigate the computational burden, we proposed a parametrization that does not rely directly on the knot vector, reducing the number of variables in the global optimization problem, with the variables that control the height of the rectangles that are obtained in closed form. Our algorithm alternates between solving a convex $L_{2}$-based problem and finding an appropriate shape $\gamma$ with a differential evolution scheme, which is the main factor in determining the computational cost. While a sparse and nonsmooth $\beta$ is usually enforced through an $L_{1}$-based penalty, we suggest that if the shape $\gamma$ has those properties, the $L_{2}$ norm is also able to recover an adequate solution, as shown in multiple simulations and two real world case studies.

\section*{Acknowledgments}
Edoardo Belli was financially supported by the ABB-Politecnico di Milano Joint Research Center through the PhD scholarship \textit{"Development and prototyping of distributed control systems for electric networks based on advanced statistical models for the analysis of complex data"}.

\appendix
\section{}

\begin{proposition}
\label{closedform1}
Let $q\in \mathbb{N}^{+}$. For any fixed $t_{0}$=$\left(t_{01},\dots,t_{0q}\right)^{\top} \in [0,1]^{q}$ and $T$=$\left(T_{1},\dots,T_{q}\right)^{\top} \in(0,2]^{q}$, consider the objective function of Problem \ref{eq:hardinit} as a function of $A$=$\left(A_{1},\dots,A_{q}\right)^{\top} \in \mathbb{R}^q$: 

\begin{equation*}
\begin{aligned}
J(A) &= \sum_{i=1}^{N} \left[ y_{i} -\bar{y} - \int_{I}x_{i}(t)\sum_{j=1}^{q} A_{j}g\left(t,t_{0j},T_{j}\right) dt \right]^{2} \\
     &= \sum_{i=1}^{N} \left[ (y_{i} -\bar{y})^{2} + \left(\sum_{j=1}^{q} A_{j}\int_{I}x_{i}(t)g\left(t,t_{0j},T_{j}\right) dt \right)^{2} - 2(y_{i} -\bar{y})\sum_{j=1}^{q} A_{j}\int_{I}x_{i}(t)g\left(t,t_{0j},T_{j}\right) dt \right]
\end{aligned}
\end{equation*}
\\

\noindent let $S_{i} = \left(
\begin{array}{c}
\int_{I}x_{i}(t)g\left(t,t_{01},T_{1}\right) dt \\
\vdots \\
\int_{I}x_{i}(t)g\left(t,t_{0q},T_{q}\right) dt \\
\end{array} \right) \in \mathbb{R}^q \:$ for $i=1,\dots,N$, then:\\

\begin{equation*}
J(A)= \sum_{i=1}^{N} \left[ (y_{i} -\bar{y})^{2} + A^{\top}S_{i}S_{i}^{\top}A -2(y_{i} -\bar{y})A^{\top}S_{i} \right]
\end{equation*}
\\

\noindent differentiating with respect to $A$ and setting the first derivative to zero results in: 

\begin{equation*}
\frac{\partial J}{\partial A}= \sum_{i=1}^{N} \left[ 2S_{i}S_{i}^{\top}A -2(y_{i} -\bar{y})S_{i} \right] = 0
\end{equation*}
\\

\noindent let $\mathbf{\Sigma}_{SS}$=$\sum_{i=1}^{N}S_{i}S_{i}^{\top} \in \mathbb{R}^{q \times q} \:$, $\: \mathbf{\Sigma}_{yS}$=$\sum_{i=1}^{N}(y_{i} -\bar{y})S_{i} \in \mathbb{R}^{q}$, we finally obtain:

\begin{equation*}
\begin{aligned}
\mathbf{\Sigma}_{SS}A &= \mathbf{\Sigma}_{yS} \\
A &= \mathbf{\Sigma}_{SS}^{\dagger}\mathbf{\Sigma}_{yS}
\end{aligned}
\end{equation*}
\end{proposition}

\newpage

\begin{proposition}
\label{closedform2}
Let $q\in \mathbb{N}^{+}$, $\lambda \in \mathbb{R}^{+}$ and $\tilde{\beta}$ the solution of Problem \ref{eq:L2} from the previous step. For any fixed $t_{0}$=$\left(t_{01},\dots,t_{0q}\right)^{\top} \in [0,1]^{q}$ and $T$=$\left(T_{1},\dots,T_{q}\right)^{\top} \in(0,2]^{q}$, consider the objective function of Problem \ref{eq:reshape} as a function of $A$=$\left(A_{1},\dots,A_{q}\right)^{\top} \in \mathbb{R}^q$:

\begin{equation*}
\begin{aligned}
J(A) &= \sum_{i=1}^{N} \left[ y_{i} -\bar{y} - \int_{I}x_{i}(t)\sum_{j=1}^{q} A_{j}g\left(t,t_{0j},T_{j}\right) dt \right]^{2} + \lambda \int_{I} \Big[ \tilde{\beta}(t) - \sum_{j=1}^{q} A_{j}g\left(t,t_{0j},T_{j}\right) \Big]^{2}dt  \\
     &= \sum_{i=1}^{N} \left[ (y_{i} -\bar{y})^{2} + \left(\sum_{j=1}^{q} A_{j}\int_{I}x_{i}(t)g\left(t,t_{0j},T_{j}\right) dt \right)^{2} - 2(y_{i} -\bar{y})\sum_{j=1}^{q} A_{j}\int_{I}x_{i}(t)g\left(t,t_{0j},T_{j}\right) dt \right] \\
     & \hspace{0.5cm} + \lambda \int_{I} \Big[ \tilde{\beta}^{\,2}(t) + \left( \sum_{j=1}^{q} A_{j}g\left(t,t_{0j},T_{j}\right) \right)^{2} - 2\tilde{\beta}(t)\sum_{j=1}^{q} A_{j}g\left(t,t_{0j},T_{j}\right) \Big]dt
\end{aligned}
\end{equation*}
\\

\vspace{1cm}

\noindent let $S_{i} = \left(
\begin{array}{c}
\int_{I}x_{i}(t)g\left(t,t_{01},T_{1}\right) dt \\
\vdots \\
\int_{I}x_{i}(t)g\left(t,t_{0q},T_{q}\right) dt \\
\end{array} \right) \in \mathbb{R}^q \:$ for $i=1,\dots,N$ \\

\vspace{1cm}

\noindent let $\,G\!\!:\![-1,1] \rightarrow \mathbb{R}^{q} \:$ such that $\,G(t) = \left(
\begin{array}{c}
g\left(t,t_{01},T_{1}\right) \\
\vdots \\
g\left(t,t_{0q},T_{q}\right) \\
\end{array} \right)$ , then:\\

\vspace{1cm}

\begin{equation*}
\begin{aligned}
J(A) &= \sum_{i=1}^{N} \left[ (y_{i} -\bar{y})^{2} + A^{\top}S_{i}S_{i}^{\top}A -2(y_{i} -\bar{y})A^{\top}S_{i} \right] \\
& \hspace{0.5cm} + \lambda \int_{I} \Big[ \tilde{\beta}^{\,2}(t) + A^{\top}G(t){G(t)}^{\top}A - 2\tilde{\beta}(t)A^{\top}G(t) \Big]dt
\end{aligned}
\end{equation*}
\\

\noindent differentiating with respect to $A$ and setting the first derivative to zero results in: 

\begin{equation*}
\frac{\partial J}{\partial A}= \sum_{i=1}^{N} \left[ 2S_{i}S_{i}^{\top}A -2(y_{i} -\bar{y})S_{i} \right] + \lambda \int_{I} \Big[ 2G(t){G(t)}^{\top}A - 2\tilde{\beta}(t)G(t) \Big]dt = 0
\end{equation*}
\\

\noindent let $\mathbf{\Sigma}_{SS}$=$\sum_{i=1}^{N}S_{i}S_{i}^{\top} \in \mathbb{R}^{q \times q} \:$, $\: \mathbf{\Sigma}_{yS}$=$\sum_{i=1}^{N}(y_{i} -\bar{y})S_{i} \in \mathbb{R}^{q}$, $\mathbf{\Sigma}_{GG}$=$\int_{I}G(t){G(t)}^{\top}dt \in \mathbb{R}^{q \times q} \:$, $\: \mathbf{\Sigma}_{\beta G}$=$\int_{I}\tilde{\beta}(t)G(t)dt \in \mathbb{R}^{q}$, we finally obtain:

\vspace{1cm}

\begin{equation*}
\begin{aligned}
\left[ \mathbf{\Sigma}_{SS}+\lambda \mathbf{\Sigma}_{GG} \right] A &= \mathbf{\Sigma}_{yS} +\lambda \mathbf{\Sigma}_{\beta G} \\
A &= \left[ \mathbf{\Sigma}_{SS}+\lambda \mathbf{\Sigma}_{GG} \right]^{\dagger} \left[ \mathbf{\Sigma}_{yS} +\lambda \mathbf{\Sigma}_{\beta G} \right]
\end{aligned}
\end{equation*}
\end{proposition}

\clearpage

\bibliographystyle{elsarticle-harv} 
\bibliography{mypaper}

\begin{thebibliography}{68}
\expandafter\ifx\csname natexlab\endcsname\relax\def\natexlab#1{#1}\fi
\providecommand{\url}[1]{\texttt{#1}}
\providecommand{\href}[2]{#2}
\providecommand{\path}[1]{#1}
\providecommand{\DOIprefix}{doi:}
\providecommand{\ArXivprefix}{arXiv:}
\providecommand{\URLprefix}{URL: }
\providecommand{\Pubmedprefix}{pmid:}
\providecommand{\doi}[1]{\href{http://dx.doi.org/#1}{\path{#1}}}
\providecommand{\Pubmed}[1]{\href{pmid:#1}{\path{#1}}}
\providecommand{\bibinfo}[2]{#2}
\ifx\xfnm\relax \def\xfnm[#1]{\unskip,\space#1}\fi
\bibitem[{Arnold and Tibshirani(2020)}]{package_genlasso}
\bibinfo{author}{Arnold, T.B.}, \bibinfo{author}{Tibshirani, R.J.},
  \bibinfo{year}{2020}.
\newblock \bibinfo{title}{genlasso: Path algorithm for generalized lasso
  problems}.
\newblock
  \bibinfo{howpublished}{\url{https://CRAN.R-project.org/package=genlasso}}.
\newblock \bibinfo{note}{R package version 1.6}.
\bibitem[{Bilgrau et~al.(2020)Bilgrau, Peeters, Eriksen, Boegsted and van
  Wieringen}]{targetedridge}
\bibinfo{author}{Bilgrau, A.E.}, \bibinfo{author}{Peeters, C.F.},
  \bibinfo{author}{Eriksen, P.S.}, \bibinfo{author}{Boegsted, M.},
  \bibinfo{author}{van Wieringen, W.N.}, \bibinfo{year}{2020}.
\newblock \bibinfo{title}{Targeted fused ridge estimation of inverse covariance
  matrices from multiple high-dimensional data classes}.
\newblock \bibinfo{journal}{Journal of Machine Learning Research}
  \bibinfo{volume}{26}, \bibinfo{pages}{1--52}.
\bibitem[{de~Boor(1973)}]{goodapproxfreeknot}
\bibinfo{author}{de~Boor, C.}, \bibinfo{year}{1973}.
\newblock \bibinfo{title}{Good approximation by splines with variable knots}.
\newblock \bibinfo{journal}{Spline Functions and Approximation Theory}
  \bibinfo{volume}{1}, \bibinfo{pages}{57--72}.
\bibitem[{Boyd et~al.(2018)Boyd, Hastie, Boyd, Recht and Jordan}]{satursplines}
\bibinfo{author}{Boyd, N.}, \bibinfo{author}{Hastie, T.},
  \bibinfo{author}{Boyd, S.}, \bibinfo{author}{Recht, B.},
  \bibinfo{author}{Jordan, M.I.}, \bibinfo{year}{2018}.
\newblock \bibinfo{title}{Saturating splines and feature selection}.
\newblock \bibinfo{journal}{Journal of Machine Learning Research}
  \bibinfo{volume}{18}, \bibinfo{pages}{1--32}.
\bibitem[{Breheny(2015)}]{groupexpolasso}
\bibinfo{author}{Breheny, P.}, \bibinfo{year}{2015}.
\newblock \bibinfo{title}{The group exponential lasso for bi‐level variable
  selection}.
\newblock \bibinfo{journal}{Biometrics} \bibinfo{volume}{71},
  \bibinfo{pages}{731--740}.
\bibitem[{Breheny(2020)}]{package_ncvreg}
\bibinfo{author}{Breheny, P.}, \bibinfo{year}{2020}.
\newblock \bibinfo{title}{ncvreg: Regularization paths for scad and mcp
  penalized regression models}.
\newblock
  \bibinfo{howpublished}{\url{https://CRAN.R-project.org/package=ncvreg}}.
\newblock \bibinfo{note}{R package version 3.12.0}.
\bibitem[{Breiman(1991)}]{pmethod}
\bibinfo{author}{Breiman, L.}, \bibinfo{year}{1991}.
\newblock \bibinfo{title}{The {I}{I} method for estimating multivariate
  functions from noisy data}.
\newblock \bibinfo{journal}{Technometrics} \bibinfo{volume}{33},
  \bibinfo{pages}{125--143}.
\bibitem[{Brumback and Rice(1998)}]{fda_smoothsplnested}
\bibinfo{author}{Brumback, B.A.}, \bibinfo{author}{Rice, J.A.},
  \bibinfo{year}{1998}.
\newblock \bibinfo{title}{Smoothing spline models for the analysis of nested
  and crossed samples of curves}.
\newblock \bibinfo{journal}{Journal of the American Statistical Association}
  \bibinfo{volume}{93}, \bibinfo{pages}{961--976}.
\bibitem[{Cai and Hall(2006)}]{fda_pred}
\bibinfo{author}{Cai, T.T.}, \bibinfo{author}{Hall, P.}, \bibinfo{year}{2006}.
\newblock \bibinfo{title}{Prediction in functional linear regression}.
\newblock \bibinfo{journal}{The Annals of Statistics} \bibinfo{volume}{34},
  \bibinfo{pages}{2159--2179}.
\bibitem[{Campbell and Allen(2017)}]{reg_exclusivelasso}
\bibinfo{author}{Campbell, F.}, \bibinfo{author}{Allen, G.I.},
  \bibinfo{year}{2017}.
\newblock \bibinfo{title}{Within group variable selection through the exclusive
  lasso}.
\newblock \bibinfo{journal}{Electronic Journal of Statistics}
  \bibinfo{volume}{11}, \bibinfo{pages}{4220--4257}.
\bibitem[{Cardot et~al.(2007)Cardot, Crambes, Kneip and
  Sarda}]{fda_splineerrors}
\bibinfo{author}{Cardot, H.}, \bibinfo{author}{Crambes, C.},
  \bibinfo{author}{Kneip, A.}, \bibinfo{author}{Sarda, P.},
  \bibinfo{year}{2007}.
\newblock \bibinfo{title}{Smoothing splines estimators in functional linear
  regression with errors-in-variables}.
\newblock \bibinfo{journal}{Computational Statistics \& Data Analysis}
  \bibinfo{volume}{51}, \bibinfo{pages}{4832--4848}.
\bibitem[{Cardot et~al.(2003)Cardot, Ferraty and Sarda}]{fda_splineflm}
\bibinfo{author}{Cardot, H.}, \bibinfo{author}{Ferraty, F.},
  \bibinfo{author}{Sarda, P.}, \bibinfo{year}{2003}.
\newblock \bibinfo{title}{Spline estimators for the functional linear model}.
\newblock \bibinfo{journal}{Statistica Sinica} \bibinfo{volume}{13},
  \bibinfo{pages}{571--591}.
\bibitem[{Crambes et~al.(2009)Crambes, Kneip and Sarda}]{fda_smoothsplines}
\bibinfo{author}{Crambes, C.}, \bibinfo{author}{Kneip, A.},
  \bibinfo{author}{Sarda, P.}, \bibinfo{year}{2009}.
\newblock \bibinfo{title}{Smoothing splines estimators for functional linear
  regression}.
\newblock \bibinfo{journal}{The Annals of Statistics} \bibinfo{volume}{37},
  \bibinfo{pages}{35--72}.
\bibitem[{Dierckx(1993)}]{book_fittingsplines}
\bibinfo{author}{Dierckx, P.}, \bibinfo{year}{1993}.
\newblock \bibinfo{title}{Curve and Surface Fitting with Splines}.
\newblock \bibinfo{publisher}{Oxford University Press}, \bibinfo{address}{New
  York, NY, USA}.
\bibitem[{Eilers and Marx(1996)}]{psplines}
\bibinfo{author}{Eilers, P.H.}, \bibinfo{author}{Marx, B.D.},
  \bibinfo{year}{1996}.
\newblock \bibinfo{title}{Flexible smoothing with {B}-splines and penalties}.
\newblock \bibinfo{journal}{Statistical Science} \bibinfo{volume}{11},
  \bibinfo{pages}{89--121}.
\bibitem[{Fan and Li(2001)}]{scad}
\bibinfo{author}{Fan, J.}, \bibinfo{author}{Li, R.}, \bibinfo{year}{2001}.
\newblock \bibinfo{title}{Variable selection via nonconcave penalized
  likelihood and its oracle properties}.
\newblock \bibinfo{journal}{Journal of the American Statistical Association}
  \bibinfo{volume}{96}, \bibinfo{pages}{1348--1360}.
\bibitem[{Ferraty and Vieu(2006)}]{book_nonparafda}
\bibinfo{author}{Ferraty, F.}, \bibinfo{author}{Vieu, P.},
  \bibinfo{year}{2006}.
\newblock \bibinfo{title}{Nonparametric Functional Data Analysis: Theory and
  Practice}.
\newblock \bibinfo{publisher}{Springer}, \bibinfo{address}{New York, NY, USA}.
\bibitem[{Friedman(1991)}]{mars}
\bibinfo{author}{Friedman, J.H.}, \bibinfo{year}{1991}.
\newblock \bibinfo{title}{Multivariate adaptive regression splines}.
\newblock \bibinfo{journal}{The Annals of Statistics} \bibinfo{volume}{19},
  \bibinfo{pages}{1--141}.
\bibitem[{Friedman and Silverman(1989)}]{flexsmoothadd}
\bibinfo{author}{Friedman, J.H.}, \bibinfo{author}{Silverman, B.W.},
  \bibinfo{year}{1989}.
\newblock \bibinfo{title}{Flexible parsimonious smoothing and additive
  modeling}.
\newblock \bibinfo{journal}{Technometrics} \bibinfo{volume}{31},
  \bibinfo{pages}{3--21}.
\bibitem[{Gervini(2006)}]{fda_freeknotsplines}
\bibinfo{author}{Gervini, D.}, \bibinfo{year}{2006}.
\newblock \bibinfo{title}{Free-knot spline smoothing for functional data}.
\newblock \bibinfo{journal}{Journal of the Royal Statistical Society. Series B
  (Statistical Methodology)} \bibinfo{volume}{68}, \bibinfo{pages}{671--687}.
\bibitem[{Hannah and Dunson(2013)}]{cap_partit}
\bibinfo{author}{Hannah, L.A.}, \bibinfo{author}{Dunson, D.B.},
  \bibinfo{year}{2013}.
\newblock \bibinfo{title}{Multivariate convex regression with adaptive
  partitioning}.
\newblock \bibinfo{journal}{Journal of Machine Learning Research}
  \bibinfo{volume}{14}, \bibinfo{pages}{3261--3294}.
\bibitem[{Hastie et~al.(2009)Hastie, Tibshirani and Friedman}]{book_esl}
\bibinfo{author}{Hastie, T.}, \bibinfo{author}{Tibshirani, R.},
  \bibinfo{author}{Friedman, J.}, \bibinfo{year}{2009}.
\newblock \bibinfo{title}{The Elements of Statistical Learning}.
\newblock \bibinfo{publisher}{Springer}, \bibinfo{address}{New York, NY, USA}.
\bibitem[{Hastie et~al.(2015)Hastie, Tibshirani and Wainwright}]{book_sls}
\bibinfo{author}{Hastie, T.}, \bibinfo{author}{Tibshirani, R.},
  \bibinfo{author}{Wainwright, M.}, \bibinfo{year}{2015}.
\newblock \bibinfo{title}{Statistical Learning with Sparsity: The Lasso and
  Generalizations}.
\newblock \bibinfo{publisher}{CRC press}, \bibinfo{address}{New York, NY, USA}.
\bibitem[{Hebiri and van~de Geer(2011)}]{smoothlasso}
\bibinfo{author}{Hebiri, M.}, \bibinfo{author}{van~de Geer, S.},
  \bibinfo{year}{2011}.
\newblock \bibinfo{title}{The smooth-lasso and other
  $\ell_1$+$\ell_2$-penalized methods}.
\newblock \bibinfo{journal}{Electronic Journal of Statistics}
  \bibinfo{volume}{5}, \bibinfo{pages}{1184--1226}.
\bibitem[{Hoerl and Kennard(1970)}]{ridge}
\bibinfo{author}{Hoerl, A.E.}, \bibinfo{author}{Kennard, R.W.},
  \bibinfo{year}{1970}.
\newblock \bibinfo{title}{Ridge regression: Biased estimation for nonorthogonal
  problems}.
\newblock \bibinfo{journal}{Technometrics} \bibinfo{volume}{12},
  \bibinfo{pages}{55--67}.
\bibitem[{Jacob et~al.(2009)Jacob, Obozinski and Vert}]{grouplassooverlap}
\bibinfo{author}{Jacob, L.}, \bibinfo{author}{Obozinski, G.},
  \bibinfo{author}{Vert, J.P.}, \bibinfo{year}{2009}.
\newblock \bibinfo{title}{Group lasso with overlap and graph lasso}, in:
  \bibinfo{booktitle}{International Conference on Machine Learning}, pp.
  \bibinfo{pages}{433--440}.
\bibitem[{Jain and Kar(2017)}]{nonconvexoptml}
\bibinfo{author}{Jain, P.}, \bibinfo{author}{Kar, P.}, \bibinfo{year}{2017}.
\newblock \bibinfo{title}{Non-convex optimization for machine learning}.
\newblock \bibinfo{journal}{Foundations and Trends\textregistered $\,$ in
  Machine Learning} \bibinfo{volume}{10}, \bibinfo{pages}{142--363}.
\bibitem[{James(2011)}]{chapter_fdasparsity}
\bibinfo{author}{James, G.}, \bibinfo{year}{2011}.
\newblock \bibinfo{title}{Sparseness and functional data analysis}, in:
  \bibinfo{editor}{Ferraty, F.}, \bibinfo{editor}{Romain, Y.} (Eds.),
  \bibinfo{booktitle}{The Oxford Handbook of Functional Data Analysis}.
  \bibinfo{publisher}{Oxford University Press}, \bibinfo{address}{New York},
  pp. \bibinfo{pages}{298--323}.
\bibitem[{James et~al.(2000)James, Hastie and Sugar}]{fda_pcasparse}
\bibinfo{author}{James, G.M.}, \bibinfo{author}{Hastie, T.J.},
  \bibinfo{author}{Sugar, C.A.}, \bibinfo{year}{2000}.
\newblock \bibinfo{title}{Principal component models for sparse functional
  data}.
\newblock \bibinfo{journal}{Biometrika} \bibinfo{volume}{87},
  \bibinfo{pages}{587--602}.
\bibitem[{James et~al.(2009)James, Wang and Zhu}]{fda_flirti}
\bibinfo{author}{James, G.M.}, \bibinfo{author}{Wang, J.},
  \bibinfo{author}{Zhu, J.}, \bibinfo{year}{2009}.
\newblock \bibinfo{title}{Functional linear regression that's interpretable}.
\newblock \bibinfo{journal}{The Annals of Statistics} \bibinfo{volume}{37},
  \bibinfo{pages}{2083--2108}.
\bibitem[{Jupp(1978)}]{freeknots}
\bibinfo{author}{Jupp, D.L.B.}, \bibinfo{year}{1978}.
\newblock \bibinfo{title}{Approximation to data by splines with free knots}.
\newblock \bibinfo{journal}{SIAM Journal on Numerical Analysis}
  \bibinfo{volume}{15}, \bibinfo{pages}{328--343}.
\bibitem[{Kim et~al.(2009)Kim, Koh, Boyd and Gorinevsky}]{trendfilter}
\bibinfo{author}{Kim, S.J.}, \bibinfo{author}{Koh, K.}, \bibinfo{author}{Boyd,
  S.}, \bibinfo{author}{Gorinevsky, D.}, \bibinfo{year}{2009}.
\newblock \bibinfo{title}{$\ell_{1}$ trend filtering}.
\newblock \bibinfo{journal}{SIAM Review} \bibinfo{volume}{51},
  \bibinfo{pages}{339--360}.
\bibitem[{Lee and Park(2012)}]{fda_sparseest}
\bibinfo{author}{Lee, E.R.}, \bibinfo{author}{Park, B.U.},
  \bibinfo{year}{2012}.
\newblock \bibinfo{title}{Sparse estimation in functional linear regression}.
\newblock \bibinfo{journal}{Journal of Multivariate Analysis}
  \bibinfo{volume}{105}, \bibinfo{pages}{1--17}.
\bibitem[{Lindstrom(1999)}]{penfreeknot}
\bibinfo{author}{Lindstrom, M.J.}, \bibinfo{year}{1999}.
\newblock \bibinfo{title}{Penalized estimation of free-knot splines}.
\newblock \bibinfo{journal}{Journal of Computational and Graphical Statistics}
  \bibinfo{volume}{8}, \bibinfo{pages}{333--352}.
\bibitem[{Locatelli et~al.(2014)Locatelli, Maischberger and Schoen}]{delocal}
\bibinfo{author}{Locatelli, M.}, \bibinfo{author}{Maischberger, M.},
  \bibinfo{author}{Schoen, F.}, \bibinfo{year}{2014}.
\newblock \bibinfo{title}{Differential evolution methods based on local
  searches}.
\newblock \bibinfo{journal}{Computers \& Operations Research}
  \bibinfo{volume}{43}, \bibinfo{pages}{169--180}.
\bibitem[{Locatelli and Schoen(2016)}]{globolocal_survey}
\bibinfo{author}{Locatelli, M.}, \bibinfo{author}{Schoen, F.},
  \bibinfo{year}{2016}.
\newblock \bibinfo{title}{Global optimization based on local searches}.
\newblock \bibinfo{journal}{Annals of Operations Research}
  \bibinfo{volume}{240}, \bibinfo{pages}{251--270}.
\bibitem[{Magnani and Boyd(2009)}]{convex_pwlin}
\bibinfo{author}{Magnani, A.}, \bibinfo{author}{Boyd, S.P.},
  \bibinfo{year}{2009}.
\newblock \bibinfo{title}{Convex piecewise-linear fitting}.
\newblock \bibinfo{journal}{Optimization and Engineering} \bibinfo{volume}{10},
  \bibinfo{pages}{1--17}.
\bibitem[{Mammen and van~de Geer(1997)}]{lars}
\bibinfo{author}{Mammen, E.}, \bibinfo{author}{van~de Geer, S.},
  \bibinfo{year}{1997}.
\newblock \bibinfo{title}{Locally adaptive regression splines}.
\newblock \bibinfo{journal}{The Annals of Statistics} \bibinfo{volume}{25},
  \bibinfo{pages}{387--413}.
\bibitem[{Marx and Eilers(1999)}]{psplinesgenreg}
\bibinfo{author}{Marx, B.D.}, \bibinfo{author}{Eilers, P.H.},
  \bibinfo{year}{1999}.
\newblock \bibinfo{title}{Generalized linear regression on sampled signals and
  curves: A {P}-spline approach}.
\newblock \bibinfo{journal}{Technometrics} \bibinfo{volume}{41},
  \bibinfo{pages}{1--13}.
\bibitem[{Matsui and Konishi(2011)}]{fda_L1reg}
\bibinfo{author}{Matsui, H.}, \bibinfo{author}{Konishi, S.},
  \bibinfo{year}{2011}.
\newblock \bibinfo{title}{Variable selection for functional regression models
  via the ${L}_1$ regularization}.
\newblock \bibinfo{journal}{Journal of Multivariate Analysis}
  \bibinfo{volume}{55}, \bibinfo{pages}{3304--3310}.
\bibitem[{Meinshausen and B{\"u}hlmann(2006)}]{varsellassograph}
\bibinfo{author}{Meinshausen, N.}, \bibinfo{author}{B{\"u}hlmann, P.},
  \bibinfo{year}{2006}.
\newblock \bibinfo{title}{High-dimensional graphs and variable selection with
  the lasso}.
\newblock \bibinfo{journal}{The Annals of Statistics} \bibinfo{volume}{34},
  \bibinfo{pages}{1436--1462}.
\bibitem[{Osborne et~al.(1998)Osborne, Presnell and Turlach}]{knotselectlasso}
\bibinfo{author}{Osborne, M.R.}, \bibinfo{author}{Presnell, B.},
  \bibinfo{author}{Turlach, B.A.}, \bibinfo{year}{1998}.
\newblock \bibinfo{title}{Knot selection for regression splines via the lasso}.
\newblock \bibinfo{journal}{Computing Science and Statistics}
  \bibinfo{volume}{30}, \bibinfo{pages}{44--49}.
\bibitem[{Paterlini and Krink(2006)}]{declust}
\bibinfo{author}{Paterlini, S.}, \bibinfo{author}{Krink, T.},
  \bibinfo{year}{2006}.
\newblock \bibinfo{title}{Differential evolution and particle swarm
  optimisation in partitional clustering}.
\newblock \bibinfo{journal}{Computational Statistics \& Data Analysis}
  \bibinfo{volume}{50}, \bibinfo{pages}{1220--1247}.
\bibitem[{Pedregosa et~al.(2011)Pedregosa, Varoquaux, Gramfort, Michel,
  Thirion, Grisel, Blondel, Prettenhofer, Weiss, Dubourg, Vanderplas, Passos,
  Cournapeau, Brucher, Perrot and Duchesnay}]{scikit}
\bibinfo{author}{Pedregosa, F.}, \bibinfo{author}{Varoquaux, G.},
  \bibinfo{author}{Gramfort, A.}, \bibinfo{author}{Michel, V.},
  \bibinfo{author}{Thirion, B.}, \bibinfo{author}{Grisel, O.},
  \bibinfo{author}{Blondel, M.}, \bibinfo{author}{Prettenhofer, P.},
  \bibinfo{author}{Weiss, R.}, \bibinfo{author}{Dubourg, V.},
  \bibinfo{author}{Vanderplas, J.}, \bibinfo{author}{Passos, A.},
  \bibinfo{author}{Cournapeau, D.}, \bibinfo{author}{Brucher, M.},
  \bibinfo{author}{Perrot, M.}, \bibinfo{author}{Duchesnay, {\'E}.},
  \bibinfo{year}{2011}.
\newblock \bibinfo{title}{Scikit-learn: Machine learning in python}.
\newblock \bibinfo{journal}{Journal of Machine Learning Research}
  \bibinfo{volume}{12}, \bibinfo{pages}{2825--2830}.
\bibitem[{Pittman(2002)}]{adasplgenetic}
\bibinfo{author}{Pittman, J.}, \bibinfo{year}{2002}.
\newblock \bibinfo{title}{Adaptive splines and genetic algorithms}.
\newblock \bibinfo{journal}{Journal of Computational and Graphical Statistics}
  \bibinfo{volume}{11}, \bibinfo{pages}{615--638}.
\bibitem[{Pittman and Murthy(2000)}]{pwlga}
\bibinfo{author}{Pittman, J.}, \bibinfo{author}{Murthy, C.},
  \bibinfo{year}{2000}.
\newblock \bibinfo{title}{Fitting optimal piecewise linear functions using
  genetic algorithms}.
\newblock \bibinfo{journal}{IEEE transactions on pattern analysis and machine
  intelligence} \bibinfo{volume}{22}, \bibinfo{pages}{701--718}.
\bibitem[{Price et~al.(2015)Price, Geyer and Rothman}]{ridgefusion}
\bibinfo{author}{Price, B.S.}, \bibinfo{author}{Geyer, C.J.},
  \bibinfo{author}{Rothman, A.J.}, \bibinfo{year}{2015}.
\newblock \bibinfo{title}{Ridge fusion in statistical learning}.
\newblock \bibinfo{journal}{Journal of Computational and Graphical Statistics}
  \bibinfo{volume}{24}, \bibinfo{pages}{439--454}.
\bibitem[{Price et~al.(2005)Price, Storn and Lampinen}]{book_de}
\bibinfo{author}{Price, K.}, \bibinfo{author}{Storn, R.M.},
  \bibinfo{author}{Lampinen, J.A.}, \bibinfo{year}{2005}.
\newblock \bibinfo{title}{Differential evolution: a practical approach to
  global optimization}.
\newblock \bibinfo{publisher}{Springer}, \bibinfo{address}{Berlin, Germany}.
\bibitem[{Ramsay and Silverman(2005)}]{book_fda}
\bibinfo{author}{Ramsay, J.O.}, \bibinfo{author}{Silverman, B.W.},
  \bibinfo{year}{2005}.
\newblock \bibinfo{title}{Functional Data Analysis}.
\newblock \bibinfo{publisher}{Springer}, \bibinfo{address}{New York, NY, USA}.
\bibitem[{Rapin and Teytaud(2018)}]{nevergrad}
\bibinfo{author}{Rapin, J.}, \bibinfo{author}{Teytaud, O.},
  \bibinfo{year}{2018}.
\newblock \bibinfo{title}{Nevergrad - a gradient-free optimization platform}.
\newblock
  \bibinfo{howpublished}{\url{https://GitHub.com/FacebookResearch/Nevergrad}}.
\bibitem[{She(2010)}]{clusteredlasso}
\bibinfo{author}{She, Y.}, \bibinfo{year}{2010}.
\newblock \bibinfo{title}{Sparse regression with exact clustering}.
\newblock \bibinfo{journal}{Electronic Journal of Statistics}
  \bibinfo{volume}{4}, \bibinfo{pages}{1055--1096}.
\bibitem[{Simon et~al.(2013)Simon, Friedman, Hastie and
  Tibshirani}]{sparsegrouplasso}
\bibinfo{author}{Simon, N.}, \bibinfo{author}{Friedman, J.},
  \bibinfo{author}{Hastie, T.}, \bibinfo{author}{Tibshirani, R.},
  \bibinfo{year}{2013}.
\newblock \bibinfo{title}{A sparse-group lasso}.
\newblock \bibinfo{journal}{Journal of Computational and Graphical Statistics}
  \bibinfo{volume}{22}, \bibinfo{pages}{231--245}.
\bibitem[{Storn and Price(1997)}]{de}
\bibinfo{author}{Storn, R.}, \bibinfo{author}{Price, K.}, \bibinfo{year}{1997}.
\newblock \bibinfo{title}{Differential evolution --- a simple and efficient
  heuristic for global optimization over continuous spaces}.
\newblock \bibinfo{journal}{Journal of Global Optimization}
  \bibinfo{volume}{11}, \bibinfo{pages}{341--359}.
\bibitem[{Swindel(1976)}]{ridgeprior}
\bibinfo{author}{Swindel, B.F.}, \bibinfo{year}{1976}.
\newblock \bibinfo{title}{Good ridge estimators based on prior information}.
\newblock \bibinfo{journal}{Communications in Statistics - Theory and Methods}
  \bibinfo{volume}{5}, \bibinfo{pages}{1065--1075}.
\bibitem[{Tian and James(2013)}]{fda_interpdimred}
\bibinfo{author}{Tian, T.S.}, \bibinfo{author}{James, G.M.},
  \bibinfo{year}{2013}.
\newblock \bibinfo{title}{Interpretable dimension reduction for classifying
  functional data}.
\newblock \bibinfo{journal}{Computational Statistics \& Data Analysis}
  \bibinfo{volume}{57}, \bibinfo{pages}{282--296}.
\bibitem[{Tibshirani(1996)}]{lasso}
\bibinfo{author}{Tibshirani, R.}, \bibinfo{year}{1996}.
\newblock \bibinfo{title}{Regression shrinkage and selection via the lasso}.
\newblock \bibinfo{journal}{Journal of the Royal Statistical Society. Series B
  (Statistical Methodology)} \bibinfo{volume}{58}, \bibinfo{pages}{267--288}.
\bibitem[{Tibshirani et~al.(2005)Tibshirani, Saunders, Rosset, Zhu and
  Knight}]{fusedlasso}
\bibinfo{author}{Tibshirani, R.}, \bibinfo{author}{Saunders, M.},
  \bibinfo{author}{Rosset, S.}, \bibinfo{author}{Zhu, J.},
  \bibinfo{author}{Knight, K.}, \bibinfo{year}{2005}.
\newblock \bibinfo{title}{Sparsity and smoothness via the fused lasso}.
\newblock \bibinfo{journal}{Journal of the Royal Statistical Society. Series B
  (Statistical Methodology)} \bibinfo{volume}{67}, \bibinfo{pages}{91--108}.
\bibitem[{Tibshirani and Suo(2016)}]{orderedlasso}
\bibinfo{author}{Tibshirani, R.}, \bibinfo{author}{Suo, X.},
  \bibinfo{year}{2016}.
\newblock \bibinfo{title}{An ordered lasso and sparse time-lagged regression}.
\newblock \bibinfo{journal}{Technometrics} \bibinfo{volume}{58},
  \bibinfo{pages}{415--423}.
\bibitem[{Tibshirani(2014)}]{polyestimtrendfilter}
\bibinfo{author}{Tibshirani, R.J.}, \bibinfo{year}{2014}.
\newblock \bibinfo{title}{Adaptive piecewise polynomial estimation via trend
  filtering}.
\newblock \bibinfo{journal}{The Annals of Statistics} \bibinfo{volume}{42},
  \bibinfo{pages}{285--323}.
\bibitem[{van Wieringen(2019)}]{genridgeinvcov}
\bibinfo{author}{van Wieringen, W.N.}, \bibinfo{year}{2019}.
\newblock \bibinfo{title}{The generalized ridge estimator of the inverse
  covariance matrix}.
\newblock \bibinfo{journal}{Journal of Computational and Graphical Statistics}
  \bibinfo{volume}{28}, \bibinfo{pages}{932--942}.
\bibitem[{Yao et~al.(2005)Yao, M{\"u}ller and Wang}]{fda_sparselongdata}
\bibinfo{author}{Yao, F.}, \bibinfo{author}{M{\"u}ller, H.},
  \bibinfo{author}{Wang, J.}, \bibinfo{year}{2005}.
\newblock \bibinfo{title}{Functional data analysis for sparse longitudinal
  data}.
\newblock \bibinfo{journal}{Journal of the American Statistical Association}
  \bibinfo{volume}{100}, \bibinfo{pages}{577--590}.
\bibitem[{Yuan and Cai(2010)}]{fda_rkhs}
\bibinfo{author}{Yuan, M.}, \bibinfo{author}{Cai, T.T.}, \bibinfo{year}{2010}.
\newblock \bibinfo{title}{A reproducing kernel {H}ilbert space approach to
  functional linear regression}.
\newblock \bibinfo{journal}{The Annals of Statistics} \bibinfo{volume}{38},
  \bibinfo{pages}{3412--3444}.
\bibitem[{Yuan and Lin(2006)}]{grouplasso}
\bibinfo{author}{Yuan, M.}, \bibinfo{author}{Lin, Y.}, \bibinfo{year}{2006}.
\newblock \bibinfo{title}{Model selection and estimation in regression with
  grouped variables}.
\newblock \bibinfo{journal}{Journal of the Royal Statistical Society. Series B
  (Statistical Methodology)} \bibinfo{volume}{68}, \bibinfo{pages}{49--67}.
\bibitem[{Zhao et~al.(2009)Zhao, Rocha and Yu}]{cap_penalty}
\bibinfo{author}{Zhao, P.}, \bibinfo{author}{Rocha, G.}, \bibinfo{author}{Yu,
  B.}, \bibinfo{year}{2009}.
\newblock \bibinfo{title}{The composite absolute penalties family for grouped
  and hierarchical variable selection}.
\newblock \bibinfo{journal}{The Annals of Statistics} \bibinfo{volume}{37},
  \bibinfo{pages}{3468--3497}.
\bibitem[{Zhao et~al.(2010)Zhao, Rocha and Yu}]{mcp}
\bibinfo{author}{Zhao, P.}, \bibinfo{author}{Rocha, G.}, \bibinfo{author}{Yu,
  B.}, \bibinfo{year}{2010}.
\newblock \bibinfo{title}{Nearly unbiased variable selection under minimax
  concave penalty}.
\newblock \bibinfo{journal}{The Annals of Statistics} \bibinfo{volume}{38},
  \bibinfo{pages}{894--942}.
\bibitem[{Zhao and Yu(2006)}]{lassoirrepr}
\bibinfo{author}{Zhao, P.}, \bibinfo{author}{Yu, B.}, \bibinfo{year}{2006}.
\newblock \bibinfo{title}{On model selection consistency of lasso}.
\newblock \bibinfo{journal}{Journal of Machine Learning Research}
  \bibinfo{volume}{7}, \bibinfo{pages}{2541--2563}.
\bibitem[{Zhou et~al.(2010)Zhou, Jin and Hoi}]{multi_exclusivelasso}
\bibinfo{author}{Zhou, Y.}, \bibinfo{author}{Jin, R.}, \bibinfo{author}{Hoi,
  S.C.}, \bibinfo{year}{2010}.
\newblock \bibinfo{title}{Exclusive lasso for multi-task feature selection},
  in: \bibinfo{booktitle}{International Conference on Artificial Intelligence
  and Statistics}, pp. \bibinfo{pages}{988--995}.
\bibitem[{Zou and Hastie(2005)}]{elasticnet}
\bibinfo{author}{Zou, H.}, \bibinfo{author}{Hastie, T.}, \bibinfo{year}{2005}.
\newblock \bibinfo{title}{Regularization and variable selection via the elastic
  net}.
\newblock \bibinfo{journal}{Journal of the Royal Statistical Society. Series B
  (Statistical Methodology)} \bibinfo{volume}{67}, \bibinfo{pages}{301--320}.

\end{thebibliography}

\end{document}